\documentclass[showpacs,preprintnumbers,amsmath,amssymb,superscriptaddress]{revtex4}
\usepackage{amssymb}
\usepackage{mathrsfs}
\usepackage{graphicx}% input figure files
\usepackage{dcolumn}% Align table columns on decimal point
\usepackage{bm}% bold math
\usepackage{mathrsfs}

\newcommand{\vesc}{v_\textrm{esc}}

\newcommand{\vq}{v_\textrm{r}}
\newcommand{\vlab}{v_\textrm{lab}}
\newcommand{\mchi}{m_\chi}

\newcommand{\dOq}{{\rm d}\Omega_\textrm{r}}

\newcommand{\uhat}{\bhat{u}}
\newcommand{\qhat}{\bhat{r}}

\newcommand{\vhat}{\bhat{v}}

\newcommand {\be} {\begin {equation}}
\newcommand {\ee} {\end {equation}}

\newcommand {\bes} {\begin {equation*}}
\newcommand {\ees} {\end {equation*}}

\newcommand{\es}[2] {\begin{equation} \label{#1} \begin{split} #2 \end{split} \end{equation}}

\newcommand{\bvec}[1]{\mathbf{#1}}
\newcommand{\bhat}[1]{\hat{\bf #1}}

\begin{document}

\title{A review of the discovery reach of directional Dark Matter detection}

\author{F. Mayet}\affiliation{Laboratoire de Physique Subatomique et de Cosmologie, Universit\'e Grenoble Alpes, CNRS/IN2P3, Grenoble, France}
\author{A. M. Green}\affiliation{School of Physics and Astronomy, University of Nottingham, University Park, Nottingham, NG7 2RD, UK}
\author{J. B. R. Battat}\affiliation{Physics Department, Wellesley College, 106 Central Street, Wellesley, MA 02481, USA}
\author{J. Billard}\affiliation{IPNL, Universit\'e de Lyon, Universit\'e Lyon 1, CNRS/IN2P3, 4 rue E. Fermi 69622 Villeurbanne cedex, France}
\author{N. Bozorgnia}\affiliation{GRAPPA Institute, University of Amsterdam,\\ 
Science Park 904, 1098 XH Amsterdam, Netherlands}
\author{G. B. Gelmini}\affiliation{Department of Physics and Astronomy, University of California, Los Angeles
475 Portola Plaza, Los Angeles, CA 90095, USA}
\author{P. Gondolo}\affiliation{Dept. of Physics and Astronomy, University of Utah, 115 South 1400 East \#201, Salt Lake City, Utah 84112-0830}
\author{B. J. Kavanagh}\affiliation{Institut de Physique Th\'eorique, Universit\'e Paris-Saclay, CNRS, CEA, F-91191 Gif-sur-Yvette, France}
\affiliation{Sorbonne Universit\'es, UPMC Univ Paris 06, UMR 7589, LPTHE, F-75005, Paris, France}
\author{S. K. Lee}\affiliation{Princeton Center for Theoretical Science, Princeton University, Princeton, NJ 08544, USA}
\affiliation{Broad Institute, Cambridge, MA 02142, USA}
\author{D. Loomba}\affiliation{Physics \& Astronomy Dept., University of New Mexico, 1919 Lomas Blvd NE, Albuquerque, NM 87131 USA}
\author{J. Monroe}\affiliation{Department of Physics, Royal Holloway, University of London, Egham Hill, Surrey, TW20 0EX, UK}
\author{B. Morgan}\affiliation{Department of Physics, University of Warwick, Gibbet Hill Road, Coventry, CV4 7AL, UK}
\author{C.~A.~J. O'Hare}\affiliation{School of Physics and Astronomy, University of Nottingham, University Park, Nottingham, NG7 2RD, UK}
\author{A.~H.~G.~Peter}\affiliation{CCAPP and  Department of Physics, The Ohio State University, 191 W. Woodruff Ave.,  Columbus, OH 43210, USA}
\affiliation{Department of Astronomy, The Ohio State University, 140W. 18$^{th}$ Ave.,  Columbus, OH 43210, USA}
\author{N. S. Phan}\affiliation{Physics \& Astronomy Dept., University of New Mexico, 1919 Lomas Blvd NE, Albuquerque, NM 87131 USA}
\author{S.~E.~Vahsen}\affiliation{Department of Physics and Astronomy, University of Hawaii, 2505 Correa Road, Honolulu, HI 96822, USA}

%
% --------------ABSTRACT ---------------------- 
%
\date{\today}% It is always \today, today,

% Edited by FM 30/11/15 : modify "inverse of the direction of Solar motion to
%fit convention for skypmaps.
% Edited by FM 26/11/15 : minor corrections  
% Edited by FM 22/11/15 : minor corrections (following comments from SV)
\begin{abstract}
Cosmological observations indicate that most of the matter in the
  Universe is Dark Matter. Dark Matter in the form of Weakly
  Interacting Massive Particles (WIMPs) can be detected directly, via
  its elastic scattering off target nuclei. Most current direct detection
  experiments only measure the energy of the recoiling nuclei. However, 
  directional detection experiments are sensitive to the direction of
  the nuclear recoil as well. Due to the Sun's motion with respect to
  the Galactic rest frame, the directional recoil rate has a dipole
  feature, peaking around the   direction of the Solar
  motion. This provides a powerful tool for demonstrating the Galactic
  origin of nuclear recoils and hence unambiguously detecting Dark
  Matter.  Furthermore, the directional recoil distribution depends on
  the WIMP mass, scattering cross section and local velocity distribution. Therefore, 
  with a large number of recoil events it will be possible to study
  the physics of Dark Matter in terms of particle and 
  astrophysical properties. We review the
  potential of directional detectors for detecting and characterizing
  WIMPs.
\end{abstract}

\maketitle

% \begin{center}
%{\bf \huge Version 11a}
%\end{center}

\newpage
\tableofcontents

%
% -------------- INTRO
%1. Introduction : F. Mayet  
\newpage
% English polished by AMG 13/01/16
% Edited by FM 7/01/16 : last sentence removed (referee request)
% Edited by FM 3/12/15 : minor corrections (following comments from SV)
% Edited by FM 30/11/15 : minor corrections 
% Edited by FM 22/11/15 : minor corrections (following comments from SV)
% Edited by FM 16/11/15 : minor corrections (following comments from AMG and GBG)

% Edited by FM (10/11/15) - taking into account all comments 
% Edited by FM and JB (13/10/15) - taking into account comments (Bradley, Gariela, Julien and James)
% Edited by FM (30/07/15) - small text on the basic idea of directional feature (dipole). signal versus bckg
% Edited by FM (17/07/15) - minor

%
% -------------- INTRO
%1. Introduction : F. Mayet  (mayet@lpsc.in2p3.fr)
\section{Introduction}
\label{sec:intro}
A multitude of astrophysical and cosmological observations, involving
different physical processes at diverse scales, provide evidence for
the existence of non-baryonic Dark Matter in the Universe and, in
particular, in the halo of our Galaxy.  These observations also
provide hints to the properties of Dark Matter.  Dark
Matter makes up 27\% of the total mass-energy
density of the Universe \cite{planck}. This precise measurement constrains the Dark Matter production
mechanism in the early Universe and its interaction cross section.  The fact that halos of
Dark Matter surround galaxies and extend an order of magnitude in
distance beyond the visible component of the galaxies implies that
Dark Matter experiences at most weak dissipative interactions
\cite{Mandelbaum:2005nx,BoylanKolchin:2012xy}.
%It may interact at best weakly with Standard Model particles, otherwise Dark-Matter halos should glow in the sky.  
Since estimates of the Dark Matter density at the epoch of the cosmic microwave background are consistent with present-day 
observations, Dark Matter must have a lifetime larger than the Hubble
time.  Together these observations suggest that Dark Matter consists
of at least one quasi-stable particle, that
is not part of  the Standard Model of particle physics.
One of the best motivated particle Dark Matter candidates is a stable, 
massive, neutral particle generically referred to as a WIMP  (Weakly Interacting 
Massive Particle)~\cite{Steigman:1984ac,Jungman:1995df,Bertone:2004pz}.\\ 

The search for Dark Matter is at the interface between astrophysics
and particle physics. In astrophysics its gravitational influence 
appears at all scales (from galaxies to large-scale structures), while extensions of the 
Standard Model of particle physics suggest the existence of a particle
with the characteristics of a WIMP. 
Several WIMP detection techniques have been proposed and are being pursued.  
Searches for laboratory-created WIMPs are ongoing at the Large Hadron
Collider \cite{Halkiadakis:2014qda}, as are attempts to detect
Standard Model particles produced by WIMP annihilation 
in our own and extragalactic Dark Matter halos \cite{Porter:2011nv}.
Direct detection experiments aim to detect Galactic halo WIMPs streaming through terrestrial detectors, via their elastic
scattering off target nuclei.\\

Direct detection experiments are now approaching 
ton-scale in mass \cite{Baudis:2012ig}.  However, the sensitivity of these experiments to WIMPs is subject 
to both reducible (environmental) and irreducible (neutrino) backgrounds \cite{Billard:2013qya}, which  may mimic WIMPs.
To unambiguously claim a WIMP detection, it will thus be necessary to
demonstrate that the observed events are not due to  
backgrounds. Electron-induced recoils may be rejected efficiently, see
Ref.~\cite{Baudis:2012ig} for a review. However discriminating
neutron and WIMP-induced nuclear recoils remains a 
challenging issue. One potential tool to disentangle the WIMP signal from the neutron background is the annual modulation
signal~\cite{Drukier:1986tm,Freese:1987wu,Freese:2012xd}, which is 
due to the Earth's orbital motion about the Sun. The WIMP speed distribution in the
lab-frame, and hence the energy spectrum of WIMP-induced recoils, varies
annually. However, the modulation amplitude is small~\cite{Lewin:1995rx} and it is possible
that backgrounds also modulate annually.\\

The interest in directional detection is based on the unique directional signature of  Galactic 
Dark  Matter,  which enables  an  unambiguous  separation  of  signal  and 
backgrounds.  Since the Sun moves, with respect to the Galactic Dark Matter halo, 
in a nearly circular orbit with velocity $v_{\rm c} \approx 220 \, {\rm km \, s}^{-1}$~\cite{Bovy:2009dr,Bovy:2012ba},
the WIMP-induced  nuclear  recoil  spectrum  is  expected  to  be  sharply  peaked 
in the direction  of motion of the Sun.  This is 
generally referred to as the dipole feature (see Sec.~\ref{sec:featuresDipole}). 
The strength of directional detection lies in 
the fact that the is no reason for the neutron-induced  nuclear
recoil  spectrum to have a dipole anisotropy.
Directional detection requires the
measurement  of both the energy and the direction of the recoiling
nucleus and is a next generation strategy  for direct detection. \\

The idea of directional detection was first proposed by D.~N.~Spergel
in 1988~\cite{Spergel:1987kx}. Since then the interest in directional
detection has grown, due to technological advances that make the
construction of a large directional detector possible in the near
future.  In this review we focus on the discovery reach of directional
detection and address the following question: {\em What unique
  contributions can directional detection make in the search for Dark
  Matter?}  This is obviously a major issue, given the current
worldwide effort aimed at building a large directional detector and
the timescales required to do so.\\

In Sec.~\ref{sec:exp} we  briefly  introduce  the  key experimental issues that must be addressed
in order to perform a directional Dark Matter search. As the focus of this review is on 
theoretical ideas for directional searches, we only include the
technical details required to place the remaining sections in proper context.  
We refer the reader to Refs. \cite{mayet.cygnus,naka.cygnus,Ahlen:2009ev} for  a complete overview of the 
instrumental side of the subject. Section \ref{sec:theo} presents the relevant theoretical 
framework for both Galactic
and particle physics. In Sec.~\ref{sec:features} we present the main
directional feature, the so-called dipole, due to 
the Sun's motion through the Galactic halo and also two secondary directional features: the ring-like and aberration
features. Section \ref{sec:limits} studies directional detection as a tool to set 
limits on the WIMP-nucleon scattering rate in the absence of a
significant signal. The main power of directional detection, however,
is in the discovery of Dark Matter. Section \ref{sec:discovery} demonstrates 
that directional detectors can confirm the Galactic origin of recoils
at high significance, and hence make a definitive discovery of Dark
Matter. In particular we discuss how directionality can overcome the
floor in the sensitivity of conventional (non-directional) direct
detection experiments due to neutrino backgrounds.  In Sec.~\ref{sec:secondarysignatures} we study the prospects of observing the ring-like and aberration features. After the convincing discovery of WIMPs by 
at least one experiment, the next goal will be to constrain the properties  of both the WIMP particle (mass and cross section) and 
the Galactic Dark Matter halo (three dimensional local WIMP velocity distribution and density). 
This topic is addressed  in Sec.\ref{sec:identification}. The capability of a directional detector to probe 
the underlying particle physics model is investigated in
Sec.~\ref{sec:probingparticlemodel}. Motivated by recent
experimental and theoretical interest in light WIMPs, the potential of
directional detectors to detect light WIMPs is studied in Sec.~\ref{sec:light}. 
The use of directional detection to constrain the astrophysical properties of 
the Dark Matter velocity distribution, including substructures, is an exciting possibility that is presented 
in Sec.~\ref{sec:probinghalo}. Finally, in Sec.~\ref{sec:decomposing} we present 
a technique to decompose the WIMP velocity distribution, enabling a model-independent analysis of 
directional data.

%To focus the review, we work in the context of a supersymmetric WIMP with a standard halo velocity model 
%unless otherwise specified.  
%Where the particle physics is relevant, we also consider non-standard WIMP scattering and astrophysical models.

%
% -------------- Experimental
%1. Introduction : F. Mayet and J. Billard 
%\newpage
% Edited by FM 10/2/16 : comments from SV, JM
% Edited by FM 10/2/16 : comments from AP, AG, SV
% Edited by FM 09/2/16 : comments of Julien have been included.
% Edited by FM 29/1/16 : new subsection on current results of directional detectors
% Edited by FM 30/11/15 : modify "inverse of the direction of Solar motion to
%fit convention for skypmaps.
% Edited by FM 30/11/15 :  too many thetas and phis --> vartheta, varphi + typos
% Edited by FM 27/11/15 : minor corrections (following comments from JB) + add definition of lambda
% Edited by FM 22/11/15 : minor corrections (following comments from SV)
%08/11/15 : FM -->taking into account comments from AG, JB, GG, BK, JB, NP, ...

% -------------- theory 
%0 Experimental framework :  J. Billard and F. Mayet  (others welcome)
\section{Experimental framework}
\label{sec:exp}

\subsection{Introduction}
In this section, we briefly address the  key experimental issues for directional WIMP  detection. 
For a detailed review of the status of directional detectors, we refer the reader 
to Refs. \cite{Ahlen:2009ev, mayet.cygnus,naka.cygnus}.\\

Directional detection aims to measure both the energy and the direction of the recoiling
nuclei. Hence, it constitutes an extension of direction-insensitive
direct detection, for which only the energy  is measured. Several directional detection strategies have been
proposed: low pressure gaseous Time Projection Chambers (TPCs) 
\cite{Ahlen:2010ub,Lopez:2013ah,Daw:2010ud,Battat:2014mka,Battat:2014oqa,Battat:2014van,Vahsen:2011qx,Santos:2011kf,Miuchi:2010hn}, 
nuclear emulsions~\cite{DAmbrosio:2014wma,Naka:2013mwa}, 
columnar recombination in high-pressure gaseous  Xenon TPCs \cite{Mohlabeng:2015efa,nygren,Gehman:2013mra}, 
solid-state detectors 
\cite{Cavoto:2016lqo,Capella,Sekiya}, DNA-based detectors \cite{Drukier:2014rea}, 
 graphene-based heterostructures \cite{Wang:2015kya} and polarized detectors \cite{Chiang:2012ze}. Not all techniques are equally mature. 
Some are at the level of concepts or of advanced R\&D,  others have led to the construction of detectors that are already operating 
underground \cite{Miuchi:2010hn,Battat:2014van,Riffard:2015rga}. As for direction-insensitive detection, the exposure, and hence the mass of the detector, is a key issue to reach low cross sections. 
Directional detectors thus face the scale-up challenge, in addition to that of directional sensitivity. 
The cubic meter scale has been operated by the DRIFT detector \cite{Battat:2014van} and DMTPC, but significantly larger directional 
detectors must be constructed to reach leading sensitivity.

In the following, we discuss the  main experimental issues that directional detection faces: the track 
reconstruction (Sec.~\ref{sec:exp.track}), the energy measurement  (Sec.~\ref{sec:exp.energy}) and 
the residual background contamination (Sec.~\ref{sec:exp.bck}). To illustrate the current state of the art, 
we give examples of difficulties and achievements, mainly focusing on directional TPCs that currently constitute 
the most mature directional technology.

\subsection{Track reconstruction} 
\label{sec:exp.track}
The recoil track may be reconstructed either in 1d, 2d, or 3d, depending on the choice of the readout technology. 
Let $\bhat{r}$ be the recoil 
direction given by
\begin{equation}
\bhat{r} = \sin\vartheta\cos\varphi \, \bhat{x} + \sin\vartheta\sin\varphi \, \bhat{y} + \cos\vartheta \, \bhat{z}\, , 
\label{eq:angles} 
\end{equation}
where the  Earth-fixed $\bhat{x}$,  $\bhat{y}$, and $\bhat{z}$ axes point toward the North, the West and the Zenith directions  
respectively.
For a 1d track reconstruction, the track is projected along the $z$-axis and hence the $\vartheta$ angle is measured.
For a 2d track reconstruction,  $\varphi$ is measured, {\it e.g.} by  projecting the 
track onto the $(x,y)$ plane of the detector.  
A 3d track reconstruction requires that both the $\varphi$ and $\vartheta$ angles are  reconstructed. 
Note, that an equatorial mount is considered for some detectors \cite{asada}.
 
Directional TPCs are operated at low pressure so that the low energy recoil track is  long  enough to be detected on the readout plane.
 3d track reconstruction can be achieved with a time-sampling of the 2d-projection of the 
ionization-induced electron cloud on a pixelized readout plane.
Conversely, low pressure limits the achievable exposure, but small scale 3d-detectors, $\sim$0.1 kg, are already in operation \cite{Santos:2011kf}. 
1d-directional detection could be achieved with liquid or even solid-state detectors \cite{Capella,Sekiya}, 
hence {\it a priori} enhancing the exposure. However, no 1d liquid/solid detector has been operated and  there is yet no  experimental  evidence that columnar recombination is measurable for keV nuclear recoils.

The choice of the readout strategy has consequences on the Dark Matter reach, as studied in  Ref. \cite{Billard:2014ewa} and presented in Sec. \ref{sec:discovery}.\\

It is also desirable to retrieve the sense of the track of the recoiling nucleus ($+\bhat{r}$ vs. $-\bhat{r}$). 
Sense recognition may be achieved  thanks to 
an asymmetry between $+\bhat{r}$ and  $-\bhat{r}$ tracks due to:
\begin{enumerate}
\item[i)] the angular dispersion of recoiling tracks. It means that the beginning of the
track is  more rectilinear than its end \cite{Billard:2012bk}. Using this feature is challenging due to the angular dispersion, which  
increases with decreasing energy.
\item[ii)] charge collection asymmetry. Indeed,  ${\rm d}E/{\rm d}x$ decreases with energy at low recoil energies. 
Consequently, more primary electrons are generated at the beginning of the track. The effect is often referred to as the 
head-tail effect. Note that to measure the head/tail charge asymmetry, a readout that gives energy versus position on the track is
required.
\end{enumerate}
Although sense recognition has been demonstrated experimentally \cite{dujmic,burgos,majewski}, it remains a challenging issue for directional detection. 
This is because the intrinsic asymmetry in ${\rm d}E/{\rm d}x$ has significant fluctuations, and with a non-negligible 
angular dispersion of the recoil nucleus and diffusion of its 
track in the detector medium the ability to determine the sense is compromised. 
As such, directional data 
 are expected to  be only partially sense-recognized, with an efficiency  strongly depending 
on the energy and the drift distance \cite{Billard:2012bk}. With no sense recognition, the expected WIMP
distribution is a double dipole (see Sec.~\ref{sec:features}) that  is closer to the expected isotropic background event distribution  
and hence harder to discriminate \cite{Green:2007at}.\\

The main goal of the track reconstruction is the estimation of the initial recoil direction, as  
this is directly linked to the kinematics of the WIMP scattering. Once the recoiling nucleus encounters 
its first elastic scattering off another nucleus from the sensitive medium, the knowledge of the initial recoil direction starts to get lost. 
Eventually, the successive scatters give rise to  a non-negligible angular 
dispersion, usually referred to as the straggling effect, which constitutes an intrinsic limitation to directional detectors. For instance,  in the case of $\rm CF_4$ gas, F recoils can suffer from an intrinsic 
angular dispersion from straggling effect of about 25 deg (rms) at 
10 keV \cite{Billard.these}.
 
For low pressure gaseous TPCs, the main feature is transverse and longitudinal diffusion of the charge cloud during 
the drift \cite{christophorou} from the scattering vertex to the readout plane. For small drift distances, it 
constitutes an intrinsic limitation to the angular resolution. For long drift distances, the charge cloud is being so diluted that the
reconstruction of the track properties is no longer feasible. For a given instrumental configuration (gain, pressure), this effect sets a
limit on the drift distance, hence 
to the mass of the detector and consequently to the achievable exposure for a single module.
The choice of the target medium is thus critical. ${\rm CS}_2$ offers the possibility to drift negative ions,  which enables very long drift distances with minimal diffusion 
and hence large volumes \cite{drift.dist1,drift.dist2,drift.dist3,drift.dist4,drift.dist5,Snowden-Ifft:2014taa,Lewis:2014poa}. Directional electron-drift TPCs are using ${\rm
CF}_4$, for the spin content of ${\rm ^{19}F}$  that enables a sensitivity to spin dependent interaction \cite{ellis.spin}, and for the properties of this gas: electron drift velocity, diffusion and 
its scintillation output, which can be used for optical readout \cite{Ahlen:2010ub,Phan:2015pda}. ${\rm SF}_6$  is also being studied, as it may combine the advantages of  ${\rm CS}_2$ and ${\rm CF}_4$, {\it i.e.} a target with a non-vanishing spin 
in an electronegative gas \cite{sf6}. Note that the angular resolution of directional detectors must be evaluated through detector commissioning, using an ion beam or neutron field 
\cite{Golabek}, and accounted for in dedicated data analyses \cite{Billard:2012bk}.
 
The reconstruction of the localization of the vertex of the WIMP-nucleus 
 scattering  in the detector volume is also important in order to discriminate WIMP-induced recoils from surface events, which 
 mostly come 
from the radioactivity of the surrounding detector material \cite{drift.bck1,drift.bck2,drift.bck3,drift.bck4,Riffard:2015rga,Billard:2012bk}.\\\

The impact of track reconstruction (angular resolution, sense recognition) on the reach of directional detection is evaluated in Secs. \ref{sec:limits} and \ref{sec:discovery}.

\subsection{Energy measurement}
\label{sec:exp.energy}
The issue of energy measurement is not specific to direction-sensitive direct detection. In particular, 
the quenching of the deposited energy   must be accounted for to
convert a measured energy, usually expressed in electron-equivalent energy ({\it e.g.} $\rm keV_{ee}$), into a recoil 
energy $E_r$, expressed in recoil energy ($\rm keV_{r}$), see {\it e.g.} Ref. \cite{guillaudin}.  
As  for direction-insensitive direct detection, the energy threshold plays a key role for directional detection. 
It is even more challenging for directional detection, as in this case the energy threshold is the lowest energy at 
which both the initial recoil direction and the energy can be measured. In a low-pressure gaseous detector, 
a keV nuclear recoil has a short track length  and encounters a rather large diffusion during the drift to the readout plane. Hence, the directional energy threshold depends on the  choice of the 
target, the gas mixture, the signal strength (gas gain)  and   the read-out strategy.\\  

In the following, we present 
  the impact of the energy threshold on the capability of a directional detector to exclude 
  (Sec.~\ref{sec:exclusion}) or discover (Sec.~\ref{sec:discovery}) Dark Matter.

%When increasing the energy threshold, the Dark Matter reach is modified due to: a reduction of the expected number of   
%  WIMP events  and a increased relative sensitivity to the most anisotropic part of the WIMP induced recoil distribution.  Note that 
%  that these two effects (reduced event numbers and increased anisotropy) have opposing impacts on the DM reach. 

\subsection{Residual background contamination} 
\label{sec:exp.bck}
As for a direction-insensitive detector, the discrimination of background electron recoils 
from nuclear recoils is a key issue for a directional detector 
\cite{Billard:2012dy,Lopez:2012zwa,Lopez:2013ah,Battat:2014van}. 
However, as neutrons lead to a nuclear recoil, just as WIMPs do, they are usually considered as an 
 irreducible background for the direct detection of Dark Matter. The strength of directional detection lies in 
the fact that the WIMP signal exhibits a unique feature, as the WIMP-induced  nuclear  recoil  spectrum  is  expected  
to  be  peaked in the direction  of motion of the Sun, see  Sec.~\ref{sec:features}. Hence, one expects a WIMP dipole 
feature in Galactic coordinates, while there is no reason for the neutron-induced  nuclear  recoil  spectrum to present the same feature. 

Moreover, the neutron background is expected to be isotropically distributed. Neutrons in underground laboratories originate either from 
spontaneous fissions and $(\alpha,n)$ reactions, or from cosmic muon interactions. Whereas the former are associated with an isotropic
neutron distribution in the laboratory frame, the latter need a careful evaluation. 
D. Mei {\it et al.} have shown that the muon-induced neutron angular distribution is mainly flat with respect to $\cos \theta$, $\theta$
being the zenith angle, with a small forward excess due to the spallation process \cite{Mei:2005gm}. As noted in Ref. 
\cite{Morgan:2004ys}, a small anisotropy in the laboratory rest-frame will be erased when considering the Galactic rest-frame, due to the
Earth rotation and orbit. Moreover, any residual anisotropy would be further smeared as directional detection is sensitive to the induced recoils, and not the neutrons themselves.  
 Under the isotropy hypothesis, a likelihood analysis of directional data is possible 
\cite{Billard:2009mf}. In fact, the neutron background only has to be known and a full Monte Carlo evaluation of the neutron angular
distribution in underground laboratories   
would be a valuable contribution to the field, to assess the level of isotropy of the neutron background in Galactic coordinates. 
Note that the situation is similar for other backgrounds ($\gamma$, electrons), but they can be discriminated 
from nuclear recoils  \cite{Billard:2012dy,Lopez:2012zwa,Lopez:2013ah,Battat:2014van}, or included in a likelihood analysis.   
Eventually, we emphasize that the isotropy of the neutron background is not a {\it sine qua non} condition for directional detection. Only the WIMP
dipole feature is mandatory and it is due to the Sun's motion with respect to the Dark Matter halo.

In the following, we show that the directional detection reach is only mildly degraded by   a small background 
contamination (Sec.~\ref{sec:discovery}), thanks to the intrinsic  difference between the neutron- 
 and WIMP-induced angular recoil spectra. To quantify the level of residual background in the data, we define the 
 WIMP fraction as 
\begin{equation}
\lambda =
S/(B+S),
\label{eq:def.lambda}
\end{equation} 
where $S$ and $B$ are the number of signal and background events respectively.

Finally, the neutrino background is considered to be the ultimate background for the direct detection of Dark Matter \cite{Monroe:2007xp,Billard:2013qya,Gutlein:2010tq,Ruppin:2014bra}. 
In Sec. \ref{subsec-nu}, we present how to overcome the neutrino floor by using directionality.

\subsection{Current experimental landscape} 
For completeness, we conclude this section with an overview of the directional TPCs that are currently being operated underground, 
or planned to start running underground in the near future, in terms of detection strategies, achievements and future developments, 
see Tab. \ref{tab:exp.landscape}. We also present a short overview of current R\&D efforts toward spatial-resolution and/or higher signal-to-noise ratio.

%thesholds
%
%Drift keVrecoil (Daw et al.)
%MIMAC recoil
%Newage
%{\bf to be done : we must specify if threshold are in keVee or keVnr} 
%
%
% 
\subsubsection{DRIFT}
The DRIFT  collaboration has developed the first low-pressure gas TPC for directional Dark Matter detection \cite{SnowdenIfft:1999hz}.  
The key feature of the DRIFT detector  is the fact that negative ions ($\rm CS_2^-$) are used to transport the ionization signal to the
readout plane composed of Multi-Wire Proportional
Counters  (MWPC)  \cite{drift.dist5}. It opens the possibility to develop TPCs with very long drift distances with minimal diffusion 
and hence large volumes \cite{drift.dist1,drift.dist2,drift.dist3,drift.dist4,drift.dist5}. The gas mixture currently used is 
73\% $\rm CS_2$ + 25\% $\rm CF_4$ + 2\% $\rm O_2$ at a pressure of $55 \ {\rm mbar}$. $\rm CF_4$ is used to enable the sensitivity to SD
interaction, while the small $\rm O_2$ component \cite{{Snowden-Ifft:2013vua}} enables the measurement of the absolute position of the interaction vertex in the
drift direction \cite{Battat:2014van}. The achieved threshold is $20 \ {\rm keV}$ \cite{Daw:2010ud} and a 
factor of 2 higher in the analysis
presented in \cite{Battat:2014van}, due to the use of minority carriers. The DRIFT collaboration 
has demonstrated, through several R\&D stages and underground runs at the Boulby mine,   stable operation 
of a $\rm m^3$ detector and has recently published the first 
background-free operation of a directional dark matter detector, thanks 
to a  full-volume fiducialisation \cite{Battat:2014van}. The collaboration aims to   develop a directional detector at the 
ton scale \cite{spooner.cygnus2013}.

\subsubsection{MIMAC}
The MIMAC collaboration is developing an electron-drift TPC \cite{Mayet:2009ee,Santos:2011xk}  that aims at 3d track reconstruction \cite{Billard:2012bk}. 
The measurement of the third dimension, along the electric field, is achieved thanks to   
sampling of the primary electron cloud that is projected onto the grid of a 
bulk micromegas \cite{Iguaz:2011yc,FerrerRibas:2011ps} with a read-out at a frequency of $50 \ {\rm MHz}$ by a self-triggered 
electronics \cite{Bourrion:2010yt,Richer:2009pi}. The gas mixture chosen is 70 \% $\rm CF_4$ + 28 \% $\rm CHF_3$ + 2 \% 
$\rm C_4H_{10}$, which enables  sensitivity to SD interaction with a sufficient gain and reduced drift 
velocity \cite{Mayet:2014xra,Billard:2013cxa}. A prototype is currently being 
operated underground, at the LSM laboratory (Modane), at a pressure of $50 \ {\rm mbar}$ with an active volume of $5.8 \ {\rm L}$. 
The achieved threshold is $2 \ {\rm keV}$ \cite{Santos:2013hpa}. 
The short-term goal is the building of a $1 \ {\rm m^3}$ detector that will be 
the demonstrator of a large TPC devoted to directional detection ($50 \ {\rm m^3}$). 
The current $5.8 \ {\rm L}$ detector is meant to be the elementary cell of the matrix structure.  
Larger micromegas detectors ($20 \times 20 \ {\rm cm^2}$) are also being studied  \cite{Santos:2013hpa}.

\subsubsection{NEWAGE}
The NEWAGE collaboration is developing an electron-drift TPC that has been operated underground at Kamioka, 
at various R\&D stages. 
The TPC is characterized by a $41 \ {\rm cm}$ drift distance, readout plane made of $\mu$-PIC (micro-pixel chamber) 
and a gas electron multiplier (GEM). The gas  is pure ${\rm CF_4}$ at $100 \ {\rm mbar}$.  The achieved threshold is 
$50 \ {\rm keV}$ \cite{Nakamura:2015iza}.  The performance of the detector has been studied in Ref. \cite{Nakamura:2015iza}. 
A $^{252}$Cf neutron source has been used to evaluate the angular resolution, which is 40$^\circ$ at 50 keV,  and the detection efficiency, which is of the order of 40\% at 50 keV and reaches 90\% at 350 keV.  
The electron rejection power, which is $2.5\times 10^{-5}$ between 50 and 100  keV, has been evaluated with a  $^{137}$Cs source. 
In 2007, the NEWAGE collaboration  published the first direction-sensitive dark matter limit  \cite{Miuchi:2010hn} 
obtained from a dedicated run in an underground
laboratory at Kamioka mine. In 2015, they improved
their result by a factor of ten with a 
31.6 day run corresponding to an  exposure of
$0.327 \ {\rm kg\cdot days}$. It leads to the first experimental recoil map, presenting the observed  background events in galactic coordinates, which is the first step
toward directional measurement of WIMP and background events.

\subsubsection{DMTPC}
The DMTPC collaboration is developing an electron-drift TPC filled with $\rm CF_4$ 
at $40-100 \ {\rm mbar}$.  The readout uses 
CCD cameras together with low noise amplifiers to measure both 
optical~\cite{Dujmic:2008ut,Ahlen:2010ub} and electronic signals from particles 
in the detector~\cite{Lopez:2012zwa,Battat:2014mka}.  
The current amplifier is sensitive to the rise time of the 
electron peak, enabling the possibility to estimate the position of the track in the direction of the 
field~\cite{Lopez:2013ah}.  DMTPC demonstrated the first measurement of the 
vector track direction using the head-tail effect~\cite{Dujmic:2007bd}, and has 
achieved 75\% efficiency for vector direction reconstruction for tracks with minimum aspect 
ratio of 3:1 (length:width)~\cite{Deaconu:2015ama,Deaconu:2015vbk}.  
A 10 L prototype set a spin-dependent WIMP scattering limit and demonstrated 40$^o$ recoil angle 
resolution at 80 keVr threshold~\cite{Ahlen:2010ub}, and was operated underground at WIPP 
for assay of intrinsic detector backgrounds~\cite{Kaboth:2012pi}.  A $1 \ {\rm m^3}$ 
detector has been built, is currently commissioning at the surface with 30 mbar CF4 pressure for 
20 ${\rm keV_r}$ threshold~\cite{Leyton:2015pr,Deaconu:2015vbk}; underground operation is planned at SNOLAB. 
It is intended to be the elementary cell of a larger detector. 
 
\setlength{\tabcolsep}{0.1cm}
\renewcommand{\arraystretch}{1.4}
\begin{table*}[t]
\begin{center}
\hspace*{-0.5cm}
\begin{tabular}{|c||c|c|c|c|c|c|c|}
\hline
     Exp.            & V ($\rm L$)  &  Gas      & P (${\rm mbar}$) &  Drift    & Threshold (${\rm keV}$) &  Location               & Ref.  \\  \hline  \hline 
    DRIFT            & 800 &  73\% $\rm CS_2$ + 25\% $\rm CF_4$ + 2\% $\rm O_2$
      & 55 &  ion, $50 \ {\rm cm}$    & 20 \cite{Daw:2010ud} &  Boulby              
     & \cite{Battat:2014van}  \\  \hline 
    MIMAC        & 5.8             &  70 \% $\rm CF_4$ + 28 \% $\rm CHF_3$ + 2 \% $\rm C_4H_{10}$ & $50$             &  $e^-$, 20 cm        
                &     2                   & Modane       &       \cite{Santos:2013hpa} \\   \hline            
     NEWAGE            & 37 &  $\rm CF_4$      & 100 &  $e^-$, $41 \ {\rm cm}$    & 50 &  Kamioka               & \cite{Nakamura:2015iza}  \\  \hline 
     DMTPC            & 1000 &  $\rm CF_4$      & 40 &  $e^-$, $27 \ {\rm cm}$    & 20 &  SNOLAB                 & 
     \cite{Deaconu:2015vbk,Leyton:2015pr}  \\  \hline

\end{tabular}
\caption{Main properties of the directional experiments that are currently being operated underground or 
start running underground in the near future.} 
\label{tab:exp.landscape}
\end{center}
\end{table*}
\renewcommand{\arraystretch}{1.1}

\subsubsection{R\&D toward improved 3d readout}
In addition to the above experiments, several groups have constructed prototype detectors to demonstrate candidate technologies for
achieving 3d readout of directional TPCs with improved spatial-resolution and/or higher signal-to-noise ratio. Examples are prototypes with
thick GEMS at the University of New Mexico \cite{Phan:2015pda}, the $D^3$ prototypes with ATLAS pixel chip readout at the University of
Hawaii \cite{Vahsen:2014fba, Lewis:2014poa}, the NITEC prototypes with CERN Timepix chip readout at INFN, and work on high-density strip
readout at Wellesley College. The overall status of readout R\&D for directional dark matter detectors is the subject of a separate, forthcoming review article \cite{Readouts}. 

\subsection{Conclusion} 
Finally, we note that the most suitable readout for a 
large dark matter detector may not  be the one with the highest performances  
as that will be the most costly choice per target volume. Hence one needs to make the best 
tradeoff between cost  and sensitivity, to maximize sensitivity per unit cost.

Eventually, it should be noted that there is a worldwide effort toward the development of large TPC for the 
directional detection of Dark Matter that aimed at combining the main advantages and achievements of 
the detectors above presented. The project is still in discussion and could be done 
either in parallel with current instrumental developments or could supersed them within the framework of 
a single large TPC \cite{cygnus2015}.

\section{Theoretical framework}
\label{sec:theo}

\subsection{Introduction}
The directional detection strategy is based on the fact that the Sun 
is orbiting the Galactic center and is currently traveling toward the constellation Cygnus. 
If the WIMP distribution is predominantly smooth, the Sun's speed and the 
velocity dispersion of the WIMPs are expected to be similar in  magnitude, and 
therefore the WIMP flux is concentrated around the direction of Solar motion. As direction-insensitive direct detection, 
directional detection is also based on the 
scattering  of  a  Dark  Matter  particle on a target nucleus of the detector. In particular, 
it determines the kinematics of the nuclear recoil and the event rate.

This section is devoted to the basic concepts, from both astrophysics 
and particle physics, needed to evaluate the double-differential spectrum, {\it i.e.} the number of events per unit recoil energy 
in a given direction. This contains the information about Dark Matter that directional detection aims to  
extract: the WIMP velocity distribution 
(Sec.~\ref{sec:theo.halo}), mass and cross section (Sec.~\ref{sec:theo.event}). Sections \ref{sec:theo.nuc} and 
\ref{sec:theo.part} introduce the parameters from nuclear and particle physics that enter the evaluation of the event rate. We first focus on 
a WIMP with a standard SUSY-type elastic interaction, then Sec.~\ref{sec:theo.NR} presents a theoretical framework that opens 
the possibility   of distinguishing different Dark Matter-nucleon interactions.

%--------------------------
 
\subsection{Physics of the Galactic halo}
\label{sec:theo.halo}
%standard model\\
%triaxial halo\\ 
%Substructures in the Galactic halo (dark disk, debris, tidal streams)\\
%Astrophysical uncertainties\\

Historically, direct detection signal calculations and data analyses
have used the Standard Halo Model (SHM) to model the WIMP velocity distribution. 
This simple model assumes that
the Milky Way (MW) halo is an isotropic, isothermal sphere with density
profile $\rho \propto r^{-2}$. In this case, the normalized velocity
distribution in the rest frame of the Galaxy takes the form of a Maxwellian distribution
\begin{equation}
f_{\rm gal}({\bf v}) = 
\begin{cases}
\frac{N}{(2 \pi \sigma_v)^{3/2}} \exp{\left( - \frac{ |{\bf v}|^2}{2
      \sigma_v^2} \right)} & \text{if $ |{\bf v}| < v_{\rm esc}$} \,, \\
0  &  \text{if $|{\bf v}| > v_{\rm esc}$}  \,,
\end{cases}
\label{eq:fgal}
\end{equation}
where $N$ is a normalization constant. In the SHM, the one-dimensional velocity dispersion, $\sigma_v$, is independent of the distance from the 
Galactic center and is
related to the circular speed, $v_{\rm c}$, via $\sigma_v=v_{\rm c}
/\sqrt{2}$. Formally, the
Maxwellian distribution extends to infinity; however, in the case of the SHM it is usually truncated by hand at the local escape speed, $v_{\rm esc}$.

%BJK - I've tried to add in a couple of references to local methods of measuring rho_0
%FM ok
%AMG I've not changed 'statistical error' to 'uncertainty' as the fact
%the uncertainties are probably larger than this a key point. Instead
%I've added a sentence explicitly mentioning systematic errors.
The canonical value of the local circular speed is $v_{\rm c} = 220 \,
{\rm km \, s}^{-1}$, with a statistical error of order
$10\%$~\cite{Kerr:1986hz,Bovy:2009dr,Bovy:2012ba}.  However the assumptions
made when modeling the MW halo, for instance that the MW rotation
curve is flat, lead to significant systematic uncertainties.
Considering various models for the MW rotation curve
leads to determinations of $v_{\rm c}$ which range from $ (200 \pm 20) \, {\rm km \,
  s}^{-1}$ to ($279 \pm 33) \, {\rm km \,
  s}^{-1}$~\cite{McMillan:2009yr}. Note that in general the
relationship between the circular speed and the velocity dispersion
depends on the density profile and the velocity anisotropy (see {\it
  e.g.} Ref.~\cite{biney}). Assuming a smooth halo in equilibrium, the
RAVE survey has recently found the escape speed to be $v_{\rm
  esc}=537_{-43}^{+59} \, {\rm km \, s}^{-1}$~\cite{Piffl:2013mla}.
The local WIMP density, $\rho_{0}$, also enters the calculation of the
recoil spectrum. The value traditionally used in WIMP direct detection
calculations is $\rho_{0} = 0.3 \, {\rm GeV \, cm}^{-3}$.  Recent
analyses, which use a range of dynamical data sets to constrain a mass
model of the Milky Way, typically find a slightly higher value,
$\rho_{0} \approx 0.4 \, {\rm GeV \, cm}^{-3}$
\cite{Catena:2009mf,McMillan:2011wd}, with a statistical error of
order $10\%$. These analyses assume a Navarro Frenk White density
profile~\cite{Navarro:1995iw} for the Dark Matter halo. Considering a
wider range of halo profiles leads to local densities in the range
$\rho_{0} = (0.2-0.6) \, {\rm GeV \,
  cm}^{-3}$~\cite{Weber:2009pt,Iocco:2011jz,Nesti:2013uwa,Bovy:2013raa}. Local
estimates of the DM density, which do not require global
mass-modeling of the Milky Way
(e.g.~Refs.~\cite{Salucci:2010qr,Garbari:2012ff}), are generally in
agreement with these values, though with larger errors.
% AMG: I haven't added the Silverwood paper Annika suggested as it
% uses Mock data (so they don't actually measure the real local DM density...)

To calculate the recoil spectrum, the velocity distribution needs to be
translated into the lab frame. Formally, this should be done using
Liouville's theorem~\cite{Griest:1987vc,Alenazi:2006wu}. However the gravitational potential of the Sun
and Earth only affect the very low speed WIMPs ($v \lesssim 40 \, \mathrm{ km} \, \mathrm{s}^{-1}$) \cite{Alenazi:2006wu}, and therefore a
Galilean transformation $f({\bf v}, t) = f_{\rm gal}({\bf v} + {\bf
    v}_{\rm lab}(t), t)$, where ${\bf v}_{\rm lab}(t)$ is the lab's velocity with
  respect to the Galactic rest frame, 
  is sufficient.  The lab's velocity is the sum of the local
  circular speed, the Sun's motion with respect to the Local
  Standard of Rest, and the Earth's velocity relative the Sun. The full, correct calculation of the transformation between the
reference frames is carried out in
Refs.~\cite{Lee:2013xxa,McCabe:2013kea},  see also Appendix A. Note
that, as pointed out
  in Ref.~\cite{Lee:2013xxa}, the commonly
  used expressions from Ref.~\cite{Lewin:1995rx} contain an error.

%BJK - added Ref. to NIHAO simulations (although they seem to contradict previous results...)
The SHM   may not to be a good approximation to the real MW halo, as
the assumptions behind it (that the MW halo is an
isotropic, isothermal sphere in dynamical equilibrium) are not
valid. In cold Dark Matter cosmologies, structures form hierarchically, leading to
DM halos which are triaxial, anisotropic and contain phase-space
substructure (see {\it e.g.} Ref.~\cite{Kuhlen:2012ft}).
The velocity distributions of high-resolution Dark Matter-only simulations of MW-like
halos deviate systematically from the Maxwellian distribution of the
SHM. There are more low speed particles and the peak of the
distribution is lower and flatter~\cite{Vogelsberger:2008qb,Kuhlen:2009vh}. Several fitting functions
which encapsulate these properties have been
proposed~\cite{Lisanti:2010qx,Mao:2013nda}. %More recent hydrodynamical simulations further complicate the picture, suggesting that the local velocity distribution may in fact have a sharper peak than expected from a Maxwellian distribution \cite{Butsky:2015pya}.
% JB: Added discussion on beta parameter.
Hydrodynamical simulations can have different predictions for the local DM distribution depending on their baryonic feedback models, and how well they can reproduce the main galaxy population properties. In particular, some simulations find that the local DM velocity distribution may be better fitted by a Tsallis distribution~\cite{Ling:2009eh}, or have a sharper peak than expected from a Maxwellian distribution~\cite{Kuhlen:2013tra, Butsky:2015pya}. Very recently, however, high resolution hydrodynamic simulations of MW-like galaxies found that the best fit Maxwellian velocity distribution describes well the local DM velocity distributions of simulated haloes~\cite{Bozorgnia:2016ogo, Kelso:2016qqj}. While Ref.~\cite{Sloane:2016kyi} found some discrepancy in their results compared to the results of Refs.~\cite{Bozorgnia:2016ogo, Kelso:2016qqj}, 
the source of these differences is 
yet to be understood. Eventually, some anisotropy in the velocity distribution can be expected. It is evaluated by the velocity anisotropy parameter $\beta(r)$ which is a function of the three velocity dispersions along the three Galactic axes $\{x, y ,z\}$ and is defined as 
\cite{biney},
  \begin{equation}
  \beta(r) = 1 - \frac{\sigma^2_{y} + \sigma^2_{z}}{2\sigma^2_x}.
  \label{eq:beta}
  \end{equation}
  At the Solar radius from the Galactic center ($R_{\odot} = 8$ kpc), N-body 
  simulations~\cite{Nezri,vogelsberger,Kuhlen:2009vh,Moore:2001vq}, with or without
  baryons,  predict a $\beta$ parameter spanning the 
  range $0$ to $0.4$ which is in favor of radial anisotropy.

In addition to uncertainties in the overall shape of the distribution, the high speed tail contains stochastic features (dubbed
``debris flow'') from incompletely phase-mixed material~\cite{Lisanti:2011as,Kuhlen:2012fz}. It is also possible that
the local DM distribution may contain streams from substructures that
are in the process of being disrupted. For instance the Dark Matter
component of one of the streams from the Sagittarius dwarf galaxy may
pass through the Solar neighborhood~\cite{Freese:2003tt,Purcell:2012sh}.

%BJK - polished a little 07/01/2016, but perhaps not much better...
The baryonic disk can also affect the DM distribution. Late merging
halos are dragged into the disk by dynamical friction, where they are
destroyed, leading to the formation of a co-rotating dark disk
(DD)~\cite{Read:2008fh,Read:2009iv}. While the contribution of the DD
to the local DM density is uncertain, it is currently expected to be small, owing to the relatively quiescent merger history of the MW~\cite{Purcell:2009yp}. Eris, a simulation
of a MW-like galaxy including baryonic physics, has a DD which
contributes less than $10\%$ of the local DM density~\cite{Kuhlen:2013tra}.  Ref.~\cite{Bozorgnia:2016ogo} finds that there may be some hints for the existence of a DD  in only 2 out of 14 MW-like galaxies in the EAGLE and APOSTLE hydrodynamic simulations, while Ref.~\cite{Kelso:2016qqj} does not find a significant DD in the two MW-like galaxies they studied from the MaGICC simulations.
Furthermore, chemo-dynamical analysis of data from the Gaia-ESO Survey
has found no evidence for the stellar component of an accreted
disk~\cite{Ruchti:2015bja}. Further constraints on the dark disk density can be obtained by comparing global and local measures of the DM density (see Sec.~5.4 of 
Ref.~\cite{Read:2014qva} for a review).

%BJK - I've tried to improve the flow here...
%FM ok
In the next section, we show how the velocity distribution enters into the calculation of the directional event rate. In Sec.~\ref{sec:identification.full} we will explore how directional detection may be used to probe the broad characteristics of the DM velocity distribution. In Sec.~\ref{sec:probinghalo} and \ref{sec:decomposing} we extend this and discuss the possibility of detecting phase-space substructure and probing more general models for the velocity distribution. For more detailed discussion of astrophysical uncertainties pertaining
to direct detection experiments see {\it e.g.} Refs.~\cite{Green:2011bv,Peter:2013aha}.

%---------------------------

\subsection{Event rate and cross section}
\label{sec:theo.event}

%BJK - Edited notation for reduced mass (I've changed it to \mu, but maybe \mu_A is better)
%FM  - we have decided that the reduced mass will be noted \mu. 
Directional detection aims at measuring both the energy ($E_r$) and the 3D direction ($\Omega_r$) 
of a recoiling nucleus ${\rm ^AX}$ following a WIMP scattering.
Since the WIMP speed $v$ in the laboratory rest frame is of the order of $\sim 300 \ {\rm km\ s^{-1}}$, the recoil energy for 
elastic scattering is  given by the
non-relativistic expression   
\begin{equation}
E_r = 2v^2\frac{\mu^2}{m_A}  \cos^2\theta_r \,,
\label{eq:collision}
\end{equation}
with $\mu = m_\chi m_A/(m_\chi + m_A)$ the WIMP-nucleus reduced mass, 
$m_\chi$ the WIMP mass, $m_A$ the mass of the target nucleus and 
$\theta_r$ the angle between the initial WIMP direction ($\bhat{v}$) and that of the recoiling nucleus ($\bhat{r}$), such that 
$\bhat{v}\cdot\bhat{r} = \cos\theta_r$. One expects low energy recoils, typically below 100 keV, depending on the particular values of  $m_\chi$ and $m_A$.\\

The recoil rate per unit detector mass as a function of both the recoil energy  and direction   is given by the following 
double-differential spectrum:
\begin{equation}
\frac{\mathrm{d}^2R}{\mathrm{d}E_r\mathrm{d}\Omega_r} = \frac{\rho_0}{m_\chi m_A} \int  \frac{\mathrm{d}^2\sigma_{\chi -A}}{\mathrm{d}E_r\mathrm{d}\Omega_r} vf(\bvec{v})\mathrm{d}^3v\,,
\label{eq:doublespectrum}
\end{equation}
with $\rho_0$ the local WIMP density,  $f(\bvec{v})$ the WIMP velocity distribution in the detector reference frame and 
$\mathrm{d}^2\sigma_{\chi - A}/(\mathrm{d}E_r\mathrm{d}\Omega_r)$ the  
double-differential WIMP-nucleus cross section given by : 
\begin{equation}
\frac{\mathrm{d}^2\sigma_{\chi - A}}{\mathrm{d}E_r\mathrm{d}\Omega_r}  = \frac{\mathrm{d}\sigma_{\chi - A}}{\mathrm{d}E_r}  \frac{1}{2\pi}  v   \,
\delta\left(\bvec{v} \cdot \bhat{r} -  v_{\rm min}\right)\,,
\end{equation}
where
\begin{equation} 
v_{\rm min} = \sqrt{E_rm_A/2\mu^2},
\label{eq:vmin}
\end{equation} 
is the minimum WIMP velocity required to produce a recoil of energy $E_r$ in an elastic scattering event. 
The Dirac delta function originates from Eq.~\ref{eq:collision} linking the recoil energy and the recoil 
angle $\theta_r$. Note that Eq.~\ref{eq:doublespectrum} assumes that the detector consists of only one nuclide. If it is not the case, the measured rate is given
by the sum over all nuclides weighted by the mass fraction.\\
Under the standard assumption of contact interactions, the WIMP-nucleus cross section is obtained 
\footnote{Section \ref{sec:theo.NR} presents a non-relativistic effective framework that enables 
studies beyond standard SI and SD interactions.} by adding coherently the spin-dependent (SD) and spin-independent (SI) contributions 
\cite{Goodman:1984dc}. In the
non-relativistic limit, this reads : 
\begin{equation}
\frac{\mathrm{d}\sigma_{\chi - A}}{\mathrm{d}E_r} = \frac{m_A}{2\mu^2 v^2}\big[\sigma^{\rm SI}F^2_{\rm SI}(E_r) + 
\sigma^{\rm SD}F^2_{\rm SD}(E_r)\big]\,,
\end{equation}
where  $\sigma^{\rm SI,SD}$ are the SI and SD WIMP-nucleus cross 
section at zero momentum transfer   and   $F_{\rm SI,SD}$   the
SI and SD form factor.\\
The  double-differential spectrum is given by 
\begin{equation}
\frac{\mathrm{d}^2R}{\mathrm{d}E_r\mathrm{d}\Omega_r}=\frac{\rho_0}{4\pi m_\chi \mu^2} \Big[ \sigma^{\rm SI}F^2_{\rm SI}(E_r)+\sigma^{\rm SD}F^2_{\rm SD}(E_r)\Big] 
\hat{f}(v_{\rm min},\bhat{r})\,,
\label{eq:d2r}
\end{equation}
where $\hat{f}$ is the three-dimensional Radon transform  of the WIMP 
velocity distribution $f(\bvec{v})$, given by 
\begin{equation}
\hat{f}(v_{\rm min},\bhat{r}) = \int  \delta(\bvec{v} \cdot \bhat{r}-v_{\rm min})f(\bvec{v}) \  \mathrm{d}^3v \,.
\label{eq:radon}
\end{equation}
Geometrically, the Radon transform is the integral of the function $f(\bvec{v})$ on a plane orthogonal to the direction 
$\bhat{r}$ at a distance $v_{\rm min}$ from the origin. Using the Fourier slice theorem, Gondolo \cite{Gondolo:2002np} 
has shown that the use of the Radon transform leads to an analytic evaluation of the event rate for most halo models.

We may then distinguish two types of direct Dark Matter search strategies: directional detection, aiming at measuring 
the double-differential spectrum, Eq. (\ref{eq:d2r}), and
direction-insensitive detection, only sensitive to the energy spectrum given by 
\begin{equation}
\frac{\mathrm{d}R}{\mathrm{d}E_r} =  \frac{\rho_0}{2m_\chi \mu^2}\big[
\sigma^{\rm SI}F^2_{\rm SI}(E_r) + \sigma^{\rm SD}F^2_{\rm SD}(E_r)\big] \int_{v_{\rm min}} \frac{f(\bvec{v})}{v}\mathrm{d}^3v \,.
\label{eq:dRdEr}
\end{equation} 
The event rate, Eqs.~(\ref{eq:d2r}) and (\ref{eq:dRdEr}), receives inputs from : 
\begin{itemize}
\item nuclear physics, via the properties of the target nucleus : $m_A, F_{SI,SD}$ as well as the spin content, see Sec.~\ref{sec:theo.nuc}
\item the particle physics model, via the parameters  $m_\chi, \sigma^{SI,SD}$, see Sec.~\ref{sec:theo.part}
\item the Galactic halo, via the parameters  $\rho_0$ and  $f(\bvec{v})$, see Sec.~\ref{sec:theo.halo}  
\end{itemize}

The WIMP-nucleus cross section, at zero momentum transfer, can be separated into a SI contribution, given by 
\cite{Engel:1992bf} 
\begin{equation}
\sigma^{\rm SI}({\rm ^AX}) = \frac{4\mu^2}{\pi}\left({\rm Z}f_p + {\rm (A-Z)}f_n   \right)^2 \,,
\end{equation}
where $f_{p,n}$ is the WIMP-proton (resp. neutron) SI  coupling constant, 
and a SD contribution, given by  \cite{Engel:1992bf}
\begin{equation}
\sigma^{\rm SD}({\rm ^AX}) = \frac{32}{\pi}G^2_F\mu^2\frac{J+1}{J}\left[a_p\langle S_p\rangle + a_n\langle S_n \rangle  \right]^2 \,,
\end{equation}
where $G_F$ is the Fermi constant, $J$ the   angular momentum of the target nucleus, $a_{p,n}$  the 
WIMP-proton (resp. -neutron) SD  coupling constant, 
and $\langle S_{p,n}\rangle$  the spin content of the target nucleus.\\
For a direct detection experiment to be sensitive to SD interaction, the target nucleus ${\rm ^AX}$ must have a 
non-vanishing spin, whereas SI interaction   favors   heavy targets, due to the $A^2$ enhancement. 
The comparison between various DM detectors, using various target nuclei, must be done at the level of the WIMP-proton 
(resp. neutron) cross sections, {\it i.e.}  $\sigma_{p,n}^{\rm SI} \propto f_{p,n}^2$ and $\sigma_{p,n}^{\rm SD} \propto a_{p,n}^2$.
It requires either model-dependent assumptions or a dedicated framework valid in the context of heavy squarks \cite{Riffard:2016oss}.

%%%%%%%%%%%%%%%%%%%%%%%%%%%%%%%%%%%%%%%%%%%%%%%%%%%%%%%%%%%%%%%%%%%%%%
\subsection{Inputs from nuclear physics}
\label{sec:theo.nuc}
Two nuclear parameters   enter the evaluation of the event rate: the nuclear form factor $F_{SI,SD}$ and the 
spin content of the target nucleus $\langle S_{p,n}\rangle$. Although not specific to directional detection, as 
this issue is shared with direction-insensitive detection, they must be evaluated with caution.\\

The nuclear form factors $F_{\rm SI,SD}(q)$ describes the loss of coherence at high momentum transfer 
$q=\sqrt{2m_AE_r}$. The effect is strong for a heavy WIMP or a heavy target nucleus. 
Several evaluations of the nuclear form factor are used in the literature, {\it e.g.} the Helm form factor \cite{Helm} or 
Hartree-Fock (HF) calculations \cite{Co:2012ht},  but the differences are small.  
For SI form factors, it has been shown in Ref. 
\cite{Co:2012ht} that the differences are $\sim 1\ \%$ for a light nucleus at 1 keV and 
$\sim 10\ \%$ for a heavy nucleus at 100 keV. The effect on the event rate is only minor for directional detection using $\rm CF_4$. 
 For $\rm ^{19}F$, the SI (resp. SD) cross section is 
lowered by $\approx 10$\% (resp.  $\approx 15$\%) for $E_r = 100 \ {\rm  keV}$ \cite{Billard.these}, when the form factors are included. 
Note that uncertainties in the determination of the 
nuclear form factor $F_{SI,SD}$ will affect the theoretical predictions for the Dark Matter event rate \cite{Cerdeno:2012ix}.\\

The spin content of the target 
nucleus $\langle S_{p,n}\rangle= \ \langle N \mid S_{p,n} \mid N \rangle$ is a key issue  for SD detection of Dark Matter. 
The WIMP couples mainly to the spin of the unpaired proton ({\it e.g.} $\rm ^{19}F$) or of the unpaired neutron 
({\it e.g.}  $\rm ^{3}He$). However, this holds true only within the framework of a single-particle shell model.   
In practice, the spin of the target nucleus is carried both by constituent neutrons and protons. 
 Detailed nuclear shell-model calculations have been developed and  the accuracy of the  $\langle S_{p}\rangle$ 
and $\langle S_{n}\rangle$ evaluation is 
assessed by comparing with experimental values (magnetic
moments and low energy spectra)  \cite{Bednyakov:2004xq}. The relative sign of $\langle S_{p,n}\rangle$ is also important for SD 
direct detection, inducing either constructive or destructive interference, depending on the 
sign of the SD amplitudes $a_{p,n}$, see {\it e.g.} Ref. \cite{Moulin:2005sx}.

%Within the framework of the Minimal Supersymmetric Standard Model (MSSM), 
%this issue has became crucial since the recent results of ATLAS and CMS collaboration that set a lower
%limit on the squark mass in the TeV region \cite{Aad:2011ib}.
%As shown in \cite{AlbornozVasquez:2012px}, this result cancels the Feynman diagram with a squark exchange in the s-channel and leaves 
% the Z exchange as the only diagram at tree level. {\it De facto}, the ratio of the two amplitudes is then  fixed, with no dependence on
% SUSY parameters, and given by  $a_p/a_n \simeq-1.14$ \cite{Belanger:2008gy}. Although specific to MSSM, this result 
% highlights the key role played by the spin contents of the nucleus that must be treated with caution owing to 
%their effect in the SD Dark Matter sector, and their strong dependence
%on nuclear-shell models. 

\subsection{Parameters from particle physics models}
\label{sec:theo.part}
WIMP candidates naturally arise from extensions of the standard model of particle
physics, {\it e.g.} Supersymmetry (SUSY),  as long as the lightest particle of the hidden sector, 
{\it e.g.} the lightest supersymmetric particle (LSP),  
is stable and massive. If its interaction with ordinary matter proceeds only through weak interaction, this 
particle becomes a good candidate for cold Dark Matter. This is the case for the neutralino which is the LSP in 
most SUSY models \cite{Jungman:1995df}.

%BJK - Added a sentence or two explaining the standard framework (could definitely be improved or moved somewhere else)
As long as the WIMP is not discovered at colliders, the particle physics model does not provide inputs for 
DM searches but only a theoretical framework to interpret the results, within {\it e.g.} supersymmetric models. The standard framework assumed for direct detection experiments is in fact inspired by models of supersymmetric DM. It assumes contact interactions (i.e.~interactions mediated by heavy particles) and includes only the leading-order interactions in the non-relativistic limit.
One of the goals of Dark Matter searches, and in particular directional detection, is to  constrain the 
parameters in this framework ($m_\chi, a_p, a_n, f_p, f_n$). 
The SI  WIMP coupling constants with proton and neutron ($f_p, f_n$) are often considered equal, 
with the assumption that scattering on sea quarks dominates \cite{Jungman:1995df}. Hence, the constraint in the SI sector 
is rather straightforward, provided the SD
interaction is neglected. On the other hand, the assumption that one SD interaction dominates must be made in order to constrain either
$a_p$ or $a_n$.

The question of the complementarity between various Dark Matter search strategies, direct detection, indirect detection and 
colliders, has received much interest in the past years 
\cite{hep-ph/0405210,arXiv:0710.0553,arXiv:1011.4514,arXiv:1111.2607,Moulin:2005sx,Mayet:2002ke,AlbornozVasquez:2012px,Arbey:2013iza,Cheung:2012qy}.
Constraints from all Dark Matter searches, as well as collider \cite{pdg} and cosmological constraints \cite{planck} must 
be accounted for when 
studying {\it e.g.} the discovery potential of forthcoming detectors or the impact of current results for a given model.

Directional detection aims
at constraining both the WIMP (mass and cross section) 
and halo properties as shown in Refs. \cite{Billard:2010jh,Lee:2012pf}, see Sec.~\ref{sec:identification}. Within the framework of neutralino Dark Matter, it has been shown that 
directional  detection is a powerful tool to constrain  MSSM and NMSSM parameter
spaces \cite{AlbornozVasquez:2012px}, as shown in Sec.~\ref{sec:probingparticlemodel}.

%BJK - 26/06/2015
\subsection{Non-relativistic effective theory}
\label{sec:theo.NR}

Beyond the standard SI and SD interactions, we can also consider interactions which are higher order in the WIMP speed $v$ and 
recoil momentum $q$. Such a non-relativistic (NR) effective theory approach to WIMP-nucleon elastic scattering was proposed 
by Fan {\it et al.}~\cite{Fan:2010gt} and extended by others in 
Refs.~\cite{Fitzpatrick:2012,Anand:2013yka,Dent:2015zpa}. Within such a framework, one writes 
down all Hermitian, Galilean and rotation-invariant interactions composed of the low-energy quantum-mechanical 
degrees of freedom of the system. These include the DM and nucleon spins $\mathbf{S}_\chi,\mathbf{S}_N$ and recoil momentum $\mathbf{q}$. The DM velocity 
appears only through the Hermitian `transverse velocity' operator,
\begin{equation}
\mathbf{v}_\perp = \mathbf{v} + \frac{\mathbf{q}}{2 \mu_n}\, ,
\end{equation}
where $\mu_n$ is the WIMP-nucleon reduced mass. It is 
so called because it lies perpendicular to the recoil momentum due to energy conservation, $\mathbf{v}_\perp \cdot \mathbf{q} = 0$.
A full list of such operators can be found in Ref.~\cite{Dent:2015zpa}. 

For operator $i$, the differential cross-section for WIMP-nucleon scattering can then be written
\begin{equation}
\frac{\mathrm{d}\sigma}{\mathrm{d}E_r} =\frac{1}{32 \pi m_A m_\chi^2 v^2} \, \sum_{N, N' = p, n} c_i^{N} c_i^{N'} F^{(N,N')}_i(q^2, v_\perp^2)\,.
\end{equation}
Here, $c_i^{p, n}$ are the couplings of operator $i$ to protons and neutrons respectively. The nuclear response functions $F^{(N,N')}_i(q^2, v_\perp^2)$ are composed of powers of $q^2$ and $v_\perp^2$ multiplied by generalised nuclear form factors. These form factors have been obtained numerically for a variety of target 
nuclei, see {\it e.g.} Refs.~\cite{Fitzpatrick:2012,Catena:2015uha}. NR operators which are typically studied lead to nuclear response functions proportional to $(v_\perp)^0$ (which is the standard case) or proportional to $(v_\perp)^2$. In the latter case, the integral over $\mathbf{v}$ leads to the `transverse' Radon Transform, given by:
\begin{equation}
\hat{f}^T(v_\mathrm{min}, \hat{\mathbf{r}}) = \int (v^2 - (\mathbf{v}\cdot \hat{\mathbf{r}})^2) f(\mathbf{v})
\,\delta(\mathbf{v}\cdot\hat{\mathbf{r}} - v_\mathrm{min}) \, \mathrm{d}^3\mathbf{v} = \int (v^2 - v_\mathrm{min}^2) f(\mathbf{v})
\,\delta(\mathbf{v}\cdot\hat{\mathbf{r}} - v_\mathrm{min}) \, \mathrm{d}^3\mathbf{v}\, .
\label{eq:trt}
\end{equation}

This NR effective theory framework has been used to interpret 
recent data from direction-insensitive detection ({\it e.g.} Ref. \cite{Schneck:2015eqa}) and its 
impact in directional detectors has been studied  in Refs.~\cite{Catena:2015vpa,Kavanagh:2015jma}. In Ref.~\cite{Catena:2015vpa}, directional rates $\mathrm{d}R/\mathrm{d}\cos\theta$ were studied for 14 different NR operators for detectors composed of 
$\mathrm{CS}_2$, $\mathrm{CF}_4$ and $^3\mathrm{He}$. This study revealed that the contribution of the 
operator $\mathcal{O}_{11} = i \mathbf{S}_\chi \cdot \mathbf{q}/ m_N$, where $m_N$ is the nucleon mass, to the scattering rate 
in $\mathrm{CF}_4$ can be comparable to the standard SI and SD interactions, despite being 
suppressed by powers of the recoil momentum. Ref.~\cite{Kavanagh:2015jma} compared the directionality of the signal for a number of NR operators, assuming equal numbers of events for each operator. The resulting signals may be more strongly directional (for operators coupling to 
$\mathbf{q}$) or more isotropic (for operators coupling to  $\mathbf{v}_\perp$) than in the standard SI/SD scenario. 
In Sec.~\ref{sec:particlemodel.NRop}, we discuss 
 how directional detection may be used to discriminate between different operators within the NR effective theory framework.

%
% -------------- features (dipole, ring, aberration)
%4. Directional features (dipole, ring, aberration) : J. Battat and N. Bozorgnia ( 
%\newpage
% Edited by FM 30/11/15 : modification + vlab and explantion of recoil maps

% -------------- features (dipole, ring, aberration)
%4. Directional features (dipole, ring, aberration) : J. Battat and 
%N. Bozorgnia (jbattat@wellesley.edu,bozorgnia@mpi-hd.mpg.de)
\section{Directional features}
\label{sec:features}

% Edited by FM 26/11/15 : minor corrections (Annika's comments)
% Edited by FM 22/11/15 : minor corrections 
% Edited by FM 16/11/15 : minor corrections (following comments from AMG and GBG)
% Edited by FM (17/07/15) - minor modification
\subsection{Introduction} 

The main observable signature for directional detectors is the dipole in Galactic coordinates of the directions of the WIMP-induced nuclear recoils.  In section~\ref{sec:featuresDipole}, we describe this dipole feature. We also review two other observable signatures unique 
to directional detectors in sections \ref{sec:featuresRing} and \ref{sec:featuresAberration}: the ring and aberration features.  The detection of ring and aberration features 
requires lower energy thresholds and more events than the detection of the dipole, but they can provide 
additional constraints on the WIMP and halo properties, see Sec.~\ref{sec:secondarysignatures}.

The Radon transform (Eq.~\ref{eq:radon}) in the laboratory frame for the truncated Maxwellian WIMP velocity 
distribution, Eq. \ref{eq:fgal}, is~\cite{Gondolo:2002np}
%
%\begin{equation}
%\label{eq:fhatTM_dipole}
%\hat{f}\!\left( v_{\rm min}, \hat{\bf r} \right)=\frac{1}{{N_{\rm esc}(2\pi \sigma_v^2)^{1/2}}}~{\left\{\exp{\left[-\frac{\left[ v_{\rm min} + \hat{\bf r} \cdot {\bf v}_{\rm lab}\right]^2}{2\sigma_v^2}\right]}-\exp{\left[-\frac{v_{\rm esc}^2}{2\sigma_v^2}\right]}\right\}},
%\end{equation}
%
%if $v_{\rm min} + \hat{\bf r} \cdot {\bf v}_{\rm lab} < v_{\rm esc}$, and zero otherwise. 

\begin{equation}
\hat{f}\!\left( v_{\rm min}, \hat{\bf r} \right)=
\begin{cases}
\frac{1}{{N_{\rm esc}(2\pi \sigma_v^2)^{1/2}}}~{\left\{\exp{\left[-\frac{\left[ v_{\rm min} + \hat{\bf r} \cdot {\bf v}_{\rm lab}\right]^2}{2\sigma_v^2}\right]}-\exp{\left[-\frac{v_{\rm esc}^2}{2\sigma_v^2}\right]}\right\}} 
& \text{if $v_{\rm min} + \hat{\bf r} \cdot {\bf v}_{\rm lab} < v_{\rm esc}$} \,, \\
0  &  \text{otherwise}  \,,
\end{cases}
\label{eq:fhatTM_dipole}
\end{equation}

Here ${\bf v}_{\rm lab}$ is the velocity of the laboratory with respect to the Galaxy (hence the average velocity of the WIMPs with respect to the detector is $-{\bf v}_{\rm lab}$), and
\begin{equation}
N_{\rm esc} = {\rm erf}{\left(\frac{v_{\rm esc}}{\sqrt{2}\sigma_v} \right)} - \sqrt{\frac{2}{\pi}}\frac{v_{\rm esc}}{\sigma_v} \exp{\left[-\frac{v_{\rm esc}^2}{2 \sigma_v^2} \right]}.
\end{equation}
The nuclear recoil direction $\hat{\bf r}$ is measured in the detector reference frame, and in order to compute $\hat{f}$ we need to evaluate $\hat{\bf r} \cdot {\bf v}_{\rm lab}$. The transformation equations for $\hat{\bf r}$ and ${\bf v}_{\rm lab}$ to go from the detector frame to the Galactic reference frame are given in Appendix~\ref{app-A}.

One can see from Eq.~\ref{eq:fhatTM_dipole} that there are two regimes of interest, depending on the value of $v_{\rm min}$
, as defined by Eq.~\ref{eq:vmin}. First, if $v_{\rm min} > v_{\rm lab}$, then the argument of the first exponential cannot be zero, but is minimized when $\hat{\bf r}$ and ${\bf v}_{\rm lab}$ are anti-parallel.  This leads to a dipole feature in the recoil angle distribution 
(Sec.~\ref{sec:featuresDipole}).  Second, if $v_{\rm min} < v_{\rm lab}$, 
{\it i.e.} for low recoil energies and large WIMP masses (see Eq.~\ref{eq:vmin}), then the argument of the first exponential can be zero, and the recoil angle distribution will exhibit a ring-like feature (see Sec.~\ref{sec:featuresRing}).

%One can see from Eq.~\ref{eq:fhatTM_dipole}  that if $v_{\rm min} > v_{\rm lab}$, then the argument of the first exponential cannot be zero, but is minimized when $\hat{\bf r}$ is parallel to $-{\bf v}_{\rm lab}$, and so the double-differential recoil spectrum (Eq.~\ref{eq:d2r}) is maximized in the direction of $-{\bf v}_{\rm lab}$.  In the case when $v_{\rm min} < v_{\rm lab}$ (i.e. for lower recoil momenta), the maximum of $\hat{f}\!\left( v_{\rm min}, \hat{\bf r} \right)$ occurs when the argument of the first exponential is zero (i.e. when $v_{\rm min} = -\hat{\bf r} \cdot {\bf v}_{\rm lab}$).  This results in ring-like features in the sky map of nuclear recoils, a topic described more fully in Section~\ref{sec:featuresRing}.  For example, if the WIMP mass is 100~GeV, then the recoils of $^{19}$F will exhibit a dipole (rather than a ring-like pattern) for recoil energies above 14~keV.  If the WIMP mass is 10~GeV, then the ring-dipole transition energy is even smaller (2.4~keV).

% Edited by FM 10/02/16: comments from AP
% Edited by FM 30/11/15 : modification + vlab and explantion of recoil maps
% Edited by FM 27/11/15 : minor corrections  
% Edited by JB 26/11/15 : addition of a discussion on target angular distribution 
% Edited by FM 22/11/15 : minor corrections 
\subsection{Dipole feature}
\label{sec:featuresDipole}
 
 A directional detector located on the Earth will experience a WIMP head-wind caused by the Earth's motion through the Galactic WIMP distribution (the halo).  The resulting WIMP-induced nuclear recoils 
 will come from the direction to which the vector ${\bf v}_{\rm lab}$ is pointing.  This dipole feature was first described by Spergel \cite{Spergel:1987kx}, who showed that the recoil rates in the forward and backward directions differed by a factor of 
%$\approx$10$^{1\pm0.5}$, 
 order 10,
depending on the recoil energy threshold.  
Because no known backgrounds can mimic this angular signature, the dipole feature, which is only accessible to directional detectors, is
generally considered to be a smoking-gun evidence for WIMP Dark Matter.

\begin{figure}[t]
\begin{center}
  \includegraphics[height=140pt]{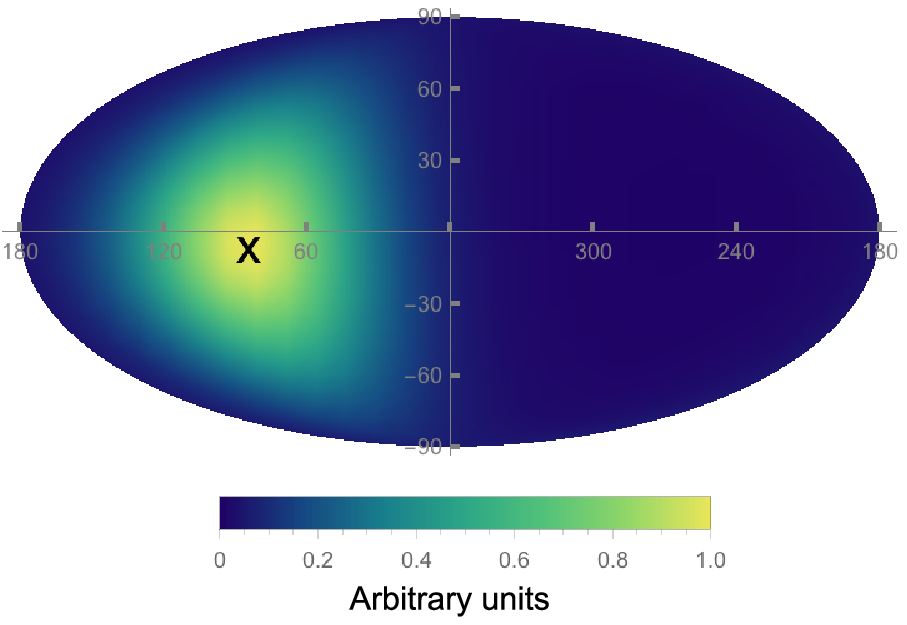}
  \includegraphics[height=140pt]{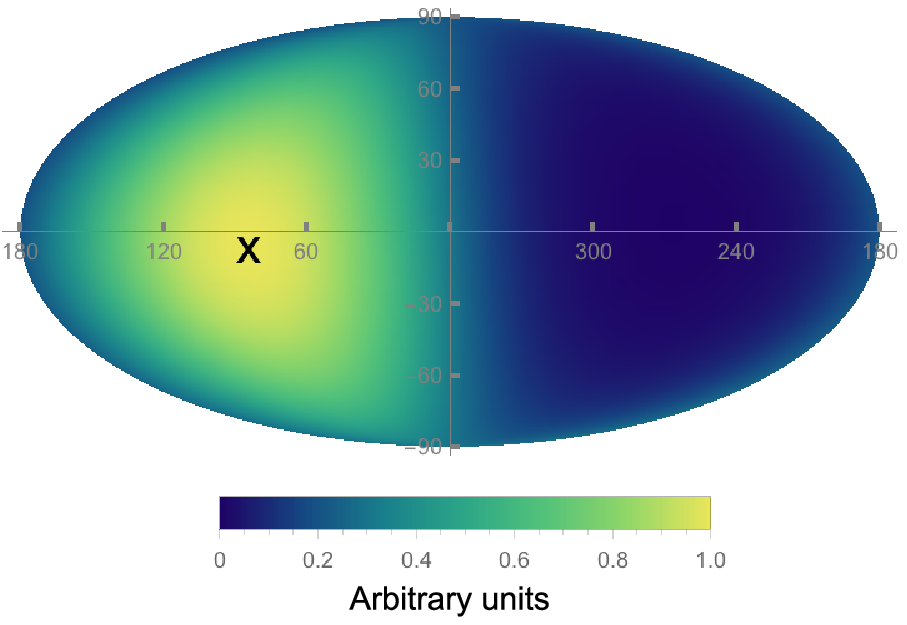}
\caption{Left: Flux of 100~GeV WIMPs moving with speeds higher than $v_{\rm min}$ as needed to produce 25 keV F recoils. 
Right: Angular distribution of the energy differential recoil rate in F for WIMP mass 100~GeV, and recoil energy of 25~keV.  
Maps are incoming direction of WIMP-induced recoils in Mollweide equal-area projections, in Galactic coordinates. 
For convenience, we present the direction of ${\bf v}_{\rm lab}$ as a cross on the maps.}
%The coordinates of the Galactic center are $(l,b)=(0,0)$. }
  \label{fig:dipoleSkymap}
\end{center}
\end{figure}

\begin{figure}[t]
\begin{center}
\includegraphics[scale=0.4]{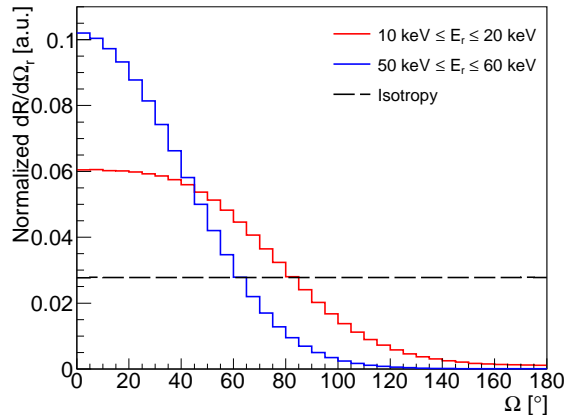}
\caption{Normalized angular spectra for F recoils in two energy ranges: 10-20~keV and 50-60~keV for a 
WIMP mass of 100\,GeV.  Here, $\Omega$ is the opening angle between the recoil direction and the lab velocity vector.  
The dipole signature is stronger for higher energy recoils.}  
%Figure from Ref. \cite{Billard:2009mf}
  \label{fig:dipolespectrum}
\end{center}
\end{figure}

As an example of the dipole feature, Figure~\ref{fig:dipoleSkymap} left shows a map of the WIMP flux 
in Galactic coordinates, assuming that the WIMP velocity distribution is Maxwellian. The incoming WIMP flux appears to come 
primarily from the direction of the Earth's motion through the Galaxy, shown as cross at position $(l,b)=(\pi/2, 0)$, where $(l,b)$ are 
Galactic longitude and latitude.
Figure~\ref{fig:dipoleSkymap} right presents the incoming direction of WIMP-induced recoils (for $E_r =25$~keV) in galactic coordinates. 
Note that in this review, the chosen convention is to present recoil maps as directions 
where the recoils originate from, in the spirit of WIMP astronomy. For convenience, the cross on 
the map presents the direction of ${\bf v}_{\rm lab}$, to highlight the fact that WIMPs are coming from the 
direction to which the laboratory's velocity vector is pointing to, which  happens to be roughly 
in the direction of the constellation Cygnus $(l,b)=(\pi/2, 0)$. 
%The recoil directions  are then clustered about the 
%$-{\bf v}_{\rm lab}$ direction, with the angular distribution of the recoils broadened relative to the incoming WIMP angular 
%distribution due to the (non-relativistic) WIMP-nucleus scattering kinematics in the laboratory frame.  
In Figure~\ref{fig:dipolespectrum}, we take advantage of the azimuthal symmetry of the recoil directions 
about ${\bf v}_{\rm lab}$ to plot the fraction of recoils per solid angle as a function of the angle between 
the recoil direction ${\bf \hat{r}}$ and $-{\bf \hat{v}}_{\rm lab}$.  From these normalized angular spectra, we see that 
higher recoil energies lead to a tighter clustering of recoil angles, 
and therefore a stronger dipole signature, but at the expense of a 
smaller event rate.  It is also clear that only modest angular resolution 
(tens of degrees) is required to resolve the dipole signature.\\

%In Figure~\ref{fig:dipolespectrum}, the form factor $F(E_r)$ was assumed to be unity.  Using the Born approximation, the form factor is given by the Fourier transform of a thin shell (valid for spin-dependent interactions), and for $^{19}$F recoils, $F^2(50\mbox{ keV})=0.9$. 

Due to the non-zero energy threshold, the WIMP induced recoil angular distribution does also depend on the target mass $m_A$. As
several directional detectors are being developed with different target nuclei, we discuss the influence of these targets. Although their
detection characteristics may be different ({\it e.g.} track length, drift velocity and straggling, see Sec.~\ref{sec:exp}), for low mass targets
($^{12}$C, $^{32}$S,  $^{19}$F) and at sufficiently low recoil energy (when the form factor can be approximated to unity), equivalent
directional signal can be found by adjusting the energy range for each target. Indeed, the directionality of the signal is encoded only in
the $\cos\theta_r$ term of Eq. (\ref{eq:collision}), then the angular distribution for a target of mass $m_{A_1}$ at a recoil energy $E_{r_1}$ is equivalent to the one of a $m_{A_2}$ target at $E_{r_2}$, using :
\begin{equation}
E_{r_2} = E_{r_1}\frac{m_{A_2}}{m_{A_1}}\left( \frac{m_\chi + m_{A_1}}{m_\chi + m_{A_2}} \right)^2.
\label{eq:equivang}
\end{equation}
Therefore, if we consider a 10 GeV WIMP mass, equivalent directional distributions are expected from a $^{19}$F target with an energy range
from 5 keV to 50 keV and from a $^{32}$S target with an energy range from 4 keV and 40 keV. As a matter of fact, 
as long as the form factor can be approximated to unity (small momentum transfer), if we consider a recoil energy range 
from 0 keV to infinity, the angular distribution of the WIMP induced recoils are independent of the WIMP mass and the 
target nucleus. Hence, all results  (skymaps, angular spectra) presented hereafter for a given target nucleus may be converted to another
one by using Eq. \ref{eq:equivang}.

%Notice that the general dipole feature in the recoil rate would be very similar to the example shown in Figure~\ref{fig:dipoleSkymap} for other targets commonly used in directional detectors, such as C or F.

%Also in Figure~\ref{fig:dipolespectrum}, we take advantage of the azimuthal symmetry of the recoil directions about $-{\bf v}_{\rm lab}$ to plot the fraction of recoils per solid angle as a function of the angle between the recoil direction ${\bf \hat{r}}$ and $-{\bf \hat{v}}_{\rm lab}$.  From these normalized angular spectra, we can see that the directionality signature is stronger for higher energy recoils, and that only modest angular resolution (tens of degrees) is required to resolve the dipole signature.

%For example, recoils with energy between 5 and 25 keV lie within XXX degrees of $(l,b)=(\pi/2,0)$, while recoils above 50 keV lie within YYY degrees.
%(XXXassumes Mwimp=100 GeV and Mtarget = 19GeVXXX)

%{\bf Include discussion of number of events/exposure needed?  Or does this to go in a different section?}

% Edited by FM 30/11/15 : modification + vlab and explantion of recoil maps

% Edited by FM 22/11/15 : minor corrections 
\subsection{Ring-like features} 
\label{sec:featuresRing}

An additional Dark Matter signature that can be searched for in directional detectors is a ring-like feature in the recoil rate. For large WIMP masses and low recoil 
energies, the recoil rate is maximum in a ring around the average WIMP arrival direction.
%FM 30/11/15 i think we do not need to specify again vlab (or -vlab)
%, $-{\bf v}_{\rm lab}$.
For example, if the WIMP mass is 100~GeV, then the recoils of $^{19}$F will exhibit a ring-like pattern (rather than a dipole) for recoil energies below 14~keV.  If the WIMP mass is 10~GeV, then the ring-dipole transition energy is even smaller (2.4~keV).
The cause of the ring-like geometry is most easily seen in the expression of the Radon transform (see Eq.~\ref{eq:fhatTM_dipole}). In that case, if $v_{\rm min} < v_{\rm lab}$, then the maximum of $\hat{ f} (v_{\rm min},\hat{\bf r })$ occurs 
when $-\hat{\bf r} \cdot {\bf v}_{\rm lab}=v_{\rm min}$, or at an angle $\gamma$ between $\hat{\bf r}$ and  
$-{\bf v}_{\rm lab}$ given by
\begin{equation}
\cos\gamma=\frac{v_{\rm min}}{v_{\rm lab}}=\sqrt{\frac{m_A E_r}{2 \mu^2 v_{\rm lab}^2}}.
\label{eq:gamma}
\end{equation}
The angular radius of the ring, $\gamma$, is larger for larger values of $v_{\rm lab}$, and can reach 90$^{\circ}$~\cite{Bozorgnia:2011vc}.  

The main uncertainty in $v_{\rm lab}$ is from the uncertainty in the Galactic rotation speed, $v_c$ (see Appendix~\ref{app-A} for the relation between ${\bf v}_c$ and ${\bf v}_{\rm lab}$). As mentioned in 
Sec.~\ref{sec:theo.halo}, there is a wide range of values for $v_c$ depending on the model used for the rotation curve. In particular 
Ref.~\cite{McMillan:2009yr} found values ranging from $v_c=(200 \pm 20)$ km s$^{-1}$ to $v_c=(279 \pm 33)$ km s$^{-1}$. In this 
Section and in Sec.~\ref{sec:secondarysignatures} we take $v_c=180$ km s$^{-1}$ and 312 km s$^{-1}$ as low and high estimates. Notice also that the times of maximum and minimum $v_{\rm lab}$ depend on the value of $v_c$, and are not exactly half a year apart due to the ellipticity of the Earth's orbit. 

\begin{figure}
\begin{center}
\begin{tabular}{cc}
  \includegraphics[height=140pt]{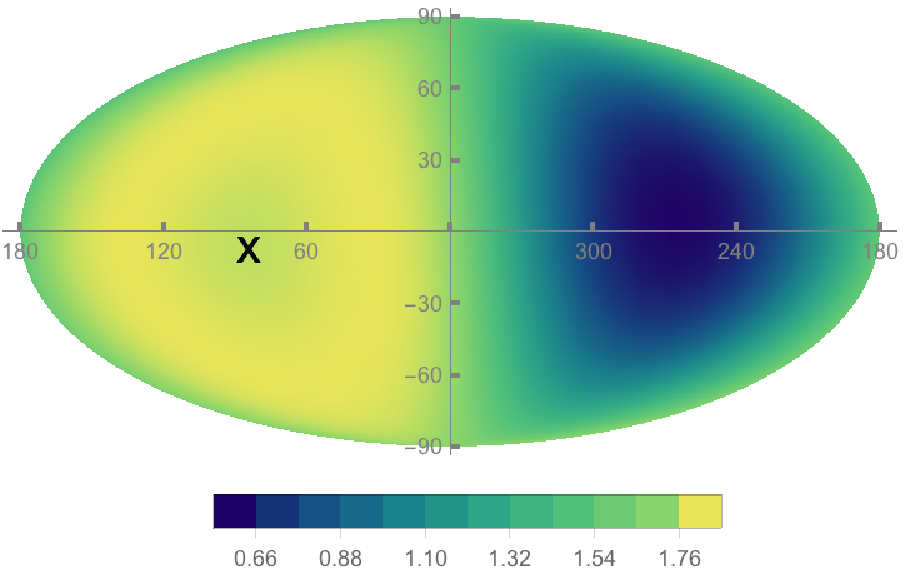} &
  \includegraphics[height=140pt]{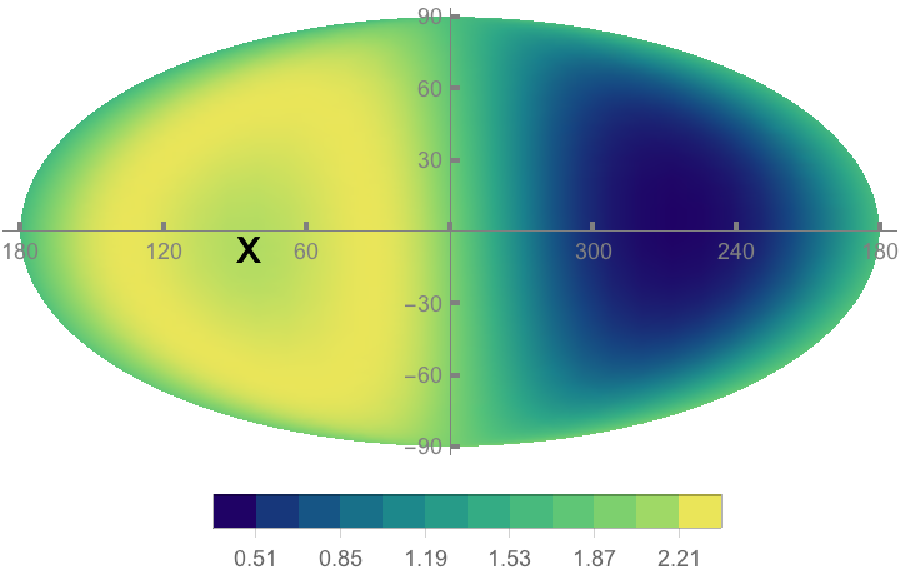}\\
  \includegraphics[height=140pt]{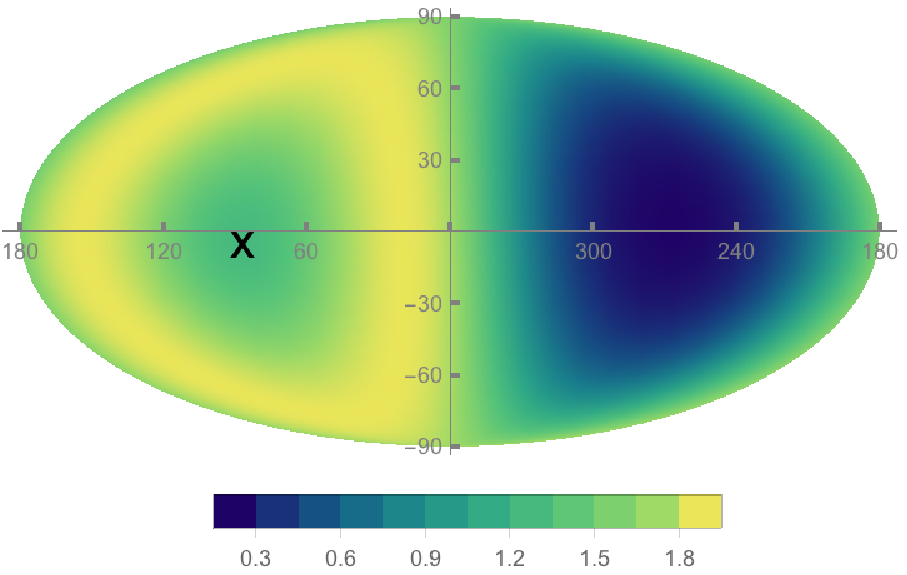} &
  \includegraphics[height=140pt]{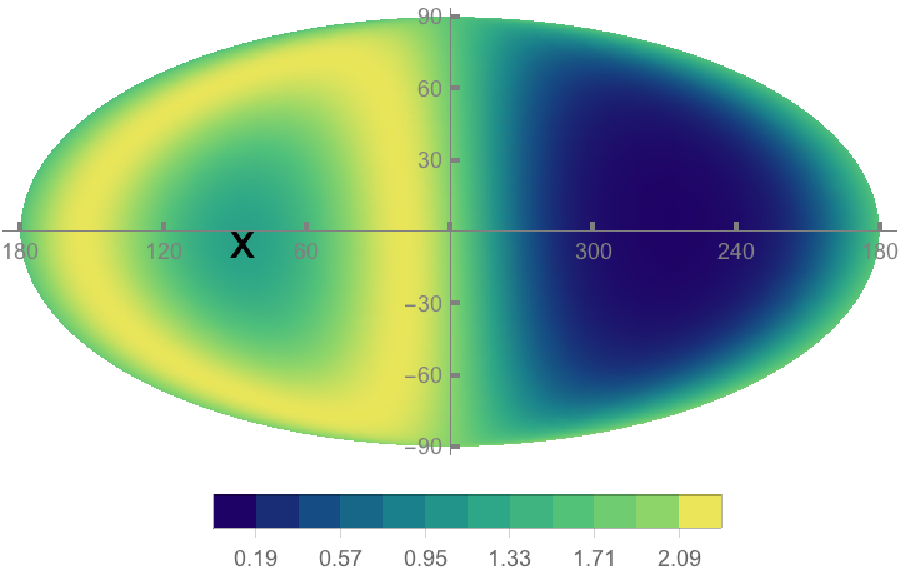}\\
  \end{tabular}
  %\vspace{-0.1cm}
 \caption{Directional differential recoil rate in F for different combinations of the Galactic rotation speed $v_c$ and the WIMP 
 velocity dispersion $\sigma_v$ shown in Mollweide equal-area projection of incoming directions in Galactic coordinates.
  Top left panel: $v_c=180$ km s$^{-1}$,  $\sigma_v=225$ km s$^{-1}$ on May 30;
  top right panel: $v_c=180$ km s$^{-1}$,  $\sigma_v=173$ km s$^{-1}$ on May 30;
  bottom left panel: $v_c=312$ km s$^{-1}$,  $\sigma_v=225$ km s$^{-1}$ on June 2;
  bottom right panel: $v_c=312$ km s$^{-1}$, $\sigma_v=173$ km s$^{-1}$ on June 2.
In all panels we assume $E_r=5$ keV,  $m_{\chi}=100$ GeV, and $v_{\rm esc}=544$ km s$^{-1}$. 
For convenience, we present the direction of ${\bf v}_{\rm lab}$ as a cross on the map.
The values corresponding to each color shown in 
the horizontal bars are given in units of $10^{-5} \times (\rho_{0.3} \sigma^{\rm SD}_{p, 40}/{\text{kg day keV sr}})$, where  $\rho_{0.3}$ is the Dark Matter density in units of 0.3 GeV cm$^{-3}$ and $\sigma^{\rm SD}_{p, 40}$ is the WIMP-proton spin-dependent cross 
section in units of $10^{-40}\;{\text{cm}}^2$ (consistent with current experimental limits from direct detection experiments). The angular radius of the ring is $\gamma=51^\circ$ and 67$^\circ$ in the top and bottom panels, respectively. }
%From Ref.~\cite{Bozorgnia:2011vc}.}
 \label{VGalRot-sigmav}
\end{center}
\end{figure}

The ring-like feature depends on $\hat{f}$, and the equations presented here refer to the case of WIMP-nucleus elastic scattering. The ring-like feature would be different for WIMP interaction types which modify the kinematics of the scattering ({\it e.g.} inelastic interaction) or change the form of the Radon transform $\hat{f}$ (Eq.~\ref{eq:fhatTM_dipole}).
To give examples of how the ring would appear in an actual recoil rate, in Fig.~\ref{VGalRot-sigmav} we show 
Mollweide equal-area projection maps of the incoming differential recoil rate in Galactic coordinates, 
assuming WIMPs with elastic SD interactions in a Fluorine detector, for $E_r=5$ keV, $m_\chi=100$ GeV, $v_{\rm esc}=544$ km s$^{-1}$, and four combinations of $v_{\rm lab}$ and $\sigma_v$. 
For convenience, we present the direction of ${\bf v}_{\rm lab}$ as a cross on the map.
%The direction of $-{\bf v}_{\rm lab}$ is shown with a cross on the maps. 
The ring of maximum recoil rate 
%FM 30/11/15 i think we do not need to specify again vlab (or -vlab)
%around $-{\bf v}_{\rm lab}$  
is clearly seen in all four panels of Fig.~\ref{VGalRot-sigmav}, but is most visible in the bottom right panel 
which has the smallest $\sigma_v$ and largest $v_{\rm lab}$ combination of the four. 

The ring is easiest to see when the rate at the ring has the largest contrast with respect 
to the rate at the center of the ring ($(l,b)=(\pi/2, 0)$). The ratio of the Radon transform at the center of the ring, $\hat{f}_{\rm center}$, to the Radon transform at the ring, $\hat{f}_{\rm ring}$, is approximately, $\hat{f}_{\rm center}/\hat{f}_{\rm ring} \simeq \exp{\left[-(v_{\rm lab}-v_{\rm min})^2/2 \sigma_v^2 \right]}$. Hence, the best chance to observe the ring is for large WIMP masses and low recoil energies (such that $v_{\rm min}$ is small), small  $\sigma_v$, and large $v_{\rm lab}$. We discuss the prospects of observing the ring-like feature 
in Sec. \ref{sec:secondarysignatures-ring}.
%
%The direction of the average WIMP velocity $-{\bf v}_{\rm lab}$ is marked with a cross. 

%BJK - 26/06/2015
We note that an enhanced ring-like feature may appear for certain non-relativistic operators \cite{Catena:2015vpa,Kavanagh:2015jma} described in 
Sec.~\ref{sec:theo.NR}. For operators coupling to $\mathbf{v}_\perp$, the recoil spectrum is governed by the transverse 
Radon Transform $\hat{f}^T(v_\mathrm{min}, \hat{\mathbf{r}})$  (Eq.~\ref{eq:trt}) which leads to a suppression of the 
rate when the incoming Dark Matter velocity is parallel to the recoil direction. This suppression leads to an enhanced 
ring, which is observable for larger values of $v_\mathrm{min}$, up to $v_\mathrm{min} \approx 2 v_\mathrm{lab}$. 
For these NR operators, then, a ring-like feature may be visible for WIMP masses as low as $20-30 \, \, \mathrm{GeV}$ or for energy 
thresholds up to $E_\mathrm{th} \sim 20 \, \, \mathrm{keV}$ for heavier WIMPs \cite{Kavanagh:2015jma}.

% Edited by FM 30/11/15 : modification + vlab and explanation of recoil maps
% Edited by FM 22/11/15 : minor corrections 
\subsection{Aberration features} 
\label{sec:featuresAberration}

  \begin{figure}
\begin{center}
  \includegraphics[scale=0.6]{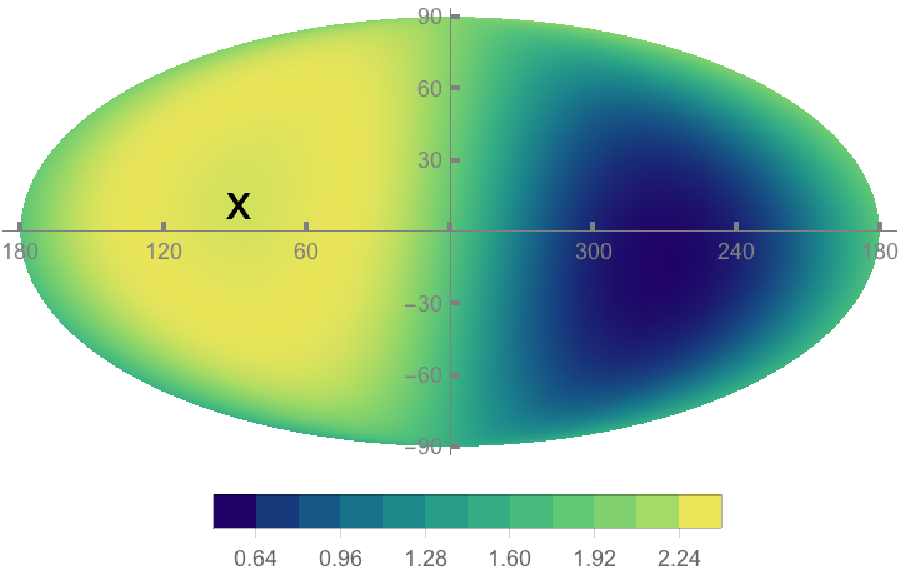} 
  \includegraphics[scale=0.6]{Sec3_May-180-173.eps} 
  \includegraphics[scale=0.6]{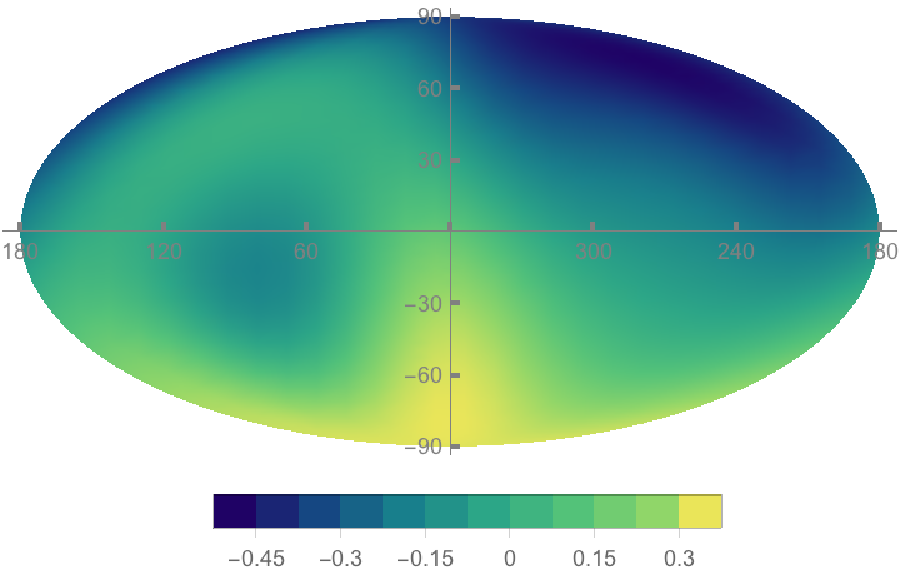}
%  \vspace{-0.1cm}
  \caption{Directional differential recoil rate at $E_r=5$ keV in F for a WIMP of mass $m_{\chi} = 100$ GeV. We assume $v_c=180$ km s$^{-1}$, $\sigma_v=173$ km s$^{-1}$ and $v_{\rm esc}=544$ km s$^{-1}$. Left panel: on December 1. Center panel: on 
  May 30 (same as the top right panel of Fig.~\ref{VGalRot-sigmav}).  For convenience, we present the direction of ${\bf v}_{\rm lab}$ as a cross on the map (in the NGH in December and in the SGH in May).
  Right panel: May $-$ Dec difference of the directional differential 
  recoil rates. The values corresponding to each color in the horizontal bars are given in units of $10^{-5} \times (\rho_{0.3} \sigma^{\rm SD}_{p, 40}/{\text{kg day keV sr}})$.}
%   From Ref.~\cite{Bozorgnia:2012eg}.}
\label{fig:aberration.map}
\end{center}
\end{figure}

Aberration features in a Dark Matter signal  are changes in the recoil direction pattern, due to the change in the direction of the arrival velocity of WIMPs on Earth caused by Earth's motion around the Sun \cite{Bozorgnia:2012eg}. Due to Earth's revolution around the Sun, the mean recoil direction will change with a period of one year. Aberration  features depend on the velocity of the Earth's revolution around the Sun and the local WIMP velocity distribution. Knowing the former, such features can be used to determine the latter.  

The annual change in the magnitude of ${\bf v}_{\rm lab}$ due to the motion of the Earth around the Sun causes the non-directional annual modulation of the rate~\cite{Drukier:1986tm}. Furthermore, the change in both magnitude and direction of ${\bf v}_{\rm lab}$ during the year causes the aberration features. As with the ring-like features, the aberration features depend on the Radon transform $\hat{f}$. Let us consider the annual variation of the directional differential recoil rate, 
\begin{equation}
\Delta\left(\frac{{\rm d}^2 R}{{\rm d}E_r {\rm d}\Omega_r}\right)=\frac{{\rm d}^2 R_2}{{\rm d}E_r {\rm d}\Omega_r}-
\frac{{\rm d}^2 R_1}{{\rm d}E_r {\rm d}\Omega_r},
\label{AnnualMod-direc}
\end{equation}
where the subscripts ``1" and ``2" refer to the times in a year when $v_{\rm lab}$ is minimum and maximum, respectively. To visualize the aberration patterns, one can plot this annual variation 
in a Mollweide map of the incoming recoil directions in Galactic coordinates.

To give specific examples, we choose WIMPs with elastic SD interactions in  a Fluorine detector. 
%In Sec.~\ref{sec:secondarysignatures-aberration} where we study the prospects of observing the aberration features, we also show examples of  WIMPs with SD interactions in CF$_4$. 
Fig.~\ref{fig:aberration.map} shows the directional  differential recoil rate in F at $E_r=5$ keV and assuming $v_c=180$ km s$^{-1}$ on December 1 (left panel) when $v_{\rm lab}$ is minimum, and  May 30 (center panel) when $v_{\rm lab}$ is maximum, plotted in Mollweide maps of the incoming recoil directions in Galactic coordinates  for a 100 GeV WIMP. The cross indicates the direction of ${\bf v}_{\rm lab}$. The difference between the December and May maps in the left and center panels is shown in the right panel, where a characteristic aberration pattern is visible.

Since the ecliptic plane is at an angle of $\sim 60^\circ$ with respect to the Galactic plane, the velocity of the Earth's revolution points 
toward the North Galactic Hemisphere (NGH) for half of the year and toward the South Galactic Hemisphere (SGH) for the other half. This causes the component of ${\bf v}_{\rm lab}$ perpendicular to the Galactic plane to point 
slightly toward the NGH in December and slightly toward the SGH half a year later (see Fig.~1 of Ref.~\cite{Bozorgnia:2012eg}). 
%Due to the change in the direction of $-{\bf v}_{\rm lab}$ from the NGH in May (WIMPs coming from SGH) to the SGH in December (WIMPs coming from NGH), 
Therefore, in the right panel of Fig.~\ref{fig:aberration.map}  the positive directional differential rate differences (lighter regions) are mostly in the SGH and the negative rate differences (darker regions) are mostly in the NGH.

The aberration effect persists in the energy and time integrated rates~\cite{Bozorgnia:2012eg}, and such patterns require a very large number of events to be observed. We study the detectability of the aberration features 
in  Sec.~\ref{sec:secondarysignatures-aberration}. Notice that Solar gravitational 
focusing~\cite{Griest:1987vc, Sikivie:2002bj, Alenazi:2006wu, Lee:2013wza, Bozorgnia:2014dqa, DelNobile:2015nua} can potentially affect the aberration features, but it is neglected in this work for simplicity.

%

%\input{features_aberration3e}

%DirectionalReview.log
% -------------- exclusion
%5. Directional exclusion : J. Monroe and S. Vahsen  
 
% Edited by FM 06/02/16: subsection inelastic DM moved to X. introduction modified
% Edited by SV 1/27/16: updated inelastic DM section as suggested by referee
% Edited by SV 1/27/16: p.d.f. --> PDF, and defined it upon first usage
% Edited by SV 1/27/16: removed commented-out text
% Edited by FM 27/11/15 : link to def. of lambda (sec. exp)
% Edited by FM 16/11/15 : minor corrections (following comments from AMG and GBG)

% Directional exclusion : J. Monroe and S. Vahsen (Jocelyn.Monroe@rhul.ac.uk, sevahsen@hawaii.edu)

\section{Setting limits on WIMP-nucleon scattering \label{sec:limits}} %Sven
\label{sec:exclusion}
\subsection{Introduction}
When the significance of a WIMP signal is insufficient to claim a discovery, one instead sets 
an upper limit on the WIMP-nucleon interaction cross section, and hence excludes a set of 
WIMP model parameter space. A directional detector by definition provides more information about 
each WIMP recoil candidate than a non-directional detector. The former measures the double-differential 
spectrum (Eq.~\ref{eq:d2r}), while the latter measures just the energy spectrum (Eq.~\ref{eq:dRdEr}). 
The additional angular information measured by a directional detector can improve the separation of 
signal and background. This is generally true for any background which has an angular 
distribution that differs from that of the WIMP recoil signal, however the advantage is 
perhaps most striking when the background itself has strong directional features, as is the case for Solar
 neutrino-induced recoils. This particular background is irreducible event-by-event except by directionality. 
 As a result, directional detectors can set stronger exclusion limits than non-directional detectors, 
 for equal exposures. Below, we   discuss several statistical methods that have been developed to 
 set improved limits with directional detectors, by incorporating both energy and angular information.

\subsection{Maximum Patch Method} %Sven & Jocelyn

The Maximum Gap Method \cite{Yellin:2002xd} has been used to set limits in non-directional Dark Matter experiments when the 
background distribution is not {\it a priori} known. The basic idea is straightforward: the largest range of energy where no events 
have been recorded ({\it i.e.} the maximum energy gap) is used to calculate the cross section limit. The method thus automatically 
excludes energy regions where unexpected backgrounds may be found after unblinding a data sample, and the corresponding statistical penalty factor is taken into account. The limit calculation assumes the theoretical prediction for the one-dimensional energy spectrum as the signal probability density function (PDF), and uses the number of events and resulting gap sizes measured in experimental data. The Maximum Patch Method \cite{Henderson:2008bn} generalizes the maximum gap method to two dimensions, by using the double-differential spectrum as the signal PDF and looking for an event-free `patch' in 
energy-recoil-angle space. 
As expected, pseudo experiments show that the maximum patch method with data from a directional detector 
is more sensitive than the maximum gap method with a non-directional detector. Both outperform the traditional Poisson method, which counts all events observed as signal candidates and lacks knowledge of the expected signal shape. 

The improvement seen with the maximum patch procedure is most dramatic when the number of background events is large, and when the WIMP mass is low.  The former because including the angle dimension can improve sensitivity for a fixed energy window (as would be selected by the maximum gap method using energy only), and the latter because of the relatively steeper shape of the recoil angle versus energy distribution for low versus high masses. This method can result in a limit that is up to one order of magnitude improved relative to the energy-only maximum gap case. Ref.~\cite{Henderson:2008bn}, however, cautions that the quantitative degree of improvement depends on the distribution of measured events. Indeed, the limit set using this method can be either better or worse than the energy-only limit, depending on the background distribution and number of background events.  This occurs because the maximum gap (and patch) methods restrict the experiment's acceptance to the fraction of parameter space in which there are zero events.  If this gap (or patch) is very small, or happens to exclude the regions of energy and/or angle parameter space that have the strongest parameter dependence of the signal distribution, then the resulting sensitivity can be worse than for an experiment with a 
larger acceptance ({\it e.g.} a lower energy threshold) with a small population of background events.  
Reference~\cite{Yellin:2002xd} shows that this can be ameliorated by choosing the ``optimum'' (rather than maximum) gap, which is the region of parameter space that maximizes sensitivity including a non-zero number of background events.  This is generally true for both directional and non-directional searches.

This maximum patch method is general, and can be used for any combination of $n$ variables with discriminating power to distinguish signal from background.  The maximum patch method has been employed beyond directional searches~\cite{zepliniii_inelasticpaper} to search for inelastic dark matter in the parameter space of energy and particle identification.  In the analysis of Ref.~\cite{zepliniii_inelasticpaper}, the resulting sensitivity is significantly improved relative to an energy-only analysis.

\subsection{Likelihood Test} %Sven

While the maximum patch method assumes that neither the angular nor the energy distribution of the background events is known, it is generally expected that the recoil angle distribution from backgrounds is isotropic in Galactic coordinates, 
{\it e.g.} in~\cite{Billard:2010gp,Grothaus:2014hja}.  Under this assumption, both the signal PDF and background PDF for recoil angle distribution are known. For example, Ref.~\cite{Billard:2010gp} uses these two distributions (only) to perform a Bayesian calculation of exclusion limits using a likelihood analysis. To remain robust against assumptions on the distributions of backgrounds, energy information is not used. This likelihood method is compared for a number of scenarios against the traditional Poisson method and the maximum gap method (using the recoil angle only), by performing pseudo-experiments and comparing the median exclusion limits set by each method.

Under the assumption that the background and signal angular distribution PDFs are known, it is found that the likelihood test generally outperforms the two other methods as soon as background events are present, and the advantage increases with increasing background contamination. Hence the likelihood method is promising for setting stringent limits with data from small, prototype directional dark matter detectors, which are not expected to have sufficient exposure to observe a WIMP signal. The improved performance of the likelihood method is consistent with what one might intuitively expect, based on the information available to each of the three methods being compared. The Poisson method only uses event counts to set limits, and completely ignores angular information available from a directional detector. The maximum gap method uses only the signal PDF, in recoil angle in Ref.~\cite{Billard:2010gp} or in recoil angle versus energy in Ref.~\cite{Henderson:2008bn}, to set limits, but assumes the background PDF is unknown. In contrast, the likelihood method assumes knowledge of the signal and background recoil angle PDFs, and uses them both to set limits. 

\begin{figure}
\begin{center}
\includegraphics[scale=0.35,angle=270]{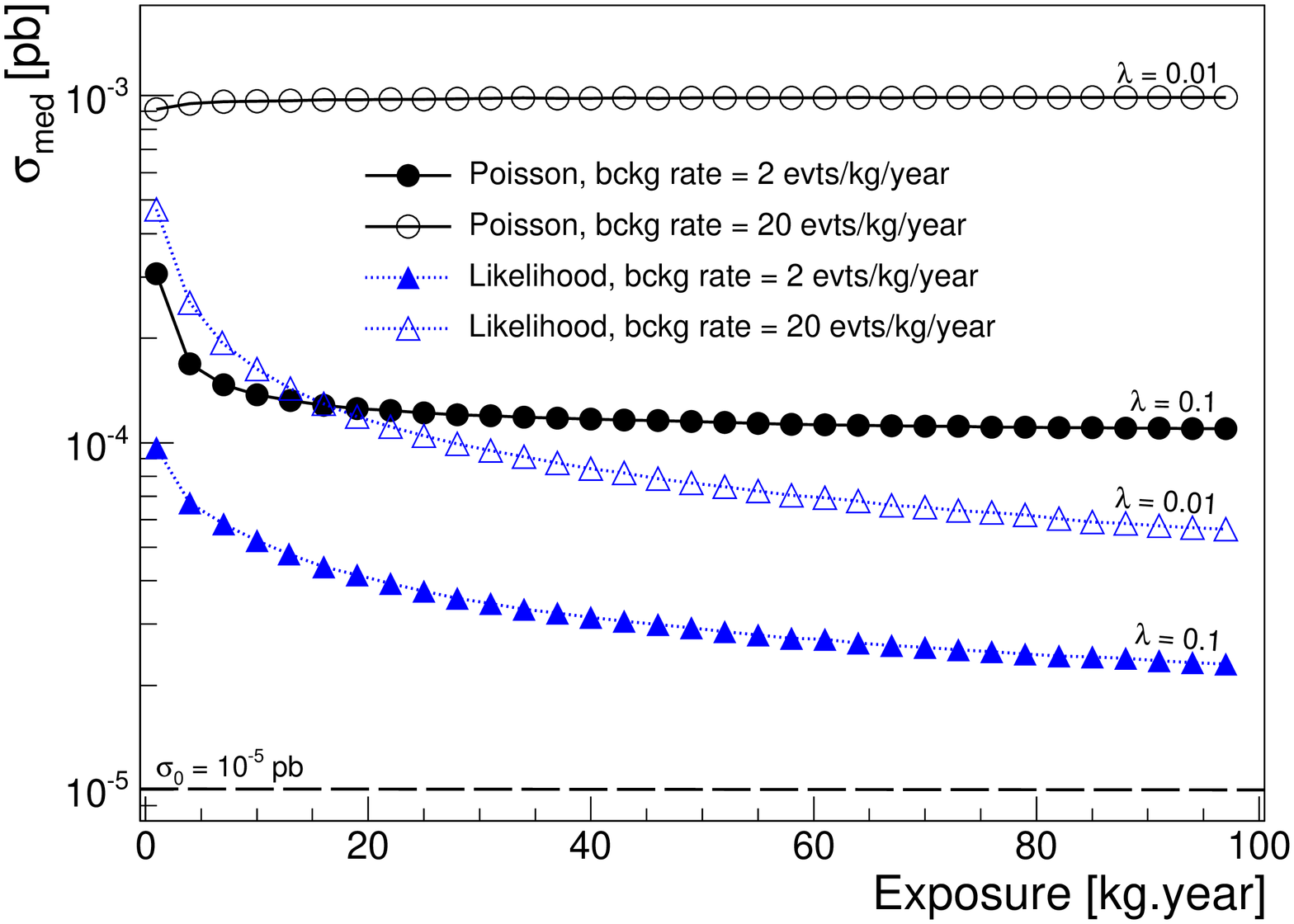}\includegraphics[scale=0.35,angle=270]{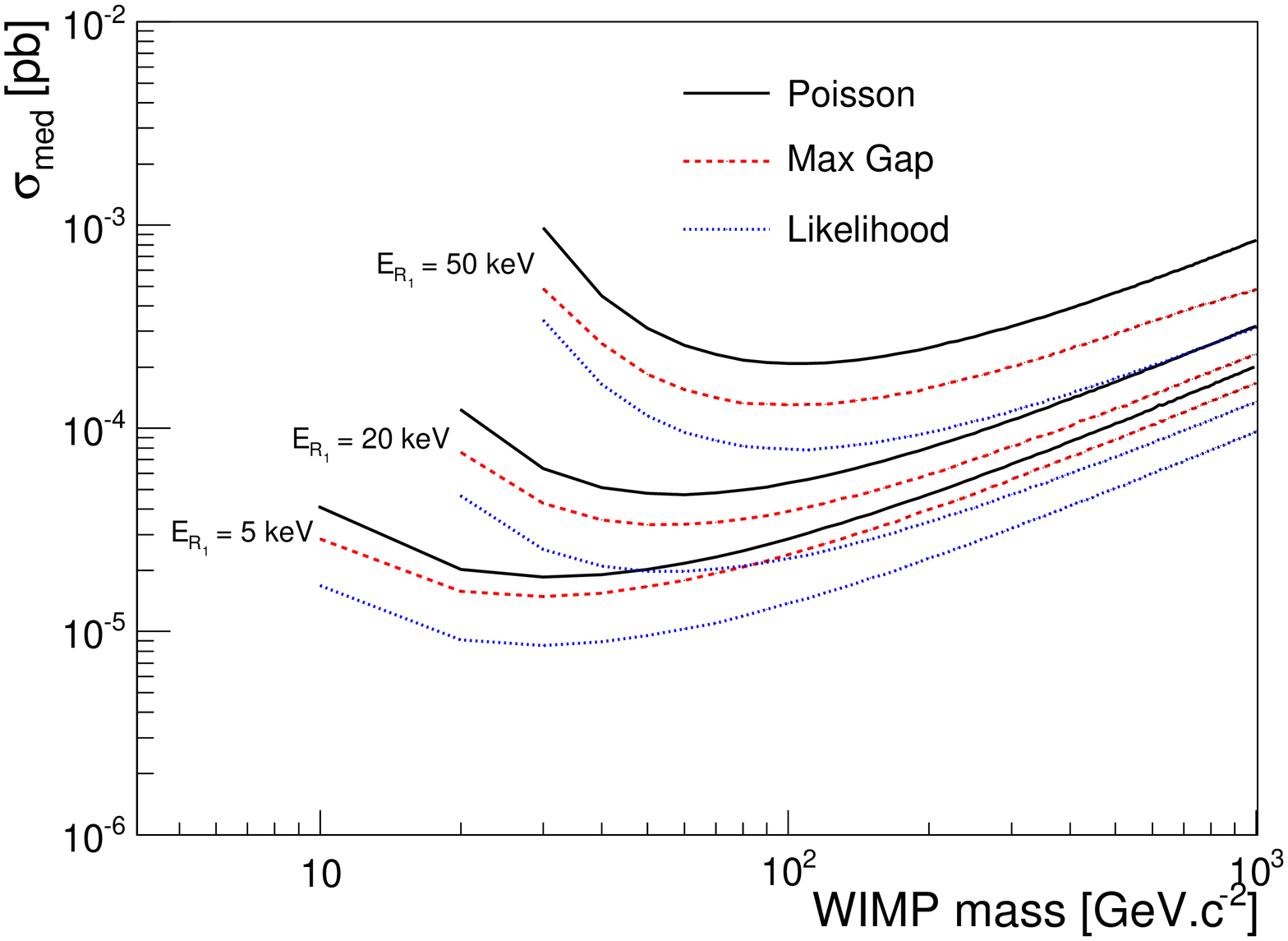}
\caption{Left: Median spin-dependent WIMP-proton cross section exclusion limits, obtained by analyzing a sample of Monte Carlo pseudo experiments with the Poisson and likelihood methods, versus simulated exposure. This particular study used a fixed WIMP mass $m_{\chi} = 100$~GeV, and an input cross section $\sigma_0 = 10^{-5}$ pb, yielding an expected WIMP signal event rate of 0.2 evts/kg/year in the detector simulated. Two different fractions of WIMP events, 
$\lambda$ (defined by Eq.~\ref{eq:def.lambda}), and two different background event rates were considered, as shown in the figure. Right: spin-dependent WIMP-proton cross section exclusion limits versus WIMP mass, for three statistical methods and three detector energy thresholds, as shown in the figure. This particular study considered the case of 10 expected background events and one expected WIMP event, an idealized detector with perfect direction reconstruction and perfect sense recognition, and included recoil energies below 200 keV. Both figures from Ref.~\cite{Billard:2010gp}.}
  \label{fig:exclusion_1}
\end{center}
\end{figure}

Quantitatively, for the case of background only, the likelihood method sets approximately three times stricter limits than the maximum gap and Poisson methods.  With five signal events present in addition to the thirty background events, the likelihood method does about a factor of 2 better than the other methods. Figure \ref{fig:exclusion_1} (left) shows the median excluded cross section obtained with the Poisson and likelihood methods as a function of the exposure, for two different background rates, and for two different assumptions on the 
WIMP event fraction (defined by Eq.~\ref{eq:def.lambda}). As the exposure increases, the Poisson-based cross section limits become asymptotic, while the limits based on the likelihood method keep approaching the actual signal cross section. The limits from the likelihood method improve as 
the inverse square root of the exposure. This is an important result: even in the presence of large, irreducible backgrounds, a directional detector will be able to set improved exclusion limits with increasing exposure.

The effect of detector performance on WIMP exclusion limits is considered in Ref.~\cite{Billard:2010gp}. While sense recognition (cf. Sec~\ref{sec:exp}.C) is known to be important for positive identification of a directional WIMP signal, it is found here that it has a factor of 2-3 effect on exclusion limits.  The effect of detector angular resolution on directional exclusion limits is also similar. The detector energy threshold is of primary importance, in particular for low WIMP masses, as shown in Fig. \ref{fig:exclusion_1} (right). The effect of astrophysical uncertainties are estimated by varying the asymptotic circular velocity $v_c$ from $170 \ {\rm km\ s^{-1}}$  to $270  \ {\rm km\ s^{-1}}$, compared to the standard value of $220 \ {\rm km\ s^{-1}}$.  This has a small (20\%-50\%) effect on exclusion limits, where the largest effects are seen for low-mass ($10 \ {\rm GeV}$) WIMPs.

Ref.~\cite{Billard:2010gp} presents expected limits using the likelihood method for a prototype detector as proposed by the 
MIMAC collaboration \cite{Santos:2011kf}. Given a 10-kg-year exposure, 10$^o$ angular resolution, and a recoil energy range of 5~-~50~keV, such a detector would be able to exclude spin-dependent WIMP-proton scattering down to cross sections of $2\times 10^{-6}$~pb for background free running. With backgrounds at the rate of 10/kg/year, cross sections down to $10^{-4}$~pb can be excluded by a detector without sense recognition. With sense recognition, the limits improve by a factor of 2-3. These limits are better than experimental spin-dependent limits at the time of writing, though the field is changing quickly.

\subsection{$\chi^2$ Test} %Jocelyn

The NEWAGE collaboration have developed a method based on a $\chi^2$ test to set dark matter limits incorporating both energy and directional information~\cite{Miuchi:2010hn}.  In fact this is the only published limit from a directional experiment that uses the measured direction in the limit setting procedure.  In this analysis the 3D track axial direction is measured, and therefore the $\cos~\theta$ distribution is folded into $|\cos~\theta|$, where $\theta$ is the measured recoil angle relative to the direction to Cygnus at the time of the event.  The limit-setting procedure performs a $\chi^2$ test between the measured and predicted $|\cos~\theta|$ distributions at each dark matter mass, in bins of recoil energy, varying the Dark Matter WIMP-nucleon cross section in order to determine the 90\% confidence level value of the cross section.  

To produce the $|\cos~\theta|$ predicted distribution, the differential rate of recoil events as a function of energy $E_r$ and angle $\cos~\theta$, $\frac{dR}{dE_r d(\cos~\theta)}$ is calculated for each dark matter mass.  To generate the angles, the simulation samples the direction to Cygnus over the live time of the detector.  The simulation accounts for the quenching of the energy loss to convert $E_r$ into visible energy $E_{vis}$ measured in electron-equivalent keV (keVee), smears the event by the measured angular and energy resolutions, and weights the event by the detection efficiency, which is a function of the energy and track angle in the local detector coordinates.  In this way, the simulation produces a $|\cos~\theta|$ distribution, in bins of visible energy, that can be compared with the measured $|\cos~\theta|$ distribution.

A $\chi^2$ test is performed between the measured and predicted $|\cos~\theta|$ distributions in order to set a limit, in a raster scan over candidate dark matter masses.  
All detected   nuclear recoil events passing cuts are considered as signal WIMP events, with no background subtraction.  Given the angular resolution (55$^o$ at 100 keVee) and the statistics, the $\chi^2$ fit is performed between measured and predicted distributions in two-bin $|\cos~\theta|$ distributions in order to find the best-fit cross section for a given dark matter mass, for each of 15 energy bins spanning the range 100-400 keVee independently.  

As presented in Ref.~\cite{Miuchi:2010hn}, the best fit cross section is 5500 pb at 100~GeV, with a $\chi^2$ per degree of freedom of 3.7, and the most stringent resulting 90\% confidence level limit is 5400 pb at 150~GeV.  This procedure is also used to test the compatibility of the data with an isotropic angular distribution.  The $\chi^2$ per degree of freedom in this case is 0.11.  In this way, the NEWAGE experiment have used the $\chi^2$ test on the measured $|\cos~\theta|$ distribution to 
disfavor the dark matter hypothesis relative to the presence of an isotropic background.

%
% -------------- discovery
%6. Discovery : J. Billard and A. Green  
%\newpage
% Edited by FM 25/1/16 : one sentence added to explain the choice of a Xe target
%for the study beyond the neutrino floor.
% Edited by FM 30/11/15 : modify "inverse of the direction of Solar motion to
%fit convention for skypmaps.
% Edited by FM 27/11/15 : link to def. of lambda (sec. exp)
% Edited by FM 16/11/15 : minor corrections (following comments from AMG and GBG)

% 1/10/15 : FM and JB : new figure
% AMG: edited on 03/05/15 to address comments from Bradley and Nguyen
% on Secs. A, B, first parts of C & D, and E 
%  VERSIOn received from JB on 20/05/14
%
%  Edited by AMG on 27/05/14
%
% -------------- discovery
%6. Discovery : J. Billard and A. Green (anne.green@nottingham.ac.uk,j.billard@ipnl.in2p3.fr)
\section{Directional discovery}
\label{sec:discovery}

\subsection{Introduction}

In the previous section we discussed the power of directional
detection to set limits on the WIMP-nucleon interactions. In this
section, we consider how directional detection can be used to confirm
the WIMP origin of recoil events, once s significant excess has been observed.

The directional recoil rate depends on both of the angles which
specify a given direction, however the strongest signal is the event
rate with respect to the angle between the recoil direction and the
direction of Solar motion~\cite{Spergel:1987kx,Alenazi:2007sy}.
Spergel found that the event rate in the forward direction is up to an
order of magnitude larger than that in the backward
direction~\cite{Spergel:1987kx}. Copi and Krauss
showed that consequently a WIMP signal could be distinguished from backgrounds
with as few as 30 events~\cite{Copi:1999pw}. 
% AMG: Have added a mention of Copi & Krauss to give appropriate
% credit (given we mention Morgan {\it et al.} first below).

The WIMP search strategy for a directional detector can be divided
into two phases. The first step is to check that the recoils are
anisotropic, and hence unlikely to be due to backgrounds, see Sec. \ref{sec:disco.aniso}. 
The next step is to measure the mean recoil direction and check
that it coincides with  the direction of Solar
motion~\footnote{As discussed in Sec.~\ref{sec:features}, for very low recoil energies the rate is maximum in a ring around the  
average WIMP arrival direction~\cite{Bozorgnia:2011tk,Bozorgnia:2011vc}.}, 
see Sec. \ref{sec:disco.meandirection}. This would provide robust 
confirmation of the Galactic origin
of the recoil events. 
In 
Sec.~\ref{subsec-nu} we discuss how directionality can overcome the
neutrino floor.

\subsection{Detecting anisotropy}
\label{sec:disco.aniso}
Most of the background contributions to directional detectors
are expected to have distributions that are close to isotropic in the Galactic frame. % AMG: is there a reference we can cite for this?
Therefore a detection of anisotropy would strongly suggest a Dark Matter origin for (at least some of)
the recoils.
% AMG: I've rewritten this, but feel free to rewrite again if you
% think what I've written is too strong.
Copi and Krauss first investigated such
detection of anisotropy within a model-dependent approach, {\it i.e.}
taking into account the expected angular distributions for both the
WIMP and background induced nuclear recoils~\cite{Copi:1999pw,Copi:2000tv}. Using an unbinned
likelihood analysis they showed that, for plausible halo models,
the angular distribution of WIMP induced recoils could be
discriminated from isotropic backgrounds with a reasonable number of
events, depending on the background contamination and the energy
threshold of the experiment considered. For example, they found that
only 50 events would be required to reject the isotropy hypothesis in
the case of the Standard Halo Model and a WIMP fraction of $\lambda=0.5$.

Morgan, Green and Spooner studied non-parametric
spherical statistics for detecting anisotropy~\cite{Morgan:2004ys}. These non-parametric
statistics have the advantage of not requiring any assumptions to be
made about the direction dependence of the WIMP recoil rate (or
equivalently the WIMP velocity distribution). As the nuclear recoils
induced by WIMPs are expected to exhibit a dipole-like angular distribution, the most powerful test for
rejecting isotropy uses the mean angle between the observed 
recoil directions
and the direction of motion of the Sun: 
\begin{equation}
\label{costheta}
\langle \cos{\theta} \rangle = \frac{\sum_{i=1}^{N}
  \cos{\theta_{i}}}{N} \,,
\end{equation}
where $\theta_{i}$ is the 3d angle between the direction of Solar
motion and the $i$-th  recoil vector and $N$ is the number of events.
For isotropic recoils $\langle \cos{\theta} \rangle $ can take values
in the range $[-1, 1]$ and, due to the central limit theorem, for a
sufficiently large number of events $\langle \cos{\theta} \rangle $
approaches a Gaussian distribution with mean zero. On the other hand for
WIMP induced recoils $\langle \cos{\theta} \rangle$ will be positive,
due to the concentration of the WIMP flux around   the direction of
Solar motion.  Using this statistic an ideal detector, with zero
background, which can measure the sense ($+\bhat{r}$ vs. $-\bhat{r}$)
of nuclear recoils in 3d with good angular resolution would require
only 10 events to reject isotropy at 95\% confidence.  
% AMG: now discussed in more detail below.
%If the senses
%can not be measured, or the directions are only measured in 2d, then
%the number of events required is increased by roughly an order of
%magnitude~\cite{Morgan:2005sq,Copi:2005ya,Green:2007at}. 
With such a model-independent method the number of events required to
reject isotropy is highly sensitive to the background
contamination as it is not subtracted. For  WIMP fractions $\lambda = 0.5$ and $0.09$, the number of WIMP induced events required increases to $27$ and
$170$ respectively.
% AMG: have changed references to background rate here, and later, to
% lambda, for consistency. 
%J.B: Good point

\subsection{Measuring the mean recoil direction}
\label{sec:disco.meandirection}
Once anisotropy of the nuclear recoil distribution has been detected,
the Galactic origin of the observed events can be established by
measuring the mean recoil direction. If the mean recoil direction is
found to be consistent, within a few degrees, with   the
direction of motion of the Sun, $(l_{\odot}, b_{\odot})$ in Galactic
coordinates, then Dark Matter will be unambiguously discovered.
%AMG: Need to define direction of Solar motion somewhere. Not sure if
%this is the best place.
%J.B: I agree, maybe in the previous chapters when we show an actual angular distribution of WIMP events?

Green and Morgan proposed using the median recoil
direction to probe the mean recoil direction~\cite{Green:2010zm}. The median direction is
defined~\cite{fle} as the direction, $\bhat{r}_{\rm med}$, which
minimizes the sum
\begin{equation}
{\cal M} = \sum_{i=1}^{N} {\rm cos}^{-1} (\bhat{r}_{\rm med} \cdot \bhat{r}_{i}) \,,
\end{equation}
where $\bhat{r}_{i}$ are the recoil directions. They found that with $\sim 30$ events an ideal
detector could demonstrate that the median recoil direction coincides
with the   direction of Solar motion at 95\%
confidence. As discussed above, as backgrounds are not
subtracted in this model-independent approach,  non-zero isotropic
backgrounds would increase this number, significantly if
the signal is subdominant.  For instance for $\lambda=0.1$ roughly  $300$ events would be required. 

\begin{figure}[t]
\begin{center}
%\hspace*{-3mm}
\includegraphics[scale=0.42,angle=0]{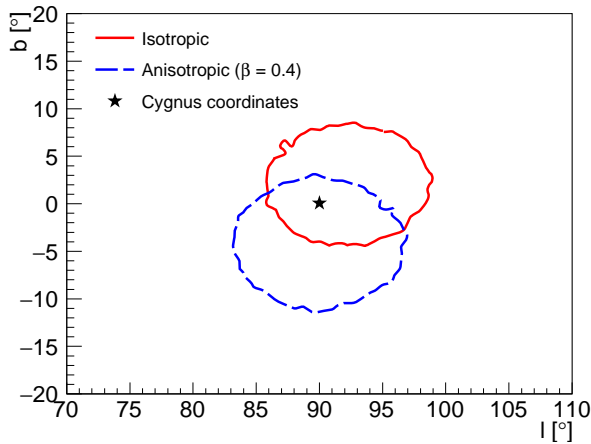}
\caption{95\% contour level in the ($\ell_{\odot},b_{\odot}$) plan  for two input halo models: Isotropic  (red solid line) and radially
anisotropic with $\beta = 0.4$ (blue solid line), see Eq. \ref{eq:beta} for a definition of $\beta$. These results have been obtained for a 50 GeV WIMP mass with a $10^{-3}$ pb SD-p cross section, a 10 kg CF$_4$ directional detector with an energy threshold of 5 keV, 3 years of exposition time and a flat background component of 10 events/kg/year. 
Figure adapted from Ref. \cite{Billard:2010jh}.}
\label{fig:discovariousmodel}
 \end{center}
\end{figure}

Billard {\it et al.}~\cite{Billard:2009mf}
investigated measuring the peak recoil direction with a map-based
likelihood method using  Poisson statistics:
 \begin{equation}
 \mathscr{L}(m_\chi,\lambda, \ell,b) = \prod_{i=1}^{N_{\rm pixels}} P(
 [(1-\lambda) B_i + \lambda S_i(m_\chi ;\ell,b) ]|M_i) \,,
 \end{equation}
 where $N_{\rm pixels}$ is the number of pixels, $B_{ i}$, $S_{ i}$
 and $M_{ i}$ are the background, WIMP-induced and observed recoil
 rates in each pixel respectively. % AMG: is this rewrite correct? J.B: Absolutely!
 The likelihood analysis has four free parameters: the WIMP mass,
 $m_\chi$, the WIMP fraction, $\lambda$ (see Eq. \ref{eq:def.lambda}), and the Galactic
 coordinates of the direction in which the WIMP directional rate is
 maximum, ($\ell$, $b$). % AMG: is this rewrite correct? J.B: Absolutely!
 Hence, $S(m_\chi;\ell,b)$ corresponds to a rotation of the
 $S(m_\chi)$ distribution by the angles ($\ell' = \ell - \ell_\odot$,
 $b' = b - b_\odot$).
% AMG: not sure I understand this bit. J.B: It's because we want to define the maximum of the distribution to coincide with the l_dot and b_dot angles.
 They used $N_{\rm pixels} = 768$ to take into
 account the rather low ($15^\circ$ FWHM) angular resolution of the
 recoil direction. As well as measuring the peak direction of the
 observed recoil distribution, and checking its consistency with the
 direction of Solar motion, this approach allows the background
 contribution to be subtracted so that the fraction of WIMP events,
 and therefore the WIMP-nucleon cross section, can be constrained as a
 function of the WIMP mass.
% I think it would be great to have a plot such as Fig. 6 from http://arxiv.org/abs/1012.3960 to illustrate both the point we are making and the robustness against halo models
% FM : done
They found that, assuming 3d read-out and sense recognition, the peak direction could be
confirmed to be within $20^{\circ}$ degrees of the direction of Solar
motion, with as few as 25 WIMP events, even with non-negligible
backgrounds. In fact even for a WIMP fraction as low as $\lambda=0.1$ the
uncertainty in the peak direction is only increased by a factor of two, which demonstrates the
robustness of this approach to background contamination of the
data. Finally, this approach also allows the Bayesian significance of
a potential WIMP signal to be estimated. For $\lambda=1/3$ only 25
WIMP events are required to reach the 3$\sigma$ significance required
for the
discovery of Dark Matter. In Ref. \cite{Billard:2012qu}, Billard {\it et al.} extended this analysis technique to 
non-standard halo models and confirmed that recovering the main incoming direction of the event direction 
is a robust signature of a positive DM detection. Indeed, Fig.~\ref{fig:discovariousmodel} shows two 95\% C.L. contours in the ($\ell$, $b$) plane derived from the likelihood analyses of two different data sets comprising 300 WIMP events and 1000 background events but considering two different halo models:  isotropic (red contour) and strongly radially 
anisotropic with $\beta = 0.4$ (blue contour), see Eq. \ref{eq:beta} for a definition of $\beta$. In both cases, the main incoming direction of the WIMP events are fully consistent with the direction of the Solar motion.
%AMG: is my rewrite OK?  If we want to talk about discovery would it
%be better to quote number of events required for 5 sigma?

%BJK - 26/06/2015
As discussed in Sec.~\ref{sec:theo.NR}, considering more general WIMP-nucleon interactions may affect the directionality of the signal and therefore the number of events required to reject isotropy or confirm the median recoil direction. Using the framework of non-relativistic (NR) effective theory, Ref.~\cite{Kavanagh:2015jma} concludes that there may be roughly a factor of 2 uncertainty on the number of events required, owing to a lack of knowledge about which operator mediates the interaction.

\subsection{Experimental considerations}

% AMG: Put this here for now. Needs to be merged with other material.
Regardless of the specific data analysis techniques used, the
directional energy threshold plays an important role in determining the
number of events, and hence exposure, required to detect an anisotropic
WIMP signal. The high energy events are more anisotropic, therefore as
the energy threshold is increased the number of events required
decreases, however the event rate above threshold also decreases significantly. In
fact only if the energy threshold is extremely low is it possible to
decrease the exposure required by slightly increasing  the energy
threshold of the experiment~\cite{Green:2006cb}.

% AMG first go at a summary of Green and Morgan (2007).
Green and Morgan investigated how the number of
events required to reject isotropy using model independent statistics
depends on the detector capabilities, in particular whether the
recoil vectors are measured in two or three dimensions and whether or
not the senses of the recoils can be measured~\cite{Green:2006cb}. 
If the recoil senses are not measured, {\it i.e.} the data is axial rather
than vectorial, then the test
statistic, Eq.(\ref{costheta}), is modified to
\begin{equation}
\langle |\cos{\theta} | \rangle = \frac{\sum_{i=1}^{N}
 | \cos{\theta_{i}}|}{N} \,,
\end{equation}
and in this case for isotropic backgrounds $\langle |\cos{\theta} |
\rangle$ takes values in the range $[0,1]$ with a mean, in the large
$N$ limit, of 0.5. For 2d read-out the most powerful test~\cite{Morgan:2005sq} is the
Rayleigh test~\cite{rayleigh} which uses the mean resultant length of the projected
recoil vectors. % AMG: don't think we want/need to go into details. J.B: I am OK with it. Some plots might help... 
With 2d axial data the standard procedure is to double the
axial angles, reduce them modulo $360^{\circ}$ and analyze the
resulting vectorial data~\cite{mardia:jupp}.

They found that the detector property which has the largest effect on
the number of events required to reject isotropy is whether or not the
sense of the recoils can be measured. If the senses cannot be measured
then the number of events is increased by one order of magnitude for
3d read-out and at least two orders of magnitude for 2d read-out. For
2d read-out this can be reduced to a factor of $\sim 30$ if the
reduced angles (with the direction of Solar motion subtracted) are
analyzed instead of the raw recoil angles. If the senses are measured then for
2d read-out in the optimal plane (which has normal perpendicular the
Earth's spin axis) the required exposure is increased by a factor of $\sim$ 3
relative to 3d read-out. Again this can be reduced, to a factor of
$\sim$ 2, by using the reduced angles. Note, however, that these
numbers assume perfect angular resolution and, due to projection effects,
angular resolution will be a more significant factor in 2d than
in 3d.

Green and Morgan also investigated the consequences of statistical,
rather than event by event, sense determination~\cite{Green:2007at}. If the probability of correctly determining
the sense is greater than 0.75, then the number of events required to
reject isotropy using non-parametric statistics is increased by at most a factor of a few. As the
probability is decreased below 0.75 then the number of events
increases sharply, and eventually it is better to discard the sense
information and instead use axial statistics. Note that in the case of a likelihood-based analysis considering also the energy information, not being able to recover the sense of the recoil only impacts the sensitivity of the experiment by about a factor 4 at high WIMP mass~\cite{Billard:2011zj}. By considering energy dependent
sense determination probabilities, they ascertained that correctly determining the sense of
the abundant, but less anisotropic, low energy recoils is most
important for minimizing the number of events required.

With model independent techniques, the background rate effectively
places a lower limit on the WIMP cross-section to which the detector
is sensitive. It would be very difficult to detect WIMPs using these
techniques if the signal rate is more than an order of magnitude below the background rate.\\

\begin{figure}[t]
\begin{center}
\includegraphics[scale=0.42,angle=0]{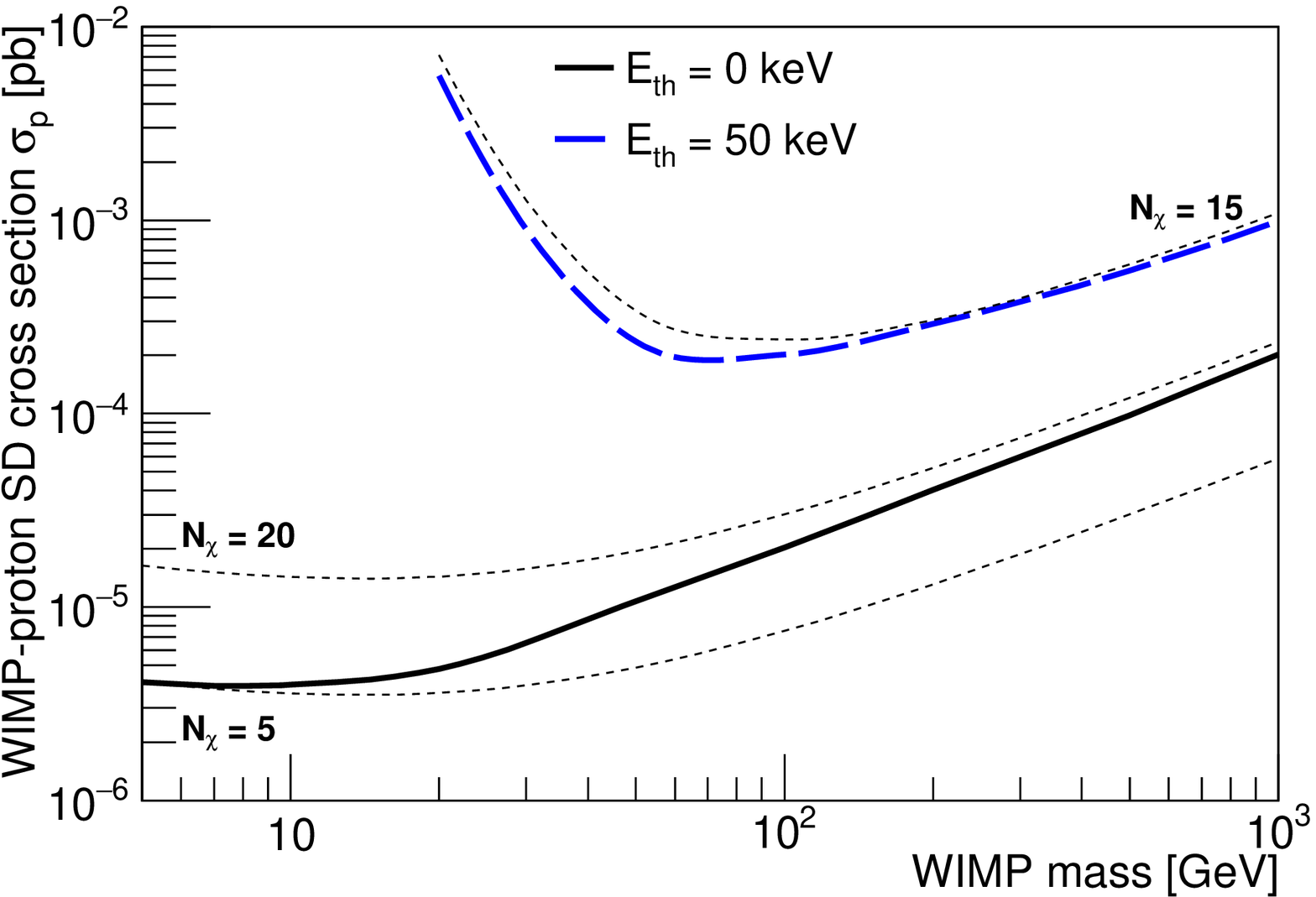}
\includegraphics[scale=0.42,angle=0]{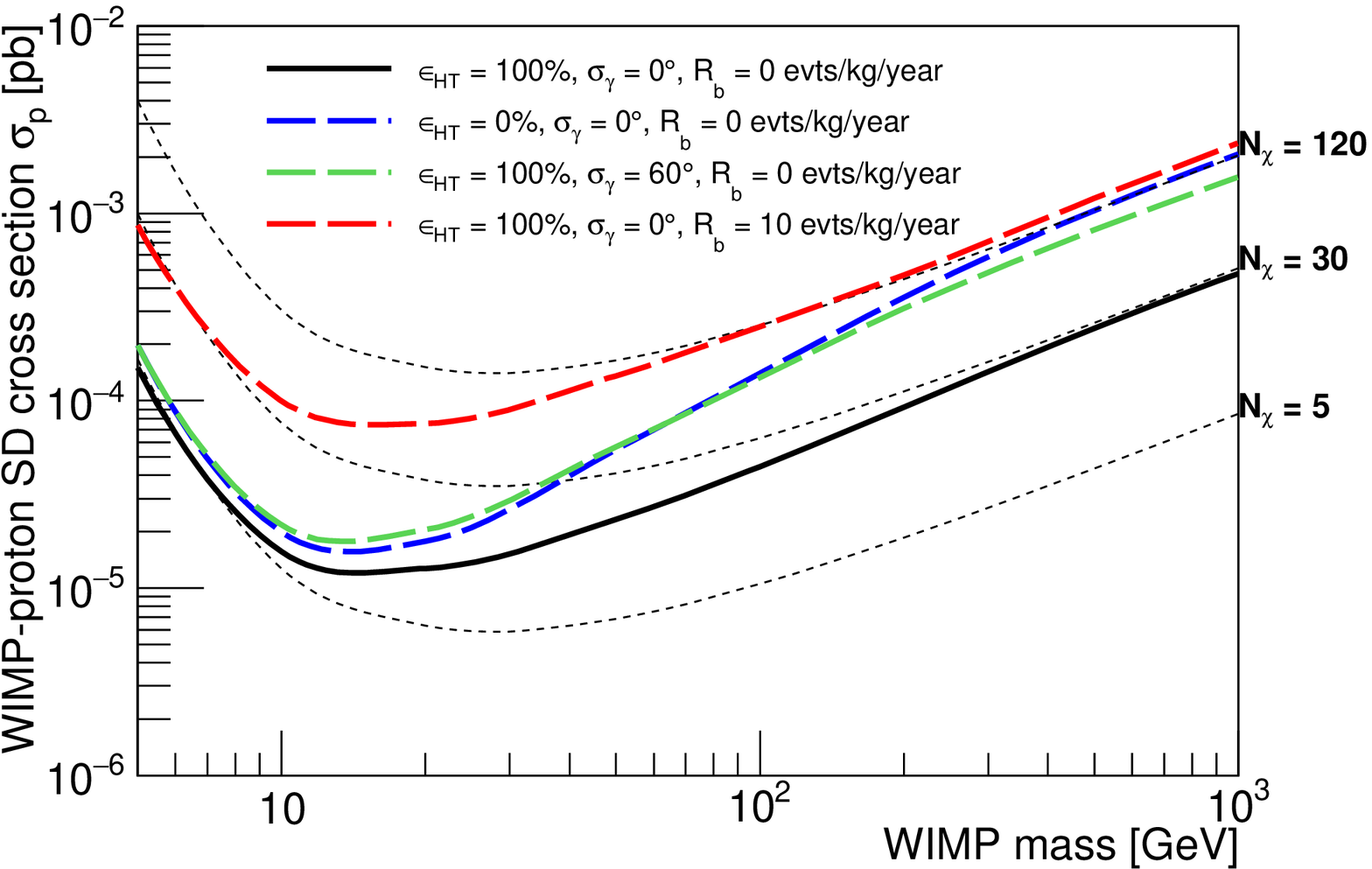}
\caption{Lower bound of the 3$\sigma$ discovery region at 90\% C.L. in the ($m_\chi,\sigma_p$) plane with 
various detector configurations. 
Left: effect   of the energy threshold ($E_{th}$). 
Black line presents  
the $E_{th} = 0$ keV   case, 
while the dashed blue line presents the same detector  
with a  $E_{th} = 50$ keV threshold.
Right: effect of the angular resolution ($\sigma_{\gamma}$), the sense recognition efficiency ($\epsilon_{HT}$) and of 
the residual background level.
Black line presents  the background free case with perfect angular resolution  $\sigma_{\gamma}=0^{\circ}$ and full 
sense recognition efficiency $\epsilon_{HT} = 100$\%. 
Dashed blue line presents the expected performance  with a lower angular resolution $\sigma_{\gamma}=60^{\circ}$.
Dashed green line presents the expected performance  with no sense recognition capability ($\epsilon_{HT} = 0$\%). 
Dashed red line presents the expected performance  with a residual background at the level of 10 background events per year per kg. 
For convenience,  the curves of iso-number of WIMP events are presented (dashed lines) 
with the corresponding number of WIMP events ($N_\chi$).} 
%Figure from Ref.~\cite{Billard:2011zj}.
\label{fig:experimentalissues}
\end{center}
\end{figure}

Billard {\it et al.} have investigated the effect of experimental limitations on the sensitivity of upcoming directional detection 
experiments in the context of model-dependent analysis techniques, namely using a profile likelihood test 
statistics~\cite{Billard:2011zj}. Considering both the angular and the energy information of the events, 
they computed the discovery potential for several different experimental configurations (see Fig.~\ref{fig:experimentalissues}). 
To remain robust against   halo parameter assumptions, they took into account most of the relevant astrophysical 
uncertainties, namely halo anisotropy and  triaxiality, as nuisance parameters in their analysis. Results from  this study considering a  CF$_4$ experiment are presented in Fig.~\ref{fig:experimentalissues} where the left panel illustrates the impact of the energy threshold only and the right panel shows the impact of limited angular resolution, sense recognition and background contamination. For convenience, the curves of iso-number of WIMP events are also presented as thin short-dashed lines. From the left panel of Fig.~\ref{fig:experimentalissues}, one can derive that as for direction-insensitive experiments, the energy threshold has a huge impact on the discovery potential and is actually the most important experimental limitation. From the right panel of Fig.~\ref{fig:experimentalissues}, we can first see that a CF$_4$ experiment with perfect sense recognition, 3d read-out, 
a threshold of 5 keV and no background contamination, only 
5 WIMP events and 30 WIMP events are required to reach a 3$\sigma$ discovery at WIMP masses 
of $5 \ {\rm  GeV}$ and $1 \ {\rm  TeV}$ respectively (black solid line). However, with a background contamination of about 300 events isotropically distributed and flat in energy (red dashed line), these required numbers of WIMP events 
are increased to 30 and 150 respectively, {\it i.e.} a decrease in discovery sensitivity by a factor of 6. 
Interestingly, they have shown that as long as the recoil direction and sense are well reconstructed, 
this result is fairly independent of the assumed energy distribution of the background, 
highlighting the interest of the directional signature of WIMP events.\\

Eventually, Billard {\it et al.} found that angular resolution and sense recognition efficiency (green and blue dashed lines) could both affect strongly (by about a factor of 4-5) the sensitivity of a directional experiments 
especially at high WIMP mass (above $100 \ {\rm  GeV}$).
On the other hand, they found that limited energy resolution only affects the experiment's sensitivity in a 
negligible way, even in extreme and unrealistic cases. 

%However, it should be emphasized that 
%detector commissioning is compulsory to derive relevant exclusion limits or discovery regions in the 
%case of such model-dependent approaches.

As a conclusion of this study, Billard {\it et al.} have shown that the most critical experimental considerations, in order of importance, are the energy threshold, background contamination, sense recognition, angular resolution and energy resolution. Interestingly, they also demonstrated that even a low 
performance directional detector with  a 30 kg-year exposure, no sense recognition capability, 50$^\circ$ angular 
resolution and a background rate of 10 events/kg/year, could  
identify WIMP events  for spin-dependent cross section on the proton above 10$^{-4}$ pb~\cite{Billard:2011zj}.

\subsection{Using directionality to overcome the neutrino floor} % Jocelyn + Ciaran
\label{subsec-nu}
%Grothaus et al 2014
%O'Hare et al 2015

\begin{figure}[t]
  \begin{center}
    \includegraphics[scale=0.5,angle=0]{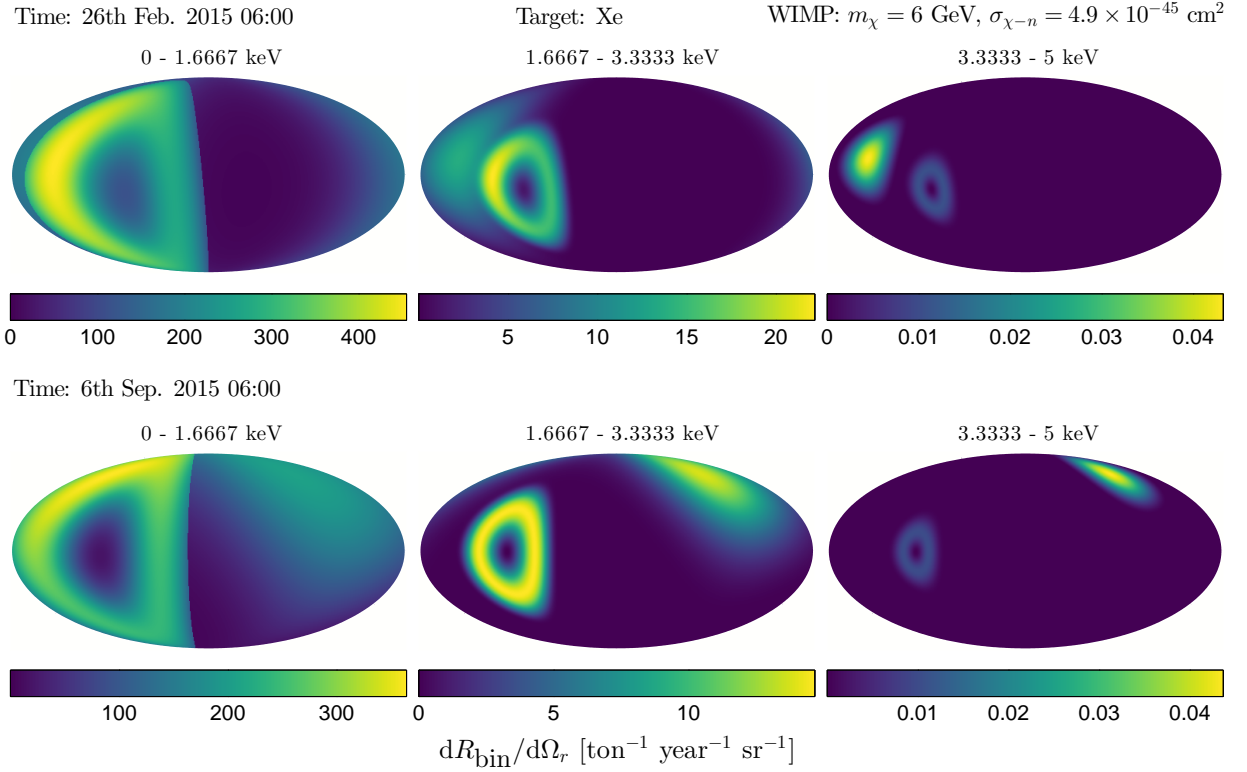}
  \caption{Mollweide projections of the WIMP plus $^8$B neutrino angular differential event rate integrated within (from left to right) three equally sized energy bins spanning the range $E_{r} = 0 \,$ to $5 \, {\rm keV}$, for a WIMP with mass $m_{\chi} = 6 \, {\rm GeV}$ and 
  $\sigma^{SI}_{n} = 4.9 \times 10^{-45} \, {\rm cm}^2$ and a ${\rm Xe}$ target. The top row shows the signal on February 26th, when the separation between the directions of the Sun and Cygnus is smallest ($\sim 60^{\circ}$), and the bottom row on September 6th, when the separation is largest ($\sim 120^{\circ}$). The WIMP contribution is to the left of the neutrino contribution on the top row and to the right on the bottom row. The Mollweide projection used in this Figure is of recoil directions in the laboratory co-ordinate system in which the horizontal axis corresponds to a plane parallel to the floor 
  and the top of the map corresponds to the zenith.}
  \label{fig:mollweideneutrino}
  \end{center}
\end{figure}

\begin{figure}[t]
\begin{center}
\includegraphics[scale=0.33,angle=0]{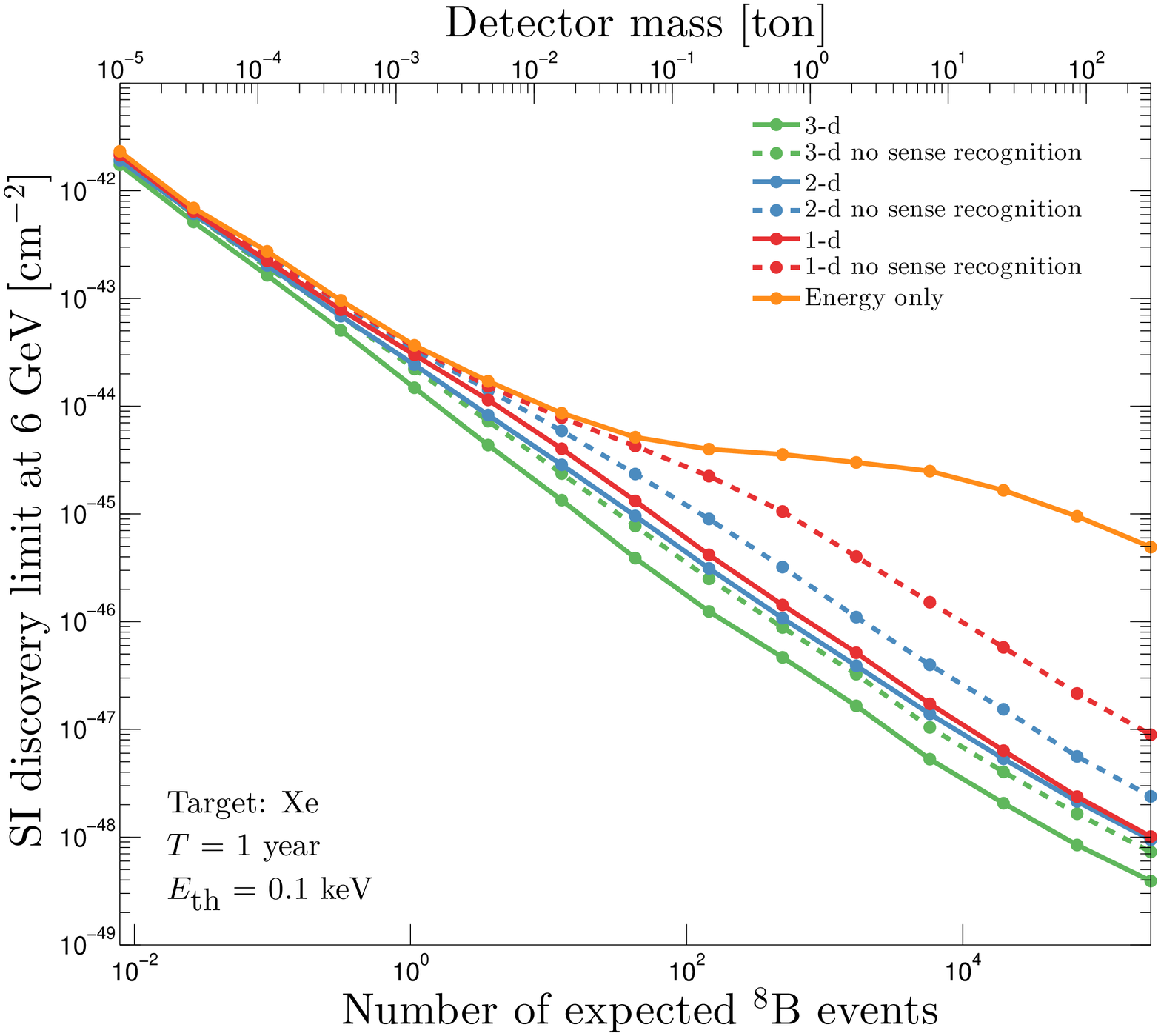}
\includegraphics[scale=0.33,angle=0]{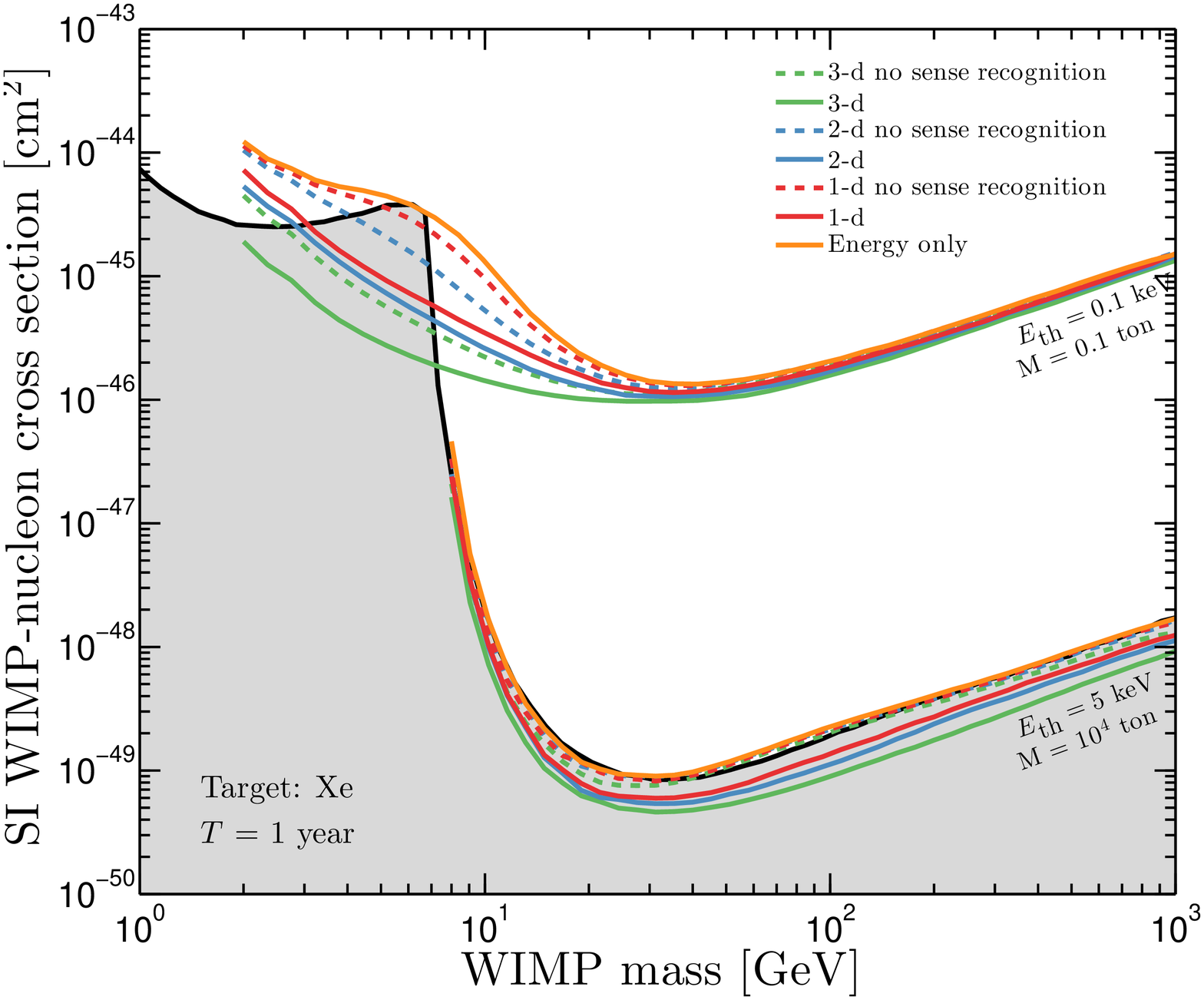}
\caption{Left: evolution of the discovery limit for a 6 GeV WIMP as a function of Xenon detector mass. The exposure time was fixed at $T = 1$ year and the energy threshold was 0.1 keV. The limits shown are for each read-out strategy, 
1d (red), 2d (blue) and 3d (green) in cases both with (solid lines) and without (dashed lines) sense recognition, the limit made by the same detector with no directional information is shown in orange. Right: the discovery limit as a function of WIMP mass for the same read-out strategies as the left panel but with fixed detector set-up. The upper set of limits are for a low threshold-low mass detector (0.1 keV, 0.1 ton) and the lower set of limits for a high threshold-high mass detector (5 keV, $10^4$ ton). The shaded region shows the neutrino floor from Ref.~\cite{Billard:2013qya} and the Figures are taken 
from Ref.~\cite{O'Hare:2015mda}.} 
\label{fig:nufloors}
\end{center}
\end{figure} 

Directional detection is particularly interesting for dealing with
coherent neutrino-nucleus scattering which represents the ultimate
background to direct Dark Matter searches. Neutrinos from the Sun,
atmospheric cosmic ray collisions and the diffuse supernova background
(DSNB) will become important backgrounds for experiments with
sensitivities to WIMP-nucleon spin-independent scattering
cross-sections around $10^{-48}$-$10^{-46}$ cm$^2$
\cite{Monroe:2007xp,Gutlein:2010tq,Billard:2013qya} and around $10^{-40}$-$10^{-46}$ cm$^2$ in the case of spin-dependent interaction~\cite{Ruppin:2014bra}. The limiting
cross-section at which the neutrino background becomes important is
known as the neutrino floor and is dependent on the systematic
uncertainty on the flux of various different neutrino contributions
for different WIMP mass ranges. For example in a Xenon detector at the
low WIMP mass range around 6 GeV the dominant neutrino background is
from $^8$B Solar neutrinos. At higher WIMP masses between 20-30 GeV
the dominant contribution is from the DSNB and around 100 GeV from
atmospheric neutrinos. Due to the low fluxes of atmospheric and DSNB
neutrinos the floor for WIMP masses higher than 10 GeV appears at much
lower cross-sections. Directional detection is a possible way to
continue Dark Matter searches below this limit as directional
information acts as a discriminator between WIMP and neutrino induced
recoils. It should be noted that the neutrino floor is not an absolute 
limit to direct detection. The floor can in fact be circumvented
without directional information in high-statistics analyses using
recoil energy alone because of the slight differences in the tails of
the recoil energy spectra of WIMPs and
neutrinos~\cite{Ruppin:2014bra}. Including additional information such
as event timing also allows the neutrino and WIMP signals to be
distinguished but this too requires very large numbers of
events~\cite{Davis:2014ama}.

The triple differential recoil rate per unit detector mass for
coherent neutrino-nucleus scattering is the convolution of the double
differential cross-section (which is well explained by the standard
model, see Ref.~\cite{O'Hare:2015mda}) and the neutrino directional flux,
\begin{equation}
\frac{\textrm{d}^3 R}{\textrm{d}E_r \textrm{d}\Omega_r \textrm{d}t} =  \mathscr{N} \int_{E_\nu^{\rm min}} \frac{\textrm{d}^2 \sigma}{\textrm{d}E_r \textrm{d}\Omega_r}\times\frac{\textrm{d}^3 \Phi}{\textrm{d}E_\nu \textrm{d}\Omega_\nu \textrm{d}t} \textrm{d}E_\nu \textrm{d}\Omega_\nu \,,
\end{equation}
where $E_\nu^{\textrm{min}}$ is the minimum neutrino energy required to generate a recoil of energy $E_r$ and $\mathscr{N}$ the number of target nuclei. The neutrino directional flux is dependent on the type of neutrino under consideration. For Solar neutrinos the flux is 
a Dirac delta function
\begin{equation}
  \frac{\textrm{d}^3 \Phi}{\textrm{d}E_\nu \textrm{d}\Omega_\nu \textrm{d}t}  = \frac{\textrm{d} \Phi}{\textrm{d} E_\nu} \, \frac{1}{\Delta t}\left[ 1 + 2\epsilon\cos\left(\frac{2\pi(t- t_\nu)}{T_\nu}\right) \right]
 \delta\left(\hat{{\bf q}}_\nu-\hat{{\bf q}}_\odot(t)\right) \,,
\label{eq:solarneutrinoflux}
\end{equation}
where $\hat{\textbf{q}}_\odot$ is a unit vector in   the
direction of the Sun.

The flux has a cosine modulation in time due to the eccentricity of
the Earth's orbit, $\epsilon = 0.016722$. The time $t_\nu = 3$ days is
the time at which the Earth-Sun distance is shorter and $T_\nu = 1$ year
is the period of modulation. The directional flux for atmospheric and
diffuse supernova neutrinos can be approximated as isotropic as the
very weak angular dependence becomes washed out in the recoil
distribution \cite{O'Hare:2015mda}.

The directional signatures of WIMP and neutrino recoils are both
unique and can be further discriminated by the correlation of these
angular patterns with time. In the sky the position of the Sun follows
the path of the ecliptic whereas the direction of motion of the Sun,
which corresponds to the constellation Cygnus, remains fixed on the
celestial sphere and these two points of origin do not
coincide. Figure~\ref{fig:mollweideneutrino} shows Mollweide
projections of the 3d angular differential event rate from a 6 GeV
WIMP plus $^8$B Solar neutrinos at the times when the separation
between the directions of the Sun and Cygnus are smallest
($60^{\circ}$) and largest ($120^{\circ}$)~\cite{O'Hare:2015mda}). Even at the time of
smallest separation, the WIMP and neutrino recoil distributions can be
distinguished as long as the angular resolution is better than a few
tens of degrees. Although Fig.~\ref{fig:mollweideneutrino} only shows the rates for $^8$B
neutrino induced recoils, the angular distributions for other Solar
neutrinos are very similar as neutrinos can only induce a recoil with
an angle in the range $(0,\pi/2)$ from their incident direction. The
event rates for atmospheric and DSNB neutrinos are isotropic so are
not shown here.

Grothaus {\it et al.} first quantitatively
explored the impact of direction-sensitivity on the neutrino bound in
direct detection ~\cite{Grothaus:2014hja}.  To do this they calculate probability
distribution functions for the Dark Matter signal and the neutrino
background, from Solar, atmospheric, and DSNB neutrinos, in the
dimensions of recoil energy, recoil direction, and event time.  A
CL${}_{\rm s}$
test~\cite{Read2000} is performed to distinguish between the neutrino
background and background + Dark Matter signal hypotheses.  This work
considers both CF$_4$ and Xe directional detectors, and includes
detector effects by smearing the probability distributions.  The
detector performance assumptions are moderately optimistic: the
angular resolution used is 30$^o$/$\sqrt{E_r}$, the energy threshold
is 5 keV in CF$_4$ and 2 keV in Xe, and the nuclear recoil detection
efficiency plateaus at 50\%. In order to set limits, the
log-likelihood ratio $Q=-2\log\widetilde{Q}$ is used, where
$\widetilde{Q}=\mathcal{L}(\vec{X},
  S+B) / \mathcal{L}(\vec{X},B)$ is the ratio of likelihoods of a set of recoils, $\vec{X}$, under signal+background and background only hypotheses respectively~\cite{Read2000}. The 90\%
confidence level limit is taken to be the cross section value at which
the overlap of the background only and signal+background distributions is 0.1.

The main results from this study are that direction-sensitivity adds
approximately an order of magnitude sensitivity beyond non-directional
searches for light Dark Matter, and depending on the target species
and energy threshold, this sensitivity can leap far beyond the Solar
neutrino bound.  Further, directionality is more helpful for lighter
targets than heavier targets; for the light target material
directional information is helpful for the complete Dark Matter mass
range, whereas for the heavy target nuclei, directional and
non-directional detectors give the same limits for heavy dark
matter. 
%This work also calculates sensitivities as a function of
%exposure, and finds that directional information removes the Solar
%neutrino discovery limits and is especially useful for light dark
%matter.

O'Hare {\it et al.} followed the work of
Ref.~\cite{Grothaus:2014hja,Billard:2014ewa} to study the effect of
direction-sensitivity on the neutrino floor for experiments with only
1d and 2d recoil track information ~\cite{O'Hare:2015mda}. Figure~\ref{fig:nufloors} shows
the discovery limits for a Xenon detector located in the Modane
underground lab, operated for one year with a range of detector
masses. The discovery limits in this work were defined as the minimum cross-section for which 90\% of hypothetical experiments can reach a $3\sigma$ discovery. They were derived using a profile
likelihood ratio test accounting for the systematic uncertainties on
the various neutrino fluxes as nuisance parameters. The left panel of
Fig.~\ref{fig:nufloors} shows the evolution of the discovery limit for
a 6 GeV WIMP in a 0.1 keV threshold detector as a function of detector
mass. The discovery limits shown are for each read-out strategy: 1d,
2d, and 3d both with and without sense recognition as
well as a comparison to a limit obtained by the same detector without
any directional information (energy only). The plateauing of the
energy-only limit when the signal becomes saturated by neutrino events
is what is commonly referred to as the neutrino floor. Including
directional information completely removes the neutrino floor at this
WIMP mass. In the case of 3d read-out with sense recognition, the
limits represent the best-case scenario with a scaling going as the inverse of detector mass
maintained even to very high neutrino event numbers. In the case of 1d
read-out without sense recognition which is the least powerful
directional read-out considered here, cross-sections below the neutrino
floor are still accessible, though require higher detector masses.

The right panel of Fig.~\ref{fig:nufloors} shows the discovery limit
for each read-out strategy now as a function of WIMP mass but for fixed
detector mass. Two sets of limits are shown, the upper set correspond
to limits obtained by a 0.1 ton detector with a 0.1 keV threshold
whereas the lower set of limits are for a $10^4$ ton detector with a 5
keV threshold. These masses and thresholds were chosen such that an
analogous non-directional experiment would have enough Solar or
atmospheric neutrino events for the discovery limit of $\sim$6 GeV and
$\sim$100 GeV WIMPs respectively to have entered the saturation
regime. This allows the benefit provided by directionality to be fully
demonstrated. In these cases the limits show that cross-sections below
the neutrino floor can be probed for the full range of WIMP masses
though the advantage is only by a factor of a few around 100 GeV where
the atmospheric neutrinos are most important. Discriminating the
isotropic atmospheric and diffuse supernova neutrino recoils from WIMP
recoils is more difficult than with Solar neutrinos as there is more
overlap between the angular dependence of the event rates. Also
because of this effect the limits obtained without sense recognition
around 100 GeV are only marginally better than the energy only limit.

To summarize, direct detection with directional information 
presents the most powerful approach for disentangling a WIMP signal
from the ultimate neutrino background. Directional information is
particularly useful for probing light masses as the Solar neutrino
events have a very different angular distribution than the WIMP signal. For heavier
WIMPs much higher exposures and detector masses are required to
distinguish WIMPs from atmospheric and supernovae neutrinos but the
floor can still be overcome by a factor of a few if sense recognition
is possible. Finally even if only partial directional information is
available such as with 1d or 2d read-outs experiments, the neutrino
floor can still be overcome.

%
% -------------- secondary features
% NEW. Discovery : N. Bozorgnia
%\newpage
% Edited by FM 25/11/15 : minor corrections  

% Edited by FM 16/11/15 : minor corrections (following comments from AMG and GBG)

%
% --------------Secondary signatures
%Secondary signatures : 

%
%-----Nassim Borzognia
%
%signatures with ring-like and aberration features
% NB : only discovery 'number of events required to...).
% The features themselves have been described in the "Directional features" section (leaders Nassim and James)

\section{Observing secondary features}
\label{sec:secondarysignatures}

\subsection{Introduction} 

In Sections \ref{sec:featuresRing} and \ref{sec:featuresAberration} we discussed two 
secondary directional features, in addition to the dipole: 
a ring of maximum recoil rate around the average WIMP arrival direction, 
and aberration features that are changes in the recoil direction pattern caused by the Earth's revolution around the Sun. 
In this section we study the prospects of observing the ring and aberration features, 
and   estimate   the number of events needed to detect them assuming no background and 
perfect energy and angular  resolutions. We also discuss using aberration features of the directional 
recoil rate to obtain information on the local WIMP velocity distribution. 

\subsection{Observing ring-like features} 
\label{sec:secondarysignatures-ring}

\begin{figure}
\begin{center}
  \includegraphics[height=147pt]{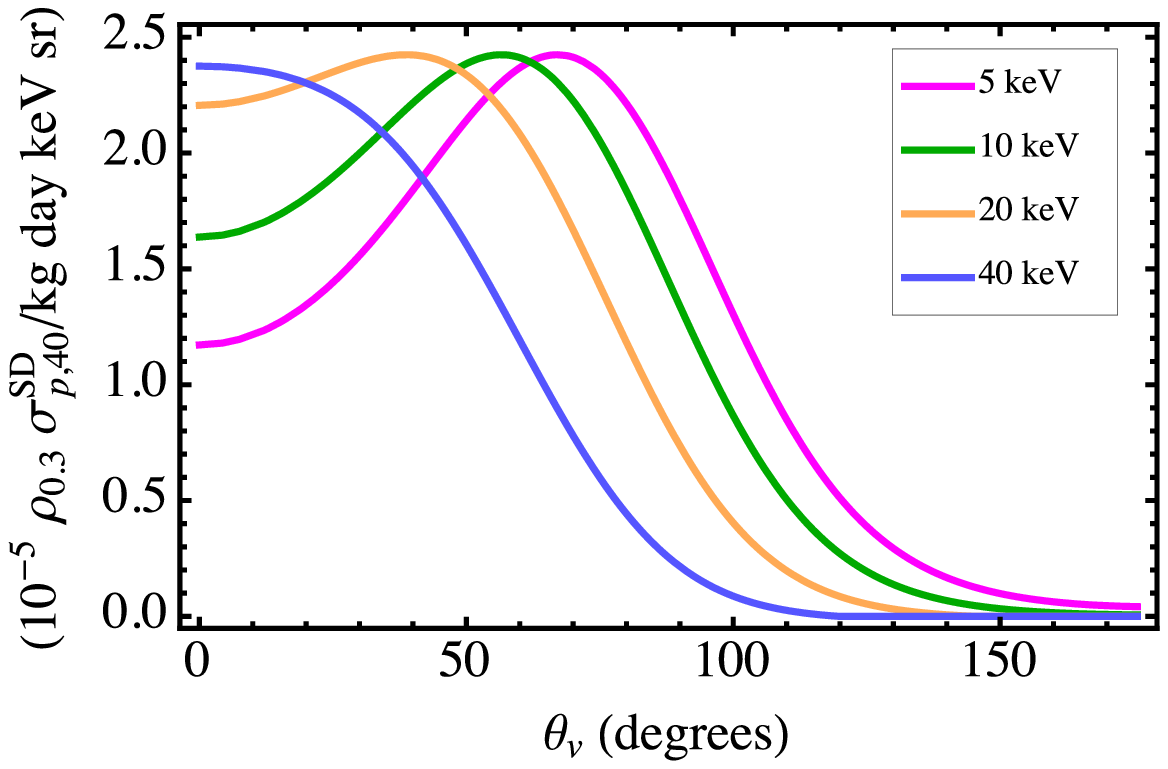}
  \hspace{5pt}\includegraphics[height=143pt]{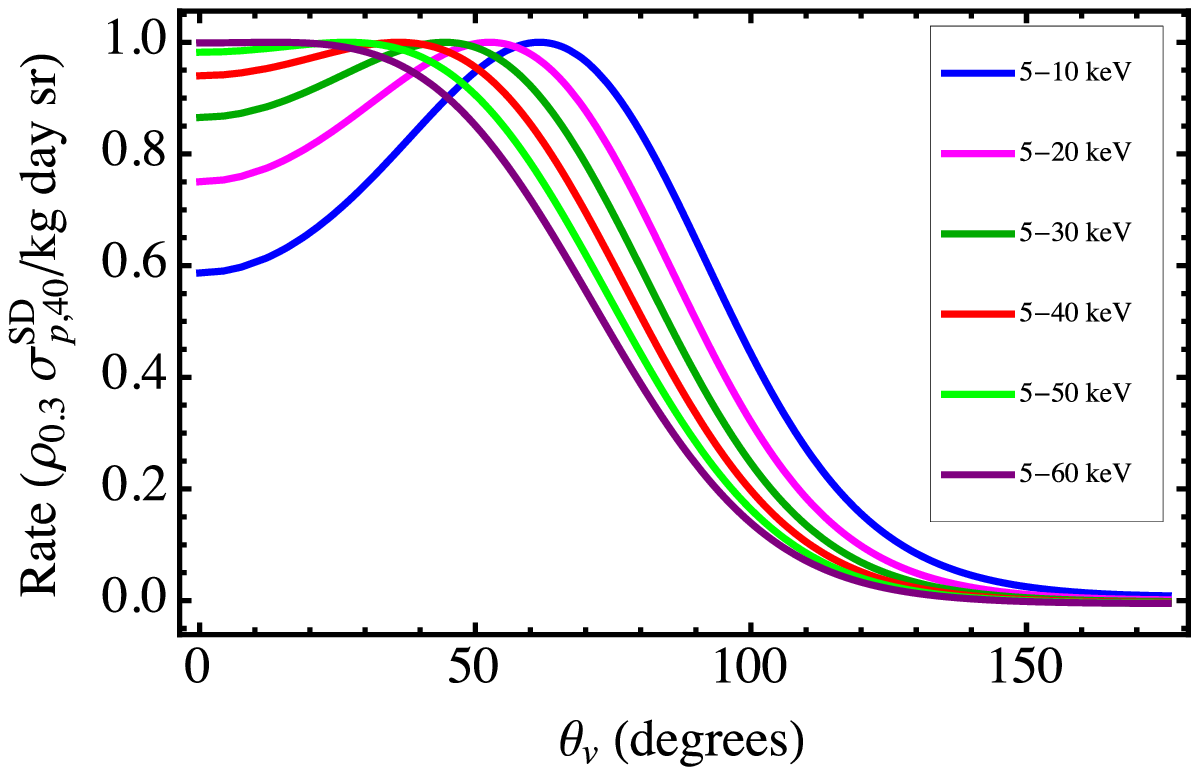}\\
  \vspace{-0.1cm}\caption{Directional differential recoil rate as a function of the polar angle $\theta_v$ at $E_r=5$, 10, 20, and 40 keV (left panel), and energy-integrated directional rate re-scaled to be between 0 and 1 for different energy intervals as a function of $\theta_v$ (right panel)  in  F, assuming $m_\chi = 100$ GeV with SD interactions, $v_c=312$ km s$^{-1}$, and  $\sigma_v=173$ km s$^{-1}$ on June 2. $\rho_{0.3}$ is the DM density in units of 0.3 GeV cm$^{-3}$ and $\sigma^{\rm SD}_{p, 40}$ is the WIMP-proton SD cross section in units of $10^{-40}\;{\text{cm}}^2$.}
  \label{RateAngle-F}
\end{center}
\end{figure}

As mentioned in Section~\ref{sec:featuresRing}, it is easiest to observe the ring-like feature when the contrast between the rate at the center of the ring and the ring is largest. The dependence of the ratio $\hat{f}_{\rm center}/\hat{f}_{\rm ring}$ on energy can be extracted from Fig.~\ref{RateAngle-F}. In the left panel of Fig.~\ref{RateAngle-F} we show the F directional differential rate as a function of the polar angle $\theta_v$ measured from the average WIMP arrival direction,  at different recoil energies for  $m_\chi=100$ GeV, $v_{\rm lab}=312$ km s$^{-1}$ and $\sigma_v=173$ km s$^{-1}$ on June 2, assuming WIMPs with SD interactions. The ring is present at all energies below 40 keV.

Since measuring the differential rate will require very large statistics, we study the prospects of observing the ring in the energy-integrated rate. The right panel of Fig.~\ref{RateAngle-F} shows the F directional recoil rate at fixed azimuthal angle, integrated over different energy intervals as a function of the polar angle $\theta_v$. To easily see that the ring-like feature exists in the recoil rate integrated over energy intervals from 5 keV up to 40 keV and lower, the energy-integrated rate is re-scaled to be between 0 and 1. From the right panel of Fig.~\ref{RateAngle-F} we can conclude that for low energy thresholds, it is possible to observe a ring in the energy-integrated rate.

The ring-like feature persists in an  anisotropic halo, but has different characteristics compared to the isotropic Maxwell-Boltzmann 
model, which was assumed so far. In the anisotropic logarithmic-ellipsoidal model of Ref.~\cite{Evans:2000gr}, the ring remains circular with constant radius at any azimuthal angle around  the average WIMP arrival direction, but the maximum value of the rate  depends on the azimuthal angle (see Fig.~14 of Ref.~\cite{Bozorgnia:2011vc}). Thus, some information on the WIMP velocity distribution can be obtained from the ring-like feature, but more work would be necessary to clarify this issue.

The ring-like feature would not exist in the background rate, and hence it  can be used as a secondary signature of Dark Matter. If directional detectors can reach low energy thresholds, only around five times more events would be required to detect the ring in the energy-integrated rate, compared with the number required to detect 
the mean incoming recoil direction~\cite{Bozorgnia:2011vc}. To detect the ring at the 3$\sigma$ level, we require that the difference between the number of events in the angular region of the ring and the center is larger than $3 \sqrt{N}$, where $N$ is the total number of events in both the ring and center regions over the duration of an experiment. With this simple statistical test it is found that for a 100 GeV WIMP, and recoil energies between 5 keV and 40 keV in 
a Fluorine detector with 3d track reconstruction, a total of 56 events over the whole sky would be needed to observe the ring in one of the most favorable cases, corresponding to an exposure of 36 kg-yr for a WIMP-proton SD cross section of $\sigma_p^{\rm SD}=10^{-40}\;{\text{cm}}^2$. For recoil energies between 10 keV and 40 keV instead, 66 events in total over the whole sky would be needed, corresponding to an exposure of 52 kg-yr for $\sigma_p^{\rm SD}=10^{-40}\;{\text{cm}}^2$. These are the minimum number of events needed to detect the ring. 

Since the ring does not exist for very small masses, the existence and the angular radius of the ring in the directional differential recoil rate (not integrated over energy) can provide some information on the WIMP mass. The number of events needed to detect the ring in the differential rate is  at least 6 times larger that those required using the energy integrated rate~\cite{Bozorgnia:2011vc}. 

In summary, if the WIMP mass is large enough, lower energy events in directional detectors produce a ring around the average WIMP arrival direction. The detailed features of the ring depend on the WIMP velocity distribution and mass. In particular, the ring could be used to obtain information on the shape of the WIMP velocity distribution. Further study is needed to find the best strategy and the highest possible precision to determine the WIMP mass using the ring-like feature.

\subsection{Observing aberration features} 
\label{sec:secondarysignatures-aberration}

\begin{figure}
\begin{center}
  \includegraphics[height=160pt]{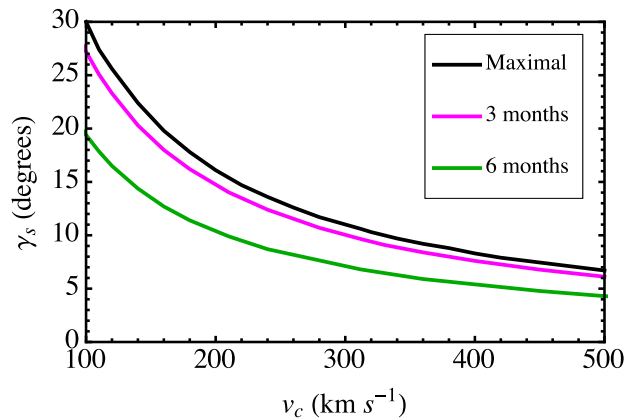}\\
  \vspace{-0.1cm}\caption{Maximal angular separation $\gamma_s$ between the directions of $-{\bf v}_{\rm lab}$ in a year (black curve), and between 
  the mean incoming recoil directions in two three months periods (magenta curve) and two six months periods (green curve), plotted as a function of $v_c$. From Ref.~\cite{Bozorgnia:2012eg}.}
  \label{AngularSep}
\end{center}
\end{figure}

If a Dark Matter signature is obtained in the future, studying the local Dark Matter velocity distribution is among the next goals. The uncertainty in the Galactic rotation speed at the position of the Sun, $v_c$, causes the largest uncertainty in ${\bf v}_{\rm lab}$. Hence to measure $v_c$, we can use the yearly variation of 
the   average WIMP velocity with respect to the Earth. We evaluate the maximal angular separation, $\gamma_s$ between the directions of $-{\bf v}_{\rm lab}$ in a year in the following simple and rather accurate way~\cite{Bozorgnia:2012eg},
\begin{equation}
\gamma_s \simeq \frac{2 v_{\rm e,rev} \sin 60^\circ}{v_c} \simeq 16.5^\circ \left(\frac{180~{\rm km~s}^{-1}}{v_c} \right),
\label{gammas}
\end{equation}
where $v_{\rm e,rev}$ is the speed of the Earth's rotation around the Sun. The maximal angular separation, $\gamma_s$ is the difference
between $-{\bf v}_{\rm lab}$ directions in March and September. Fig.~\ref{AngularSep} shows the maximal $\gamma_s$ as well as the angular
separations between the mean incoming recoil directions in two six months periods and two three months periods centered in March and September,
respectively. One can read off from Fig.~\ref{AngularSep}  the error $\Delta v_c$ in determining the Galactic rotation speed using
$\gamma_s$. From the simple estimate, $\gamma_s \sim v_c^{-1}$, we have $\Delta v_c/v_c \simeq \Delta\gamma_s/\gamma_s$. Therefore we need
$\Delta\gamma_s$ to be a few degrees at most. On the order of 1000 events are needed to measure the mean incoming recoil direction with an error of
a few degrees~\cite{Billard:2010jh, Green:2010zm}. If the data were split into two sets (centered in March and September, respectively),
double this number of events would be required to determine the six-month-averaged directional separation $\gamma_s$, while the
three-month-averaged   $\gamma_s$ would require four times this number of events. If we assume an error of $\Delta\gamma_s \simeq
3.5^\circ$, the measured $v_c$ would be $180^{+60}_{-30}$ km s$^{-1}$ or $312^{+190}_{-70}$ km s$^{-1}$ using the three-month-averaged 

Therefore, a few thousand events would be needed to measure $\gamma_s$ as implied by Refs.~\cite{Green:2010zm} and \cite{Billard:2010jh}. We discuss next the possibility to detect the annual modulation of the rate integrated over Galactic hemispheres with that same number of events.

The change in $v_{\rm lab}$ during a year causes the annual modulation  of the differential rate~\cite{Drukier:1986tm}. The differential rate is maximum (minimum) at high energies when $v_{\rm lab}$ is maximum (minimum). At low energies the phase is inverted and the differential rate is maximum when  $v_{\rm lab}$ is minimum. We integrate the directional differential recoil rate over direction to compute the annual modulation amplitude of the energy differential rate,
\begin{equation}
\Delta\left(\frac{{\rm d}R}{{\rm d}E_r}\right)=\frac{{\rm d}R_{\textrm{max}}}{{\rm d}E_r}-\frac{{\rm d}R_{\textrm{min}}}{{\rm d}E_r}=
\int{\left(\frac{{\rm d}^2R_{\textrm{max}}}{{\rm d}E_r {\rm d}\Omega_r}-\frac{{\rm d}^2 R_{\textrm{min}}}{{\rm d}E_r {\rm d}\Omega_r}\right)} 
{\rm d}\Omega_r,
\label{AnnualMod}
\end{equation}
where the subscripts ``max'' and ``min'' refer to the maximum and minimum differential rates during a year. 

Alternatively, we can integrate the directional differential rate over the recoil directions pointing to half of the sky and define the Galactic Hemisphere Annual Modulation (GHAM)~\cite{Bozorgnia:2012eg}. For specific hemispheres, the GHAM amplitude is larger than the usual annual modulation amplitude, and therefore easier to detect. The Galactic hemispheres which are divided by planes perpendicular to the Earth's orbit around the Sun have the largest GHAM, since the total orbital velocity is in the direction of one hemisphere at some time and away from it half a year later. The two hemispheres with the maximum GHAM are complimentary: in one the rate is maximum when $v_{\rm lab}$ is maximum, while in the other the rate is maximum when $v_{\rm lab}$ is minimum.

  \begin{figure}
\begin{center}
  \includegraphics[height=140pt]{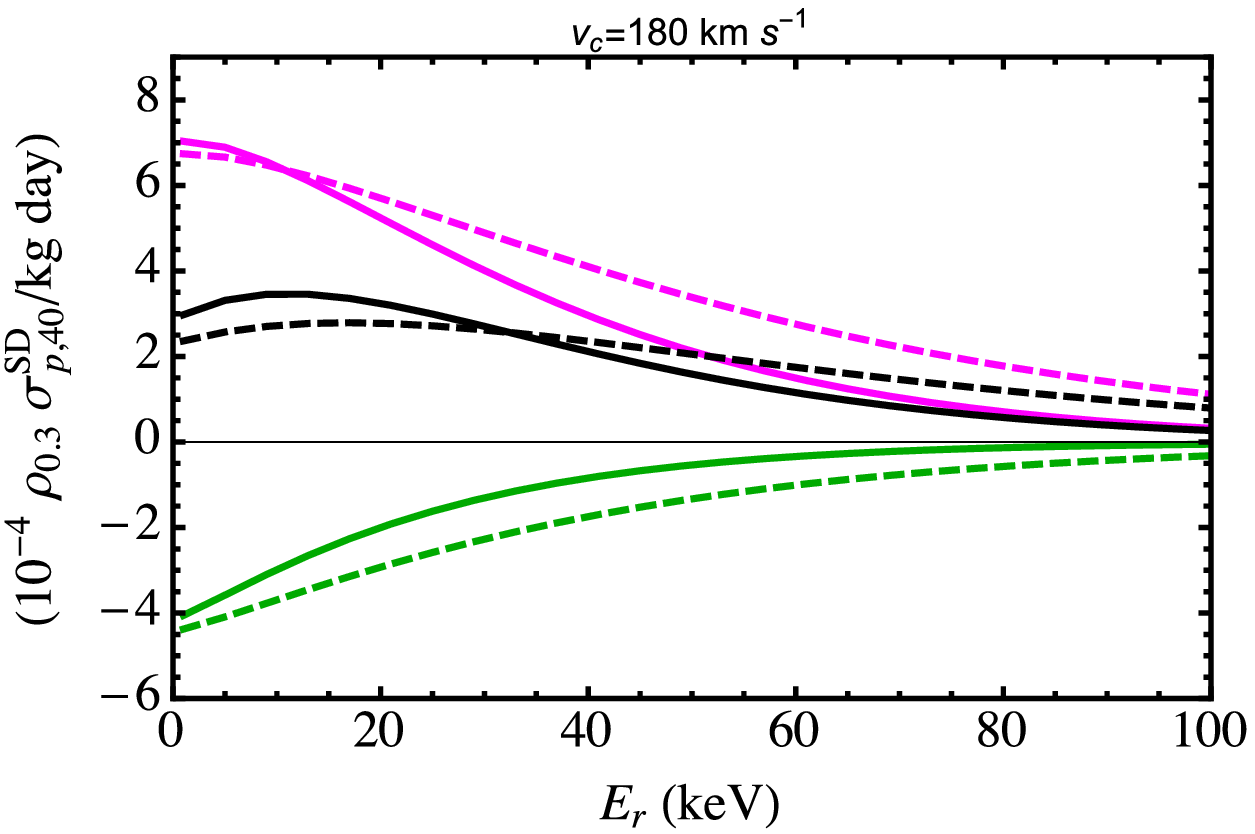} 
  \includegraphics[height=140pt]{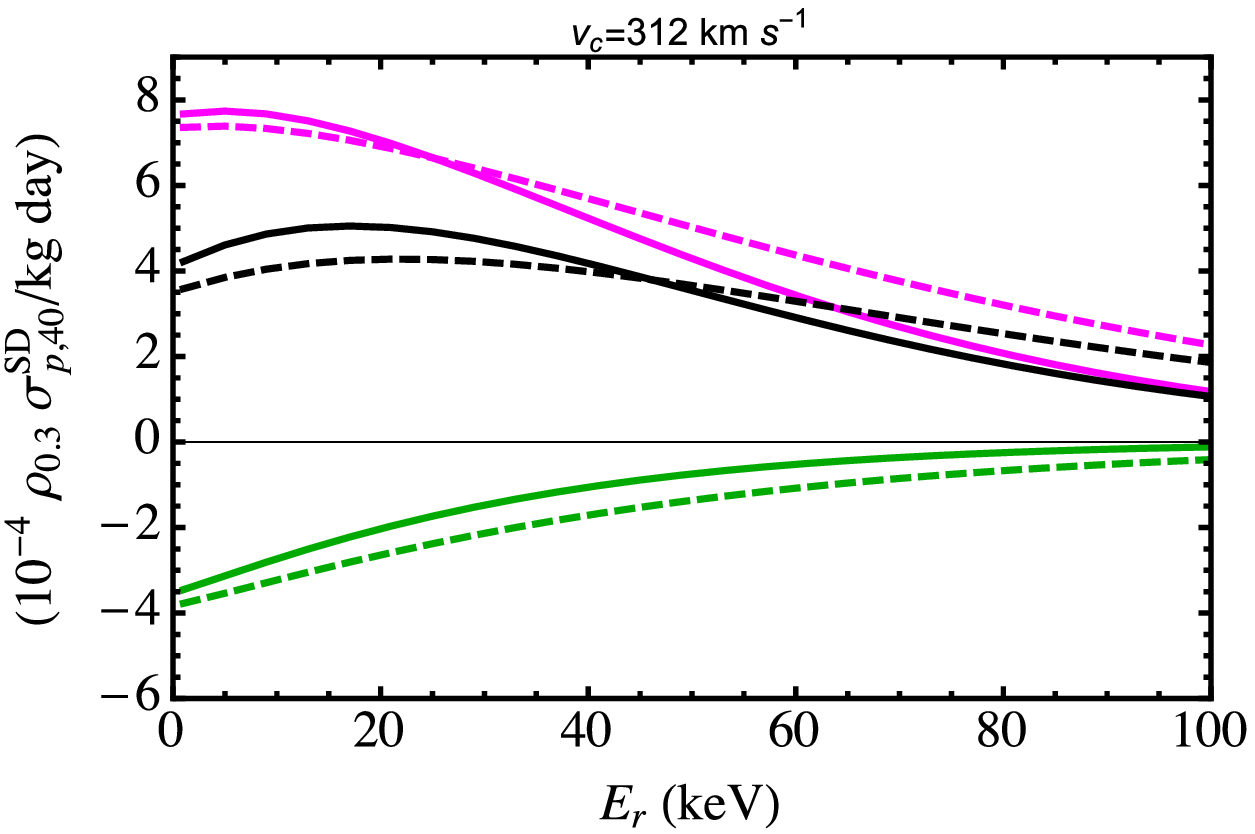}\\
  \vspace{-0.1cm}\caption{Annual modulation amplitude of the energy-integrated recoil rate in a Fluorine detector as a function of the recoil energy $E_r$ and integrated above $E_r$ and over the two hemispheres with the largest GHAM amplitude at high (magenta curves) and low (green curves) energies, and over the total sky (black curves). 
In the left  and right panels we assume $v_c=180$ km s$^{-1}$ and 312 km s$^{-1}$, respectively. The solid and dashed curves 
correspond to $\sigma_v=173$ km s$^{-1}$ and 225 km s$^{-1}$, respectively. In both panels we assume $m_\chi=100$ GeV,  
$v_{\rm esc}=544$ km s$^{-1}$, and SD interactions.}
\label{GHAM-EInt}
\end{center}
\end{figure}

We next study the GHAM in the energy-integrated rate. The annual modulation amplitude of the energy-integrated rate is $\Delta R = R_{\rm max} - R_{\rm min}$, where $R_{\rm max}$ and $R_{\rm min}$ are the maximum and minimum energy-integrated rates during a year integrated over an energy interval [$E_1$, $E_2$]. The annual modulation amplitudes of F recoil rates integrated above $E_r$, as a function of $E_r$, for the two hemispheres with the maximum GHAM amplitude at high (magenta curve) and low (green curve) energies, as well as over the total sky (black curve) are plotted in Fig.~\ref{GHAM-EInt}, for a 100 GeV WIMP with SD interactions and four different  combinations of $v_c$ and $\sigma_v$. We show the difference between the rates at dates when $v_{\rm lab}$ is maximum and minimum. Therefore, the amplitudes at high and low energies ($E_r$ greater/less than $\sim 10$ to 20 keV depending on the value of $v_c$ and $\sigma_v$) are shown as positive and negative, respectively, while the sum of the two is the usual non-directional annual modulation. At low energies, the magenta GHAM amplitude is two to three times larger than the non-directional modulation amplitude. One can see from the two panels of Fig.~\ref{GHAM-EInt} that the magnitude and shape of the GHAM amplitudes as a function of recoil energy  strongly depend on $v_{\rm lab}$ and $\sigma_v$. The annual modulation amplitude which is approximately the annual average rate times $(v_{\rm e,rev}/v_{\rm lab})$, is larger for smaller $v_{\rm lab}$. Furthermore, for smaller  $\sigma_v$ the recoil rate is more anisotropic, and thus the rate difference is larger. Hence, one obtains the largest GHAM amplitudes of the energy differential rates for the smallest $v_{\rm lab}$ and  $\sigma_v$. However, 
more work is needed to quantify the dependence of the GHAM amplitudes on $v_{\rm lab}$ and $\sigma_v$ in the energy-integrated rate.  

\begin{figure}[t]
\begin{center}
  \includegraphics[height=150pt]{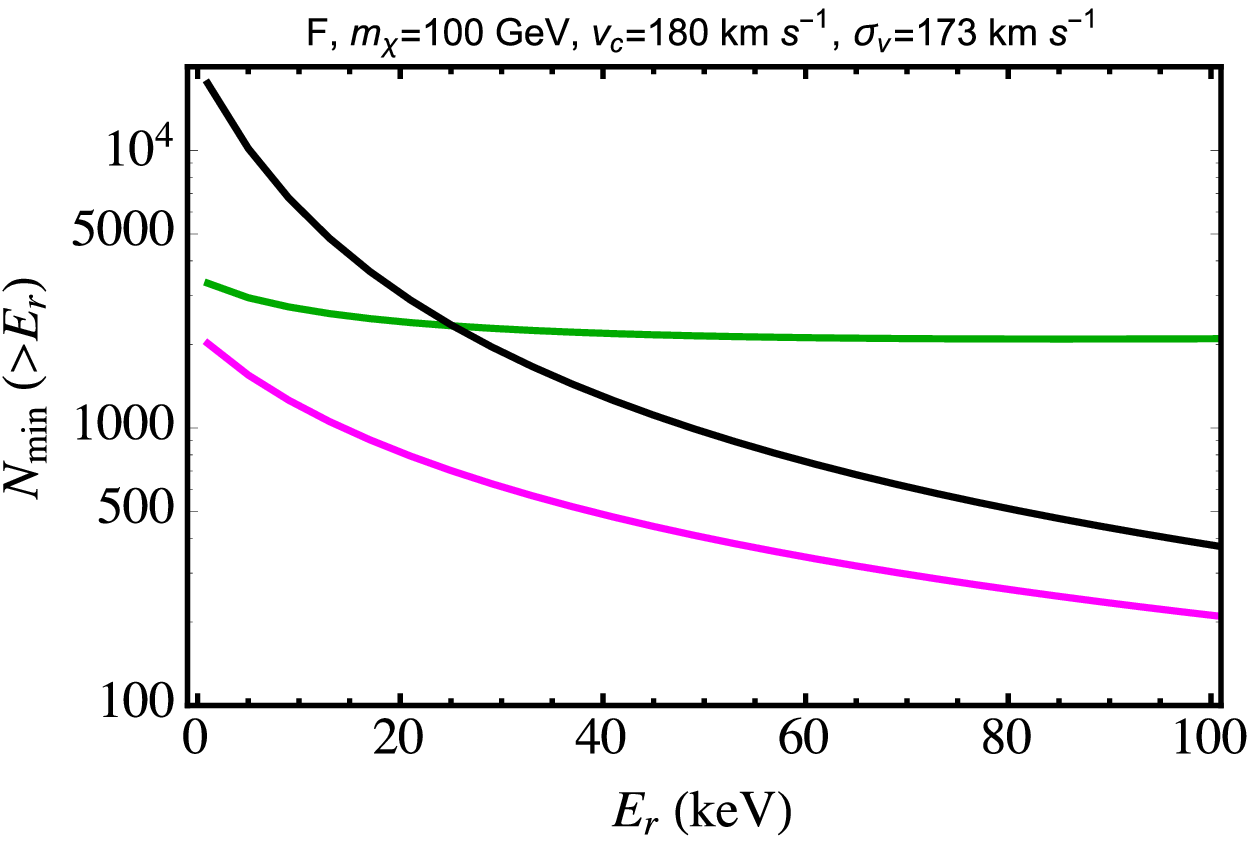} 
  \hspace{0.6cm}
  \includegraphics[height=155pt]{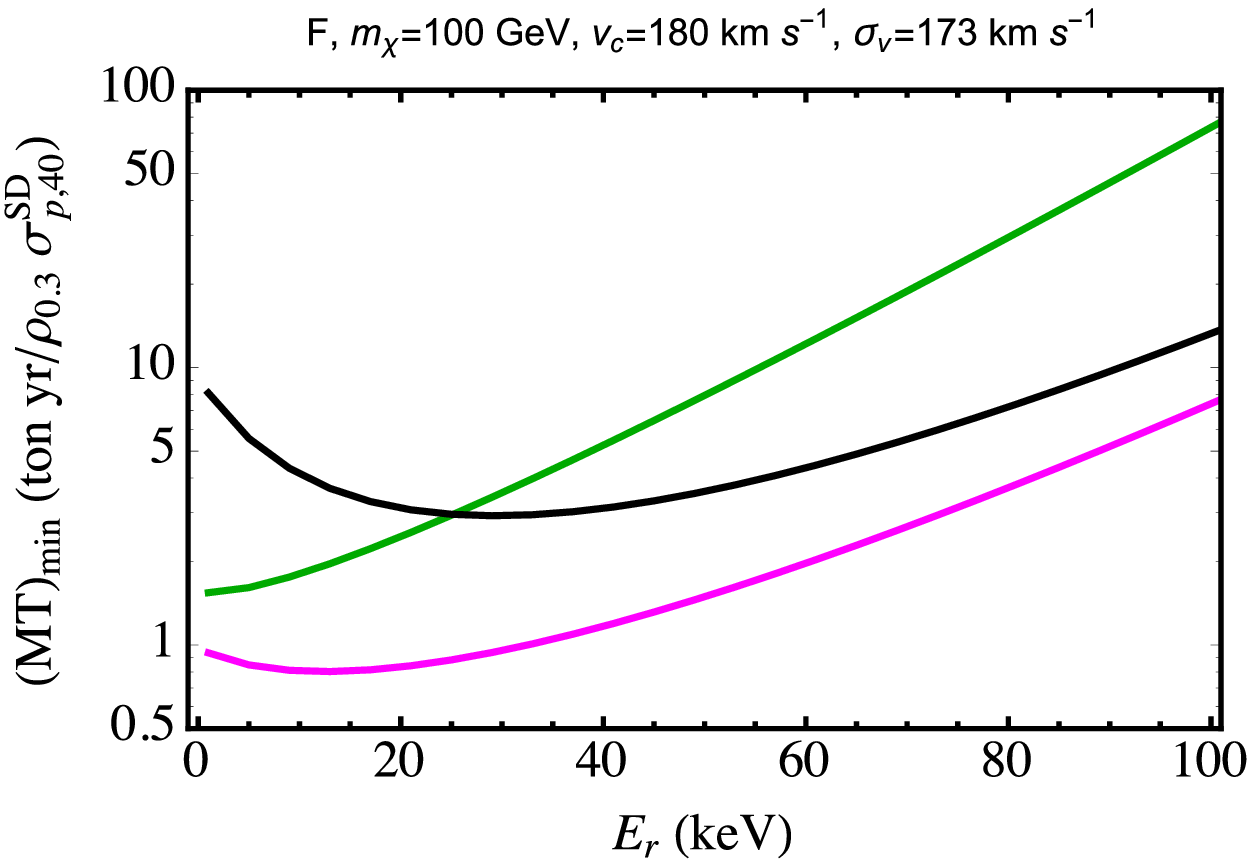}\\
  \vspace{-0.01cm}\caption{Minimum number of events (in all directions) with energy larger than $E_r$ (left panel), and the exposure $MT$ (in ton-yr) corresponding to the minimum number of events (right panel) needed to detect (at the $3\sigma$ level) the usual non-directional annual modulation (black curve) and the largest GHAM amplitude at low (green curve) and high (magenta curve) energies in an F detector, assuming SD interactions. The parameters used are $m_\chi=100$ GeV, $v_{\rm esc}=544$ km s$^{-1}$, $v_c=180$ km s$^{-1}$ and  $\sigma_v=173$ km s$^{-1}$.}
  \label{Nmin}
\end{center}
\end{figure}

To observe the annual modulation at the 3$\sigma$ level, the difference between the number of events during the annual half-cycles with high and low rates must be greater than $3 \sqrt{N_{\rm tot}/2}$, where $N_{\rm tot}$ is the number of events in both cycles. Using this statistical test, one can estimate the minimum number of events necessary to detect the GHAM~\cite{Bozorgnia:2012eg}. The left panel of Fig.~\ref{Nmin} shows the minimum number of events with energy larger than $E_r$ (in all directions), needed to detect the total annual modulation (black curve), and the maximum GHAM amplitude at low (green curve) and high (magenta curve) energies at the $3\sigma$ level  
in a Fluorine detector. We assume $m_\chi=100$ GeV with SD interactions, and the  combination of low $v_c$ and  $\sigma_v$, 180 km s$^{-1}$ and 173 km s$^{-1}$, respectively, which gives the smallest minimum number of events. A minimum of a few thousand events are needed to observe the largest GHAM amplitudes, and at least 10 times more is required to observe the usual non-directional annual modulation. The right panel of Fig.~\ref{Nmin} shows the minimum exposure needed to detect the non-directional annual modulation (black curve), and the maximum GHAM amplitude at low (green curve) and high (magenta curve) energies in the rate integrated above $E_r$.

  \begin{figure}
\begin{center}
  \hspace{0.6cm}
  \includegraphics[height=150pt]{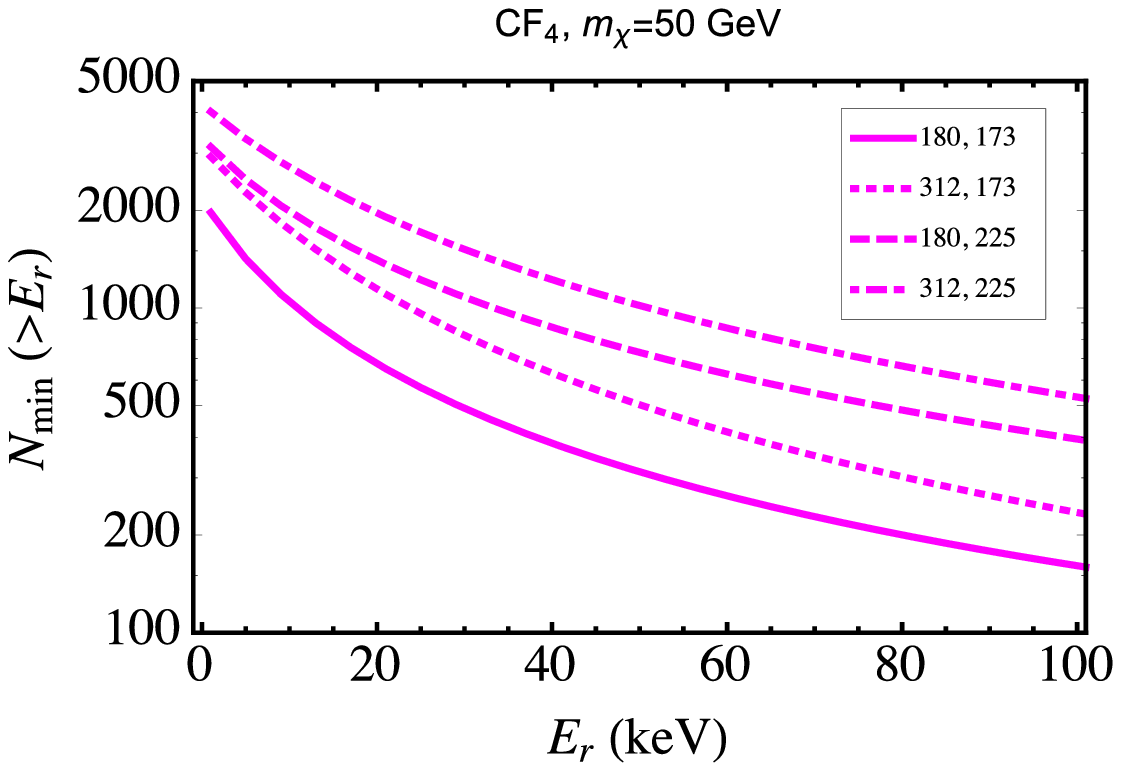} 
  \hspace{-0.4cm}
  \includegraphics[height=150pt]{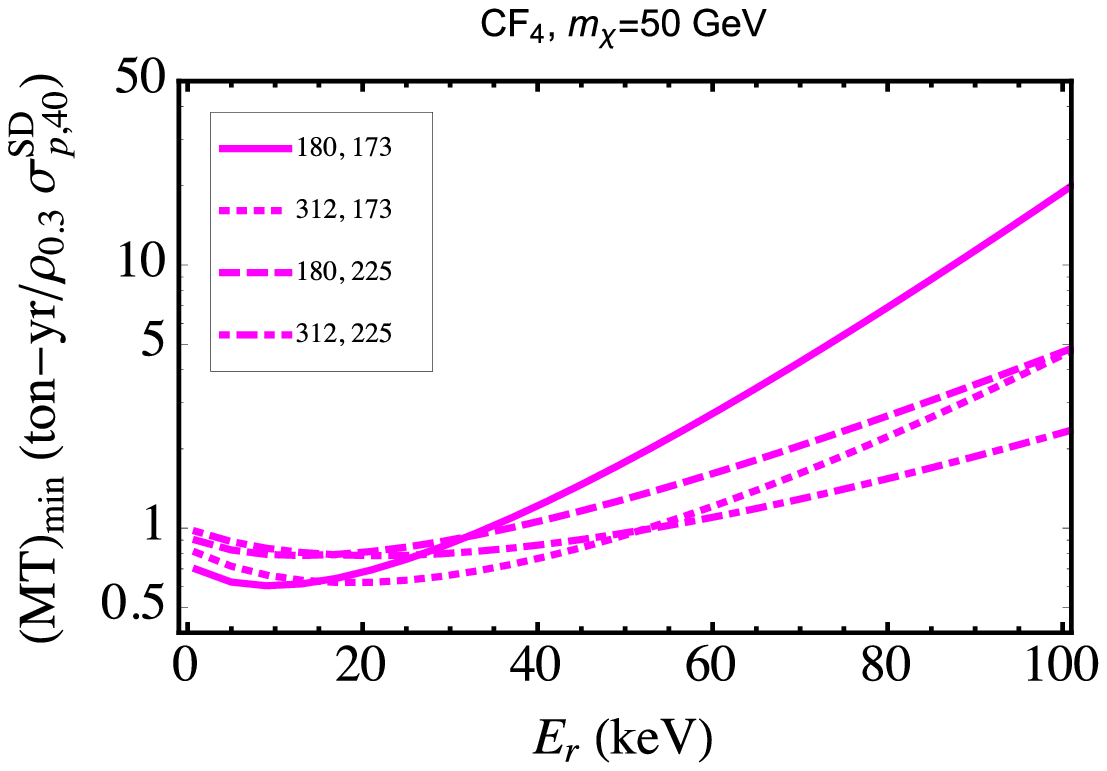}\\
  \vspace{-0.1cm}\caption{Minimum number of events in all directions (left panel), and minimum exposure $MT$ (right panel) needed to detect (at the $3\sigma$ level) the largest GHAM amplitude at high energies in the SD rate integrated above $E_r$ in CF$_4$. Different curves correspond to different combinations of $v_c$ and $\sigma_v$, as indicated by the first and second numbers in the plot legends, respectively. From Ref.~\cite{Bozorgnia:2012eg}.}
  \label{N-MTmin-SD}
\end{center}
\end{figure}

The minimum number of events with energy above $E_r$  in all directions and the minimum exposure needed to detect the largest GHAM amplitude at the $3\sigma$ level at high energies assuming $m_\chi=50$~GeV with SD interactions in a CF$_4$ detector are plotted in the left and right panels of Fig.~\ref{N-MTmin-SD}. The four curves in Fig.~\ref{N-MTmin-SD} correspond to four different combinations of $v_c$ and $\sigma_v$. One can see from Fig.~\ref{N-MTmin-SD} that between a few hundred and a few thousand events are needed to observe the largest GHAM amplitude, and the minimum exposure required to detect the largest GHAM assuming $\sigma_p^{\rm SD}=10^{-40}\;{\text{cm}}^2$ is between 0.6 ton-yr and a few tens of ton-yr, depending on the value of $v_c$ and $\sigma_v$ and the energy interval. 

In summary, aberration features can provide information on the WIMP velocity distribution, 
as well as being a way to confirm the Galactic origin of WIMPs. We would need extremely large exposures to observe the full aberration pattern, but observing the annual 
change in the mean incoming recoil direction or the number of events  over specific solid angles may be possible with moderately large exposures. Moreover, given the energy spectrum of the events over specific solid angles, one can obtain some information on $v_{\rm lab}$ and $\sigma_v$, since the magnitude and shape of the GHAM amplitudes as a function of recoil energy strongly depend on $v_{\rm lab}$ and $\sigma_v$, as well as any anisotropy in the DM velocity distribution. More work is needed to determine the best way of extracting the characteristics of the local WIMP velocity distribution using aberration features.

%
% -------------- identification
%7. DM Identification : J. Billard and F. Mayet  
%\newpage
% Edited by FM 25/11/15 : minor corrections  

% Edited by FM 16/11/15 : minor corrections (following comments from AMG and GBG)
% Edited by FM (10/11/15) - include comments received after internal review
% Edited by FM (17/07/15) - major modification
%
% -------------- identification
%8. DM Identification : J. Billard and F. Mayet (billard@lpsc.in2p3.fr,mayet@lpsc.in2p3.fr)
\section{Dark matter identification}
\label{sec:identification}

\subsection{Introduction}
\label{sec:identification.intro}

%%% Mettre en intro une description du Mimac-like experiment???

Constraining the properties of Dark Matter is the main goal 
of upcoming detectors, once conclusive evidence in favor of a discovery is observed by 
at least one experiment.  It concerns the properties of both the WIMP particle (mass and cross section) and 
the Galactic Dark Matter halo (three dimensional local WIMP velocity distribution and density). This topic may also be dealt with via 
either  indirect detection \cite{bernal2} and direct detection separately 
\cite{bernal2,Green:2011bv,green.masse2,Drees:2007hr,drees.masse,Shan:2010hr,Kavanagh:2013wba,Kavanagh:2013eya}, or within a combined analysis 
\cite{Kavanagh:2014rya,Arina:2013jya}
or in combination with collider data~\cite{Bertone:2010rv} 
or with the measurements of halo star kinematics \cite{Strigari:2009zb}. However, directional detection offers a unique opportunity to constrain Dark Matter properties (particle and halo) with the results of a single experiment, thanks to the measurement of the double-differential spectrum 
${\mathrm{d}^2R}/{\mathrm{d}E_r\mathrm{d}\Omega_r}$.  The idea  \cite{Billard:2010jh} is to constrain the WIMP properties with the help of a high dimensional multivariate analysis and within 
the framework of a general halo model, {\it i.e.} with a large number of parameters.  Several studies have been carried out on 
the subject \cite{Billard:2009mf,Billard:2010jh,Lee:2012pf} and the free parameters considered may be:
\begin{itemize}
\item $m_{\chi}$, the WIMP mass,
\item $\sigma_N$, the WIMP-nucleon cross section (assumed either SI or SD),
\item $\vlab$, the lab velocity with respect to the halo,
\item $\vesc$, the escape speed,
\item $\sigma_v$, the WIMP velocity dispersion of the standard halo model,  
\item ($\sigma_{x}$, $\sigma_{y}$, $\sigma_{z}$)  the three velocity dispersions  in the case of a triaxial halo,    
\item  ($\ell_{\odot}$, $b_{\odot}$), the main incoming direction of the recoiling nuclei,  
\item  $R_b$, the  background event rate in a given energy range. 
\end{itemize}
The ultimate goal of directional detection is to extract the posterior PDF of all parameters  with the results of a single experiment. 
However, depending on the assumptions or on the goal of the study, a subset of parameters may be considered as fixed. 
In the following, we first present in Sec.~\ref{sec:identification.proof} a study 
that highlights the power of directional detectors over experiments with sensitivity only to the energy spectrum~\cite{Lee:2012pf}. 
Then, Sec.  \ref{sec:identification.full} demonstrates that a single directional detection experiment 
may be used to constrain both the WIMP and halo parameters within the framework of a dedicated Markov Chain Monte Carlo analysis 
(MCMC)~\cite{Billard:2010jh}.

%
%as an introduction, show why directional detection is better
%
\subsection{Comparing directional and direction-insensitive detection}
\label{sec:identification.proof}
In order to compare direction-sensitive and direction-insensitive detection,  we first present a study of their 
ability to constrain the WIMP velocity distribution \cite{Lee:2012pf}.  To do so, the WIMP mass is
assumed to be known, in order to study the effect on the velocity distribution only. 
The standard halo model (SHM) is considered, given by a 
Maxwellian velocity distribution and characterized by: the WIMP velocity dispersion $\sigma_v$, the relative velocity of the 
lab with respect to the halo $v_{\rm lab}$, and the escape velocity $v_{\rm esc}$. The simulated data set is: 100 recoil events, for a $50 \ {\rm GeV}$ WIMP,  observed with a zero-background experiment. 
All parameters are supposed to be known exactly, except 
$v_{\rm lab}$ and  $\sigma_v$, which are to be
constrained by data.

\begin{figure}[t]
\centering
$\begin{array}{c}
\parbox[c]{5in}{\includegraphics[width=5in,trim=0 255pt 12pt 0, clip]{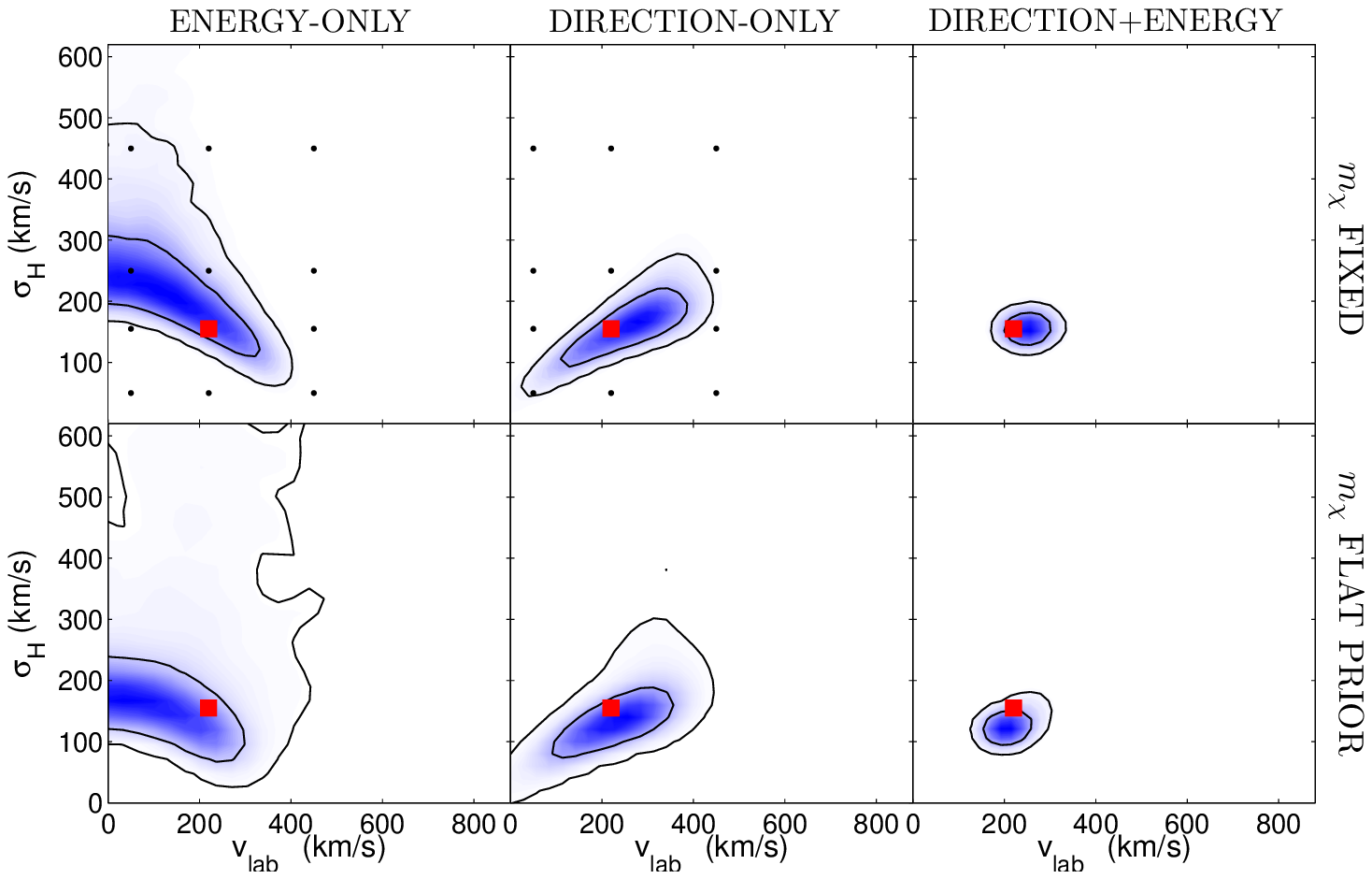}}\\
\parbox[c]{5in}{\includegraphics[width=5in,trim=0 0 12pt 129pt, clip]{halo-vlab-sigma-contours.eps}}
\end{array}$
\caption{Posterior probability distribution (68\% and 95\% CL contours) for the WIMP-wind speed $\vlab$ and velocity dispersion 
$\sigma_v$ parameters of a Maxwellian velocity distribution, for a data set consisting of 100 simulated events and assuming 
$m_\chi = 50 \ {\rm GeV}$ and a MIMAC-like experiment ($\rm CF_4$ with an energy range 5-50 keV, ).  Red square markers indicate the fiducial parameter values assumed in simulating the data.   Analyses using energy-only, direction-only, and direction+energy information on the distribution 
of recoil events are shown from left to right respectively. Fig.  from Ref. \cite{Lee:2012pf}.}  
\label{fig:halo-contours}
\end{figure}

Figure \ref{fig:halo-contours} presents the 68\% and 95\% CL contours obtained with a mock experiment measuring
the energy only (left), the direction only (middle) and the double-differential spectrum $d^2R/(dE_rd\Omega_r)$ (right). The complementarity of the energy-only and direction-only information is  easily understood by considering how the energy 
 spectrum and directional distribution of events depend on the parameters 
 $\vlab$ and $\sigma_v$.  For example, smaller values of $\vlab$ result 
 in a softer energy spectrum, as do smaller values of $\sigma_v$.  In contrast, smaller 
 values of $\vlab$ yield a more isotropic directional distribution, as do larger values of 
 $\sigma_v$. Thus, the  direction-only and energy-only measurements provide orthogonal sets of  information on these parameters. A 
 straightforward conclusion is that directional detection provides an uncorrelated estimation of 
 these two parameters, as can be seen on Fig.~\ref{fig:halo-contours}. 
Such parameter-estimation studies nicely demonstrate the power of directional detectors over experiments with sensitivity only to the energy spectrum, 
as illustrated in Fig.~\ref{fig:halo-contours}. 
 
%Relaxing the assumption that the WIMP mass is known leads to the conclusion that the analysis of directional data are relatively 
%insensitive to the lack of knowledge of the WIMP mass, which is not the case with direction-insensitive data \cite{Lee:2012pf}. 

%
% 

%halo properties 
\subsection{Constraining halo and WIMP properties} 
\label{sec:identification.full}

%Directional detection may be used to constrain the properties of both the WIMP particle and the Galactic halo with the result of 
%a single experiment \cite{Billard:2010jh}. Moreover,  directional detection can easily break parameter degeneracies 
 %that remain even with an ensemble of non-directional experiments (with varying target nuclei) \cite{Lee:2012pf,Peter:2013aha}. 

In this section, we study how directional detection can overcome the WIMP parameter degeneracies, from both the halo 
and particle physics models considering a benchmark input model described in Ref.~\cite{Billard:2010jh}: a standard Dark Matter halo (multivariate Gaussian isothermal sphere) composed 
of  $50 \ {\rm GeV}$ WIMPs with a  WIMP-nucleon SD cross section $\sigma_N = 10^{-3}$ pb.  Data are   simulated for a 
10 kg $\rm CF_4$  detector with a three-year-exposure time and a background rate of 10 events/kg/year   described by a 
flat energy spectrum and an isotropic angular distribution. Perfect energy and angular resolutions are considered.  Tab.~\ref{tab:modelinput} presents the constraints on the free parameters obtained 
from the analysis of the simulated events. Several conclusions may be drawn:
\begin{itemize}
\item the free parameters are strongly  constrained with no bias with respect to their input values,
\item the proof of discovery is given by the reconstruction of the main incoming direction $(\ell_{\odot},b_{\odot})$, 
as in Ref.~\cite{Billard:2009mf}, which corresponds to the direction of the Sun's velocity vector within 2.5 degree.
\item constraints are obtained on both the Galactic halo, via the velocity dispersions 
(Sec.~\ref{sec:identification.halo}) as well as on  the WIMP itself ($m_{\chi}, \sigma_N$), see  Sec.~\ref{sec:identification.particle}.
\end{itemize}
 
Note that the background rate is also a free parameter that is also constrained in the analysis, mainly   
by the angular part of the spectrum, in the hemisphere opposite to the constellation Cygnus  where the expected number of 
WIMP events is very small.  The parameter estimation  is thus not 
affected by the residual background~\cite{Billard:2010jh}.   
We now explore in more detail how well the halo and particle physics properties of Dark Matter are constrained.

\setlength{\tabcolsep}{0.1cm}
\renewcommand{\arraystretch}{1.4}
\begin{table*}[h]
\begin{center}
\hspace*{-0.5cm}
\begin{tabular}{|c||c|c|c|c|c|c|c|c|c|}
\hline
  & $m_{\chi} \ {\rm (GeV)}$ &  $\log_{10}(\sigma_N \ {\rm (pb))}$ & $ \ell_{\odot} \ {\rm (^\circ)}$ & $b_{\odot} \ {\rm (^\circ)}$ 
 & $\sigma_{x} \ {\rm (km.s^{-1})}$ & $\sigma_{y} \ {\rm (km.s^{-1})}$ & $\sigma_{z} \ {\rm (km.s^{-1})}$ & $\beta$ & $R_b \ 
 {\rm (kg^{-1}year^{-1})}$ \\ \hline \hline
Input & 50    &   -3  & 90 & 0 & 155 & 155 & 155 & 0 & 10  \\ \hline
Output &  $51.8^{+5.6}_{-19.4}$    & $-3.01^{+0.05}_{-0.08}$  &  $92.2^{+2.5}_{-2.5}$ & $2.0^{+2.5}_{-2.5}$ & $158^{+15}_{-17}$ & $164^{+27}_{-26}$ & $145^{+14}_{-17}$ & $-0.073^{+0.29}_{-0.18}$ & 
$10.97 \pm 1.2 $   \\ \hline   
\end{tabular}
\caption{Comparison between the input values and the values obtained with the MCMC analysis, {\it  i.e.} 
mean values and 68 \% CL uncertainties. Table from Ref.~\cite{Billard:2010jh}.}
\label{tab:modelinput}
\end{center}
\end{table*}
\renewcommand{\arraystretch}{1.1}

\subsubsection{Constraining Dark Matter halo properties} 
\label{sec:identification.halo}
In order to constrain the properties of the Dark Matter halo, 
a multivariate Gaussian WIMP velocity \cite{Evans:2000gr} is considered. It corresponds 
to the simplest triaxial generalization of the standard isothermal sphere with a density profile 
$\rho(r)\propto 1/r^2$. In this case, the three velocity dispersions ($\sigma_{x}$, $\sigma_{y}$, $\sigma_{z}$) are the parameters   
  to be constrained. However,  
they are slightly correlated with each other, with the WIMP properties ($m_{\chi},  \sigma_N$) and 
with the background rate $R_b$, see Fig. 3 in \cite{Billard:2010jh}. For instance,  increasing the velocity dispersions leads to an increase in the number of 
expected WIMP events and to a wider WIMP event angular distribution. The latter can be compensated by
decreasing the WIMP mass as it leads to a tighter angular distribution \cite{Billard:2009mf,Billard:2010gp}. The posterior  PDF of the $\beta$ parameter is computed from  Eq.~\ref{eq:beta} and 
 corrected \cite{Billard:2010jh} to account for 
 the fact that flat prior for the three velocity dispersions does not correspond to a flat prior for $\beta$.

\begin{figure*}[t]
\begin{center}
\includegraphics[scale=0.4,angle=0]{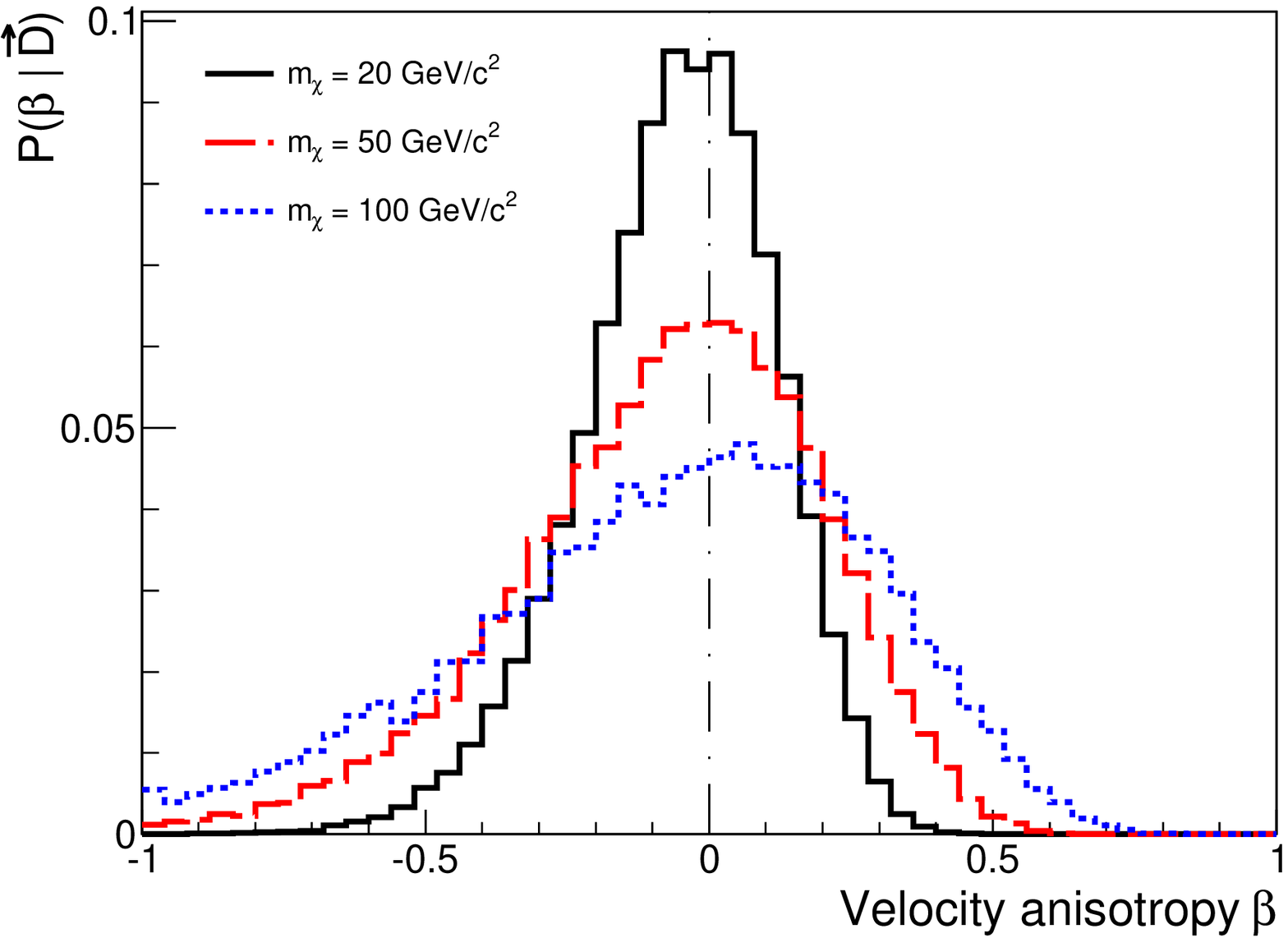}
\includegraphics[scale=0.4,angle=0]{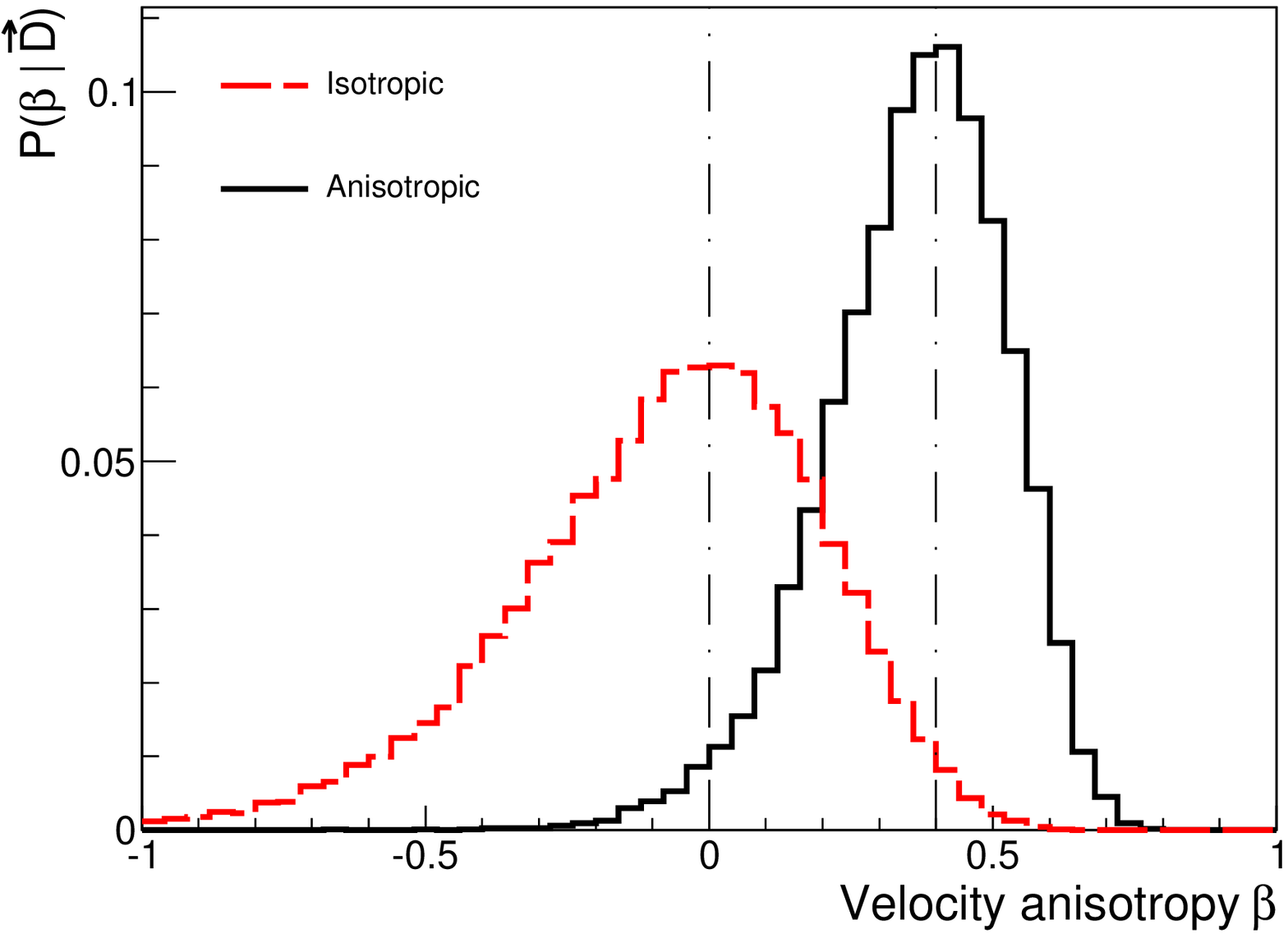}
\caption{Left: posterior PDF distribution of the $\beta$ parameter for the isotropic input halo model ($\beta=0$) and for a WIMP mass 
equal to 20, 50 and 100 $\rm GeV$. Right: posterior PDF distribution of the $\beta$ parameter,  for a 50 $\rm GeV$ WIMP 
and for two input halo models : isotropic ($\beta=0$) and triaxial ($\beta=0.4$). Fig.  from \cite{Billard:2010jh}.}   
\label{fig:identification.halo}
%\end{center}
%\end{figure*}
\end{center}
\end{figure*}

Mock data have been simulated for an isotropic ($\beta=0$) Dark Matter halo (tab. \ref{tab:modelinput}). Using directional detection, 
it is possible to evaluate the value of $\beta$. Billard 
{\it et al.} \cite{Billard:2010jh} found in this case: $\beta=-0.073^{+0.29}_{-0.18} \ (68\% \ {\rm CL})$, which is in good agreement with isotropy.
As shown on figure  \ref{fig:identification.halo} (left), this result holds true when varying the input WIMP mass, 
as the constraint on $\beta$ remains consistent with the input value. 
However, a larger input WIMP mass leads to larger  error bars.  
In addition,  
even for an extremely triaxial halo model ($\beta=0.4$), directional detection would still enable a reconstruction of $\beta$. In that
case  the constraint is $\beta = 0.38^{+0.2}_{-0.1}$.  In fact, the constraint gets even stronger with 
 increasing input value of $\beta$, see Fig.~\ref{fig:identification.halo} (right), in agreement with the expected 
   decrease of the degeneracy  between the 3 velocity dispersions with increasing departure from isotropy.

In conclusion, data from directional detection may be used to constrain 
the WIMP velocity dispersions, and hence the properties of the Dark Matter halo, without any assumption on the properties of the Dark
Matter particle, as its mass and cross section are considered as free parameters that are also constrained  by the analysis, 
as discussed in the next subsection.
 
%particle properties  
\subsubsection{Constraining WIMP properties}
\label{sec:identification.particle}
The evaluation of  the WIMP mass and cross section gives information on the properties of the Dark Matter particle that in turn may be used
to constrain the theoretical particle model (see Sec. \ref{sec:probingparticlemodel}). As shown in Tab.~\ref{tab:modelinput}, directional detection
would enable to constrain  the WIMP parameters, with a   small dispersion and no bias.  
The combined use  of angular and energy  information allows us to obviate 
 bias in the determination of the WIMP properties as degeneracies amongst the free  parameters are removed. We refer the reader to Ref. \cite{Green:2011bv}  for a detailed discussion about the effect of halo model uncertainties on 
 allowed regions.

\begin{figure*}[t]
\begin{center}
\includegraphics[scale=0.4,angle=0]{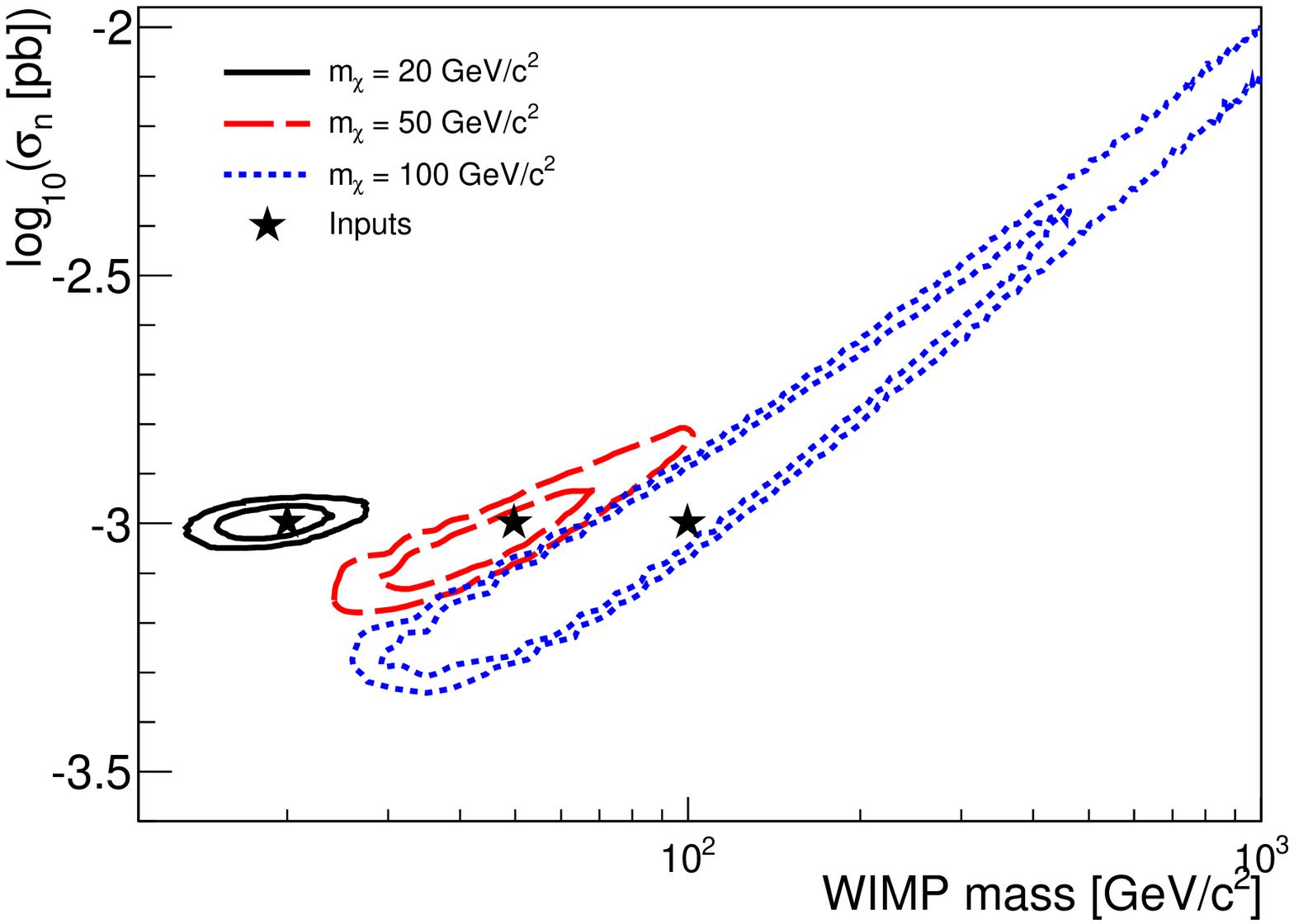}
\includegraphics[scale=0.4,angle=0]{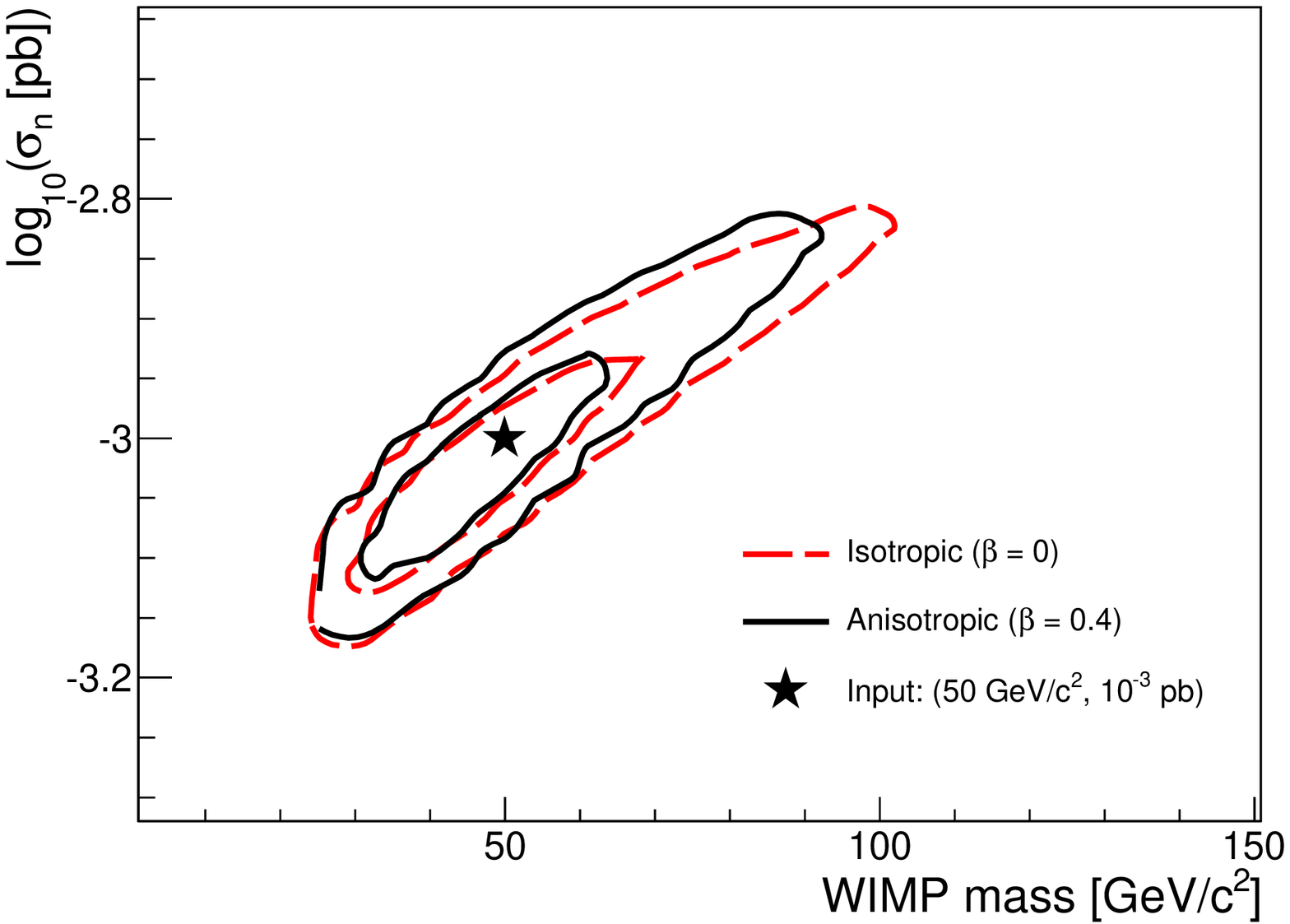}
\caption{Left: 68\% and 95\% contour level in the ($m_{\chi},\sigma_N$) plane, for the isotropic input model and for a WIMP mass 
equal to 20, 50 and 100 $\rm GeV$. 
Right: 68\% and 95\% contour level in the ($m_{\chi},\sigma_N$) plane, for a 50 $\rm GeV$ WIMP and for two input models: isotropic ($\beta=0$) and 
triaxial ($\beta=0.4$). Fig.  from \cite{Billard:2010jh}.} 
\label{fig:identification.wimp}
\end{center}
\end{figure*}

 Figure~\ref{fig:identification.wimp} (left) presents the constraint on the mass and cross section parameters. A 
 strong correlation is observed and explained  by  the fact that the event rate  scales   
 with $\sigma_N/m_{\chi}$ for low mass target nuclei. 
 It can be seen that directional detection would enable us to constrain the WIMP mass and cross section, 
 for any input WIMP mass \cite{Billard:2009mf}. However,   
the constraints  depend strongly on the input value of the WIMP mass. Indeed, the constraints on ($m_{\chi},\log_{10}(\sigma_N)$) 
  become wider  for increasing WIMP mass and only a lower limit 
may be deduced for heavy WIMPs ( $m_{\chi} \geq 100 \ {\rm GeV}$), in the case studied {\it i.e.}  
a fluorine target and a recoil energy in the range  [5,50] keV.\\ 
As shown  in Fig.~ \ref{fig:identification.wimp} (right),     the two halo models ($\beta=0$ and $\beta=0.4$) give   similar 
constraints that are   
consistent with the input values. This shows that the bias induced by an incorrect halo model assumption is avoided 
thanks to the fact that the halo model may also be constrained when analyzing directional data.
This is a key advantage of directional detection compared to direction-insensitive 
detectors \cite{Green:2011bv}.\\
In conclusion, directional detection provides a unique opportunity 
to constrain, with a single experiment,   the WIMP mass and the WIMP-nucleon cross section  within the framework of a high 
dimensional multivariate analysis. No {\it a priori} assumptions on the isotropy of the halo is needed. This is of great interest in the context of phenomenological efforts 
 \cite{bernal2,Green:2011bv,green.masse2,Drees:2007hr,drees.masse,Chou:2010qt,Shan:2010hr,Shan:2010qv,Shan:2009ym,Bertone:2010rv} trying 
 to  constrain the WIMP parameters ($m_{\chi}, \sigma_N$) with upcoming Dark Matter experiments.     It is of course possible to include 
 external data as nuisance parameters,
 {\it e.g.} measurement of the local Dark Matter density $\rho_0$ \cite{pdg,salucci} or the 
 escape velocity (taken as infinity in this study).  
% For instance, all constraints on  the WIMP-nucleon cross section  can be relaxed  and turned into  constraints on $\rho_0 \times \sigma$.

%
% -------------- particle
%8. Probing particle models (SUSY) : F. Mayet and B. Kavanagh  
%\newpage
% Edited by FM 10/02/16: comments from SV, AG taken into account
% Edited by BJK 08/02/16: minor corrections
% Edited by FM 06/02/16 : comments of the referee taken into account. idm included. figure removed
% Edited by FM 30/11/15 : corrections (following comments of JB)
% Edited by FM 26/11/15 : mdofication eq. 28 (From Bradley)
% Edited by FM 26/11/15 : major corrections  - focus on identification case.
% Edited by FM 25/11/15 : minor corrections  

% Sec. IX C edited by BJK (02/10/2015) - replaced figure and corrected text
% Edited by FM (17/07/15) - heavy modification
% most of the section has been modified.
% work of BJK is now included
% name of the section changed 
%
% -------------- susy
%9. Probing particle models : F. Mayet (mayet@lpsc.in2p3.fr) (SUSY) and B. Kavanagh (NR operators) 
% and Sven Vahsen (IDM subsection)
%
\section{Probing the   interaction of Dark Matter}
\label{sec:probingparticlemodel}

%
%FM: new version on 06/02/16
%
\subsection{Introduction}
For the standard SD and SI interaction,  directionality brings  potentially improved sensitivity as it is 
more robust against backgrounds than direction-insensitive experiments. In particular, measuring 
the mean recoil direction has been proven to be a powerful tool to reach a 3$\sigma$ discovery of Dark Matter, 
even with a low number of WIMP events and a high number of background events. As shown in Ref.~\cite{AlbornozVasquez:2012px},  
 a 30 kg-year $\rm CF_4$ directional detector would enable a $3 \sigma$ discovery for 
 supersymmetric WIMPs  below $m_\chi\leq 200$ GeV.  
 However, the constraints on model parameters, derived from directional or non-directional 
  experiments facing irreducible backgrounds, are of similar nature. Regardless of directional-sensitivity, if an experiment enables a measurement of the WIMP mass and  the WIMP-nucleon cross section  
(see Sec. \ref{sec:identification}), conclusions may be drawn on the underlying particle physics model ({\it e.g.}~supersymmetry,   see for instance Ref.~\cite{AlbornozVasquez:2012px}). 
Note also that the complementarity  with other Dark Matter search strategies is not an issue specific to directional detection. Indeed, 
it relies on the choice of the target nucleus, sensitive to either SD, SI interactions or both. Hence, all previous 
comparisons between direct, indirect and colliders searches for Dark Matter apply, 
{\it e.g.} Refs.~\cite{Moulin:2005sx,Mayet:2002ke,Ruppin:2014bra,Bergstrom:2010gh,
Jungman:1995df,arXiv:0903.0555}.\\

In this section, we  discuss how directional detectors may have additional benefits 
if one considers Dark Matter models with non-standard WIMP nucleon scattering. 
The case of inelastic Dark Matter is presented in Sec. \ref{sec:idm}.
In Sec. \ref{sec:particlemodel.NRop} we present  a  general framework (see Sec. \ref{sec:theo.NR}), which includes all non-relativistic (NR) operators describing the WIMP-nucleon interaction. We focus on 
the importance of directional detection in distinguishing between different operators and thus probing the interactions of 
Dark Matter. Note that directional detection has also been proposed as a tool to study and constrain stable
bound state of two asymmetric dark matter particles (darkonium) \cite{Laha:2015yoa}.

\subsection{Using directionality to constrain inelastic Dark Matter} % Sven
\label{sec:idm}
Inelastic Dark Matter (iDM) \cite{TuckerSmith:2001hy} is interesting as an example of how directional detectors can have enhanced sensitivity to exotic Dark Matter scenarios that go beyond the standard assumptions of elastic WIMP-nucleon scattering.  In the originally proposed iDM scenario, the WIMP has two states whose masses differ by $\delta$, and interacts with nuclei by transitioning from the lighter to the heavier state. As a result, only WIMPs with sufficient kinetic energy to create the heavier state can scatter. This corresponds to the requirement 
\begin{equation}
\delta < \frac{v^2 m_\chi m_A}{2(m_\chi+m_A)},
\end{equation}
where $v$ is the WIMP velocity in the lab frame, $m_\chi$ the mass of the lighter WIMP state, and $m_A$ the mass of the target nucleus. This means that slow WIMPs cannot scatter. Furthermore, the minimum velocity required for scattering is higher for lighter target nuclei, so that experiments with light target nuclei are at a particular disadvantage. When the original iDM model was proposed, a mass difference of order $\delta=100$~keV thus made it possible to reconcile the putative signal seen by DAMA/LIBRA with otherwise conflicting limits from CDMS~\cite{Savage:2008er}, 
as the latter experiment has a lighter target nucleus. Although this particular iDM model has since been ruled out~\cite{Aprile:2011ts}, the general concept of inelastic Dark Matter remains viable.
  
Of particular interest for directional detection, the velocity threshold in iDM models means that more recoils occur 
near threshold, and therefore are more correlated with the direction of incoming WIMPs, 
which can enhance the signal discrimination power of a directional detector. 
This was studied by Finkbeiner {\it et al.}~\cite{Finkbeiner:2009ug} and Lisanti and Wacker \cite{Lisanti:2009vy}. Finkbeiner {\it et al.}~\cite{Finkbeiner:2009ug} concluded that a directional detector with a heavy target gas, such as Xenon, would have been able to probe iDM models (that were viable at the time the paper appeared) with a surprisingly modest exposure. 
For a 50 keVr detector energy threshold, an exposure of 1000 kg-days was sufficient to exclude or support iDM benchmark 
models consistent with DAMA. In the case of iDM models, this exposure also would have allowed a significant measurement 
of the parameter $\delta$ and provided an order-of-magnitude measurement of $m_{\rm \chi}/\sigma_n$, the ratio of the 
WIMP mass to the WIMP-nucleon cross section.  
Following up on this work, Lisanti and Wacker \cite{Lisanti:2009vy} have shown how,  
in models with both elastic and inelastic contributions 
to scattering, directional detectors can set limits on or separate 
the two components, and hence shed light on the underlying dynamics of more involved Dark Matter models.

%BJK - 02/10/2015
\subsection{Distinguishing non-relativistic operators}
\label{sec:particlemodel.NRop}

Directional sensitivity also provides a tool for distinguishing between different interactions within the framework of non-relativistic Dark Matter-nucleon operators (described in Sec.~\ref{sec:theo.NR}). For mock events in a directional detector, generated assuming a particular non-relativistic operator, Ref.~\cite{Kavanagh:2015jma} examined how many events would be required to reject standard SD-only interactions at a given significance. This was achieved using a likelihood ratio test 
with 10000 mock data sets for each operator. In each case, the mock data were used to fit the WIMP mass $m_\chi$ and $A$, the ratio of standard (SD) to non-standard (non-relativistic) interactions. 

\begin{figure}[t]
\begin{center}
\includegraphics[width=0.48\textwidth]{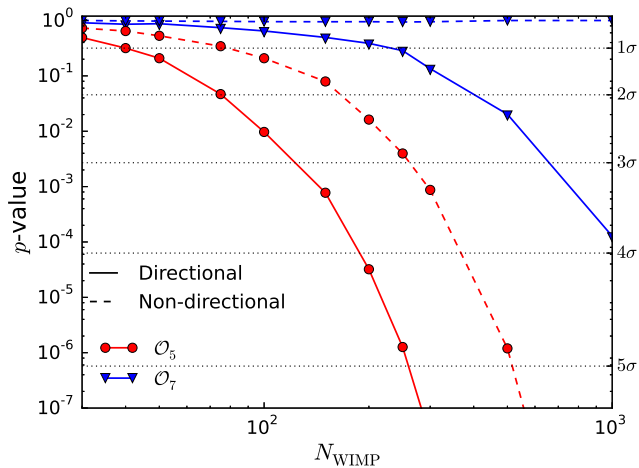}
\caption{The $p$-value obtained in 95\% of experiments as a function of the number of signal events $N_\mathrm{WIMP}$ for the null
hypothesis of standard SD-only interactions. We also show the corresponding confidence level with which standard interactions can be
rejected (in units of $\sigma$). The signal is distributed according to the non-relativistic operators $\mathcal{O}_{5}$ (red circles) or $\mathcal{O}_{7}$ (blue triangles) 
\cite{Fitzpatrick:2012}, with an input WIMP mass of 50 GeV. A Fluorine-based detector with an energy threshold of 20 keV is
assumed. Experiments with and without directional information are shown as solid and dashed lines respectively. Note that the curve for
$\mathcal{O}_{7}$ using only non-directional information lies above $p > 0.9$ for all values of $N_\mathrm{WIMP}$ considered.}
\label{fig:distinguish}
\end{center}
\end{figure}

The results for an idealised $\mathrm{CF}_4$ detector with 20 keV energy threshold and perfect angular resolution are shown in Fig.~\ref{fig:distinguish}. Dashed lines show the discrimination significance achieved in 95\% of experiments when only event energies are known, while solid lines show the results when directional information is included. The two operators considered are
\begin{equation}
\mathcal{O}_5 = i \mathbf{S}_\chi \cdot \left( \frac{\mathbf{q}}{m_N} \times \mathbf{v}^\perp \right)  \, \textrm{ and }\, \mathcal{O}_7 = \mathbf{S}_N \cdot \mathbf{v}^\perp \,.
\end{equation}
The first of these operators couples to the recoil momentum as well as the transverse velocity, while the second couples only to 
the transverse velocity.  Using only energy information, it is not possible to reject standard SD interactions if the 
underlying interaction is due to 
$\mathcal{O}_7$, even with 1000 events. This is because, assuming a SHM velocity distribution, the energy spectrum for SD interactions is indistinguishable from that of $\mathcal{O}_7$ (or any NR operator for which the cross section is proportional to $v_\perp^2$). However, including directional information (solid blue line) allows standard interactions to be rejected; roughly 400 events would be required for rejection at 95\% confidence. In contrast, the operator 
$\mathcal{O}_{5}$  can be distinguished from the standard case even without directional information (owing to its distinctive energy spectrum, which rises with increasing $q^2$). Including directional information, however, substantially reduces the number of events required for a rejection, with a $5\sigma$ rejection of SD-only scattering possible with roughly 300 ideal events. This highlights the importance of directional detection in probing the interactions of Dark Matter and distinguishing between different operators, which may not be possible with non-directional experiments alone.

%
% -------------- light wimps
%9. Light WIMPs : D. Loomba and Anne Green  
%\newpage
% Edited by FM 10/02/16: comments from AP

% Edited by FM 25/11/15 : minor corrections  

%%%%%%%%%%%%%%%%%%%%%%%%%
% Edited by FM 16/11/15 : minor corrections (following comments from AMG and GBG)
% edited by AMG on 4th Aug
% version from DL on 3rd Aug
%  Edited by FM on 17th July (minor)
%  Edited by FM on 15th May
%  Editted by AMG on 14th May
%%%%%%%%%%%%%%%%%%%%%%%%%
%
% -------------- light wimps
%9. Light WIMPs : D. Loomba and Anne Green (dloomba@unm.edu,anne.green@nottingham.ac.uk)
\section{Searching for light WIMPs}
\label{sec:light}

\subsection{Introduction}
\label{light-intro}
Event rate excesses and annual modulations in various direct detection
experiments in recent years have provoked interest in light WIMPs. The CDMS
II-Si~\cite{Agnese:2013rvf}, CoGeNT~\cite{Aalseth:2010vx} and
CRESST~\cite{Angloher:2011uu} experiments saw excesses above
expected backgrounds, while CoGeNT~\cite{Aalseth:2011wp} and
DAMA/LIBRA~\cite{Bernabei:2010mq} observed annual
modulations. While these {\em hints} could each individually be interpreted as due
to the elastic scattering of light ($5-10 \, {\rm GeV}$)
WIMPs, there is no model which is consistent with all of the
available data (see {\it e.g.} Ref.~\cite{Arina:2012dr}), in particular the
recent tight exclusion limits from the LUX
experiment~\cite{Akerib:2013tjd}. Furthermore, light WIMPs have a strongly declining energy 
spectrum, with the majority of events lying just above the threshold energy, where background rejection can 
be difficult.  More recent data from CRESST
%, using an upgraded detector, 
are inconsistent with a light WIMP interpretation of their previous excess~\cite{Angloher:2014myn} and 
it has been argued that the CoGeNT excess is due to surface ({\it i.e.} background) events~\cite{Davis:2014bla}.

Futhermore,  light WIMPs are theoretically a viable Dark Matter candidate, as 
light neutralinos can arise  in the Minimal Supersymmetric Standard
Model~\cite{Vasquez:2010ru,AlbornozVasquez:2012px,Bottino:2002ry}, 
in the Next-to-Minimal Supersymmetric
Standard Model ({\it e.g.} Ref.~\cite{AlbornozVasquez:2012px,Bottino:2002ry,Gunion:2005rw,Das:2010ww})
and asymmetric Dark Matter (for a review see Ref.~\cite{Zurek:2013wia}).

It is therefore important to pursue experiments that are sensitive to
the full range of possible WIMP masses. The differential event rate is
quasi-exponential with a characteristic energy scale~\cite{Lewin:1995rx} 
\begin{equation}
E_{\rm c} =  \frac{2 m_{\rm A} m_{\chi}^2 v_{\rm c}^2 }{(m_{\rm A} +
  m_{\chi})^2}  \propto
\begin{cases}
\frac{m_{\chi}^2}{m_{\rm A}} & \text{ if $m_{\chi} \ll m_{\rm A} $} \,, \\
m_{\rm A} & \text{ if $m_{\chi} \gg m_{\rm A} $}  \,.
\end{cases}
\end{equation}
Consequently light targets, such as those typically used in
directional detection experiments, are most suited to detecting light WIMPs. Furthermore, as discussed in Sec.~\ref{sec:discovery},
% [AMG: WOULD BE BETTER TO REFERENCE SEC. VI E, BUT THIS DOESN'T CURRENTLY HAVE A LABEL.], 
directional experiments can probe cross-sections below the neutrino floor, which occurs at $\sigma_{\rm p}^{\rm SI} \sim 10^{-42} \, {\rm cm^2}$ for light WIMPs. Below this cross-section non-directional detectors cannot distinguish between WIMP induced recoils and those due to $^8$B Solar neutrinos, due to their extremely similar energy spectra.

In Sec.~\ref{light-theory} we review theoretical work examining the potential for directional detection of light WIMPs.
Sec.~\ref{light-exp} then covers experimental prospects. In particular, we evaluate the optimal pressure required for a gaseous directional
TPC to be sensitive to light WIMPs.  We then discuss the challenges and 
issues for operating a directional detector at low pressure before concluding in Sec.~\ref{light-conc}.

\subsection{Theoretical studies}
\label{light-theory}

Theoretical work has demonstrated that directional detection
experiments are potentially sensitive to light WIMPs.  Morgan and
Green~\cite{Morgan:2012sv} determined the number of events and
exposures required for a $5\sigma$ model-independent anisotropy
detection of light WIMPs, in the mass range $5-10 \, {\rm GeV}$,
for ${}^{3}$He, C, F and S targets with directional energy thresholds
in the range $(5-20) \, {\rm keV}$. For vectorial data, where the
senses ($+\bhat{r}$ versus $-\bhat{r}$) of all recoils are measured,
they used the Rayleigh-Watson statistic~\cite{Watson,mardiajupp}. For axial data, they used the Bingham statistic, which is based on the
scatter matrix of the data~\cite{mardiajupp}. See
Sec.\ref{sec:discovery} for further discussion of these
statistics. Apart from the
possibility of axial data, they assumed that the recoil directions
were perfectly measured in 3d. Therefore their results provide a lower
limit on the number of events/exposure required by a real detector.

They found that between $10$ and $200$ events are required to reject
isotropy, depending on the WIMP and target masses and the energy threshold. 
The number of events has the advantage of being independent of the (unknown)
WIMP elastic scattering cross section. However, as discussed in
Sec.~\ref{sec:discovery},  quoting sensitives in terms of numbers of
events can be misleading. The rare high energy recoils are more anisotropic than the
more common low energy recoils (see Fig.~\ref{fig:dipolespectrum} in Sec.~\ref{sec:features}). An experiment with a high energy threshold
might only need a small number of events to reject isotropy, however
the event rate above a high threshold is small so that a large exposure
would be required to accumulate these events.
Therefore the exposure provides a more meaningful measure of the amount of
data required than the number of events.

For light WIMPs, unless the energy threshold is very low, only the
high-speed tail of the WIMP distribution will produce recoils above
threshold.  As discussed in Sec.~\ref{sec:theo.halo}, there are
significant uncertainties in the shape of the high-speed tail of the
speed distribution, and in particular the value of the escape speed
(above which particles escape from the Milky Way). Therefore for light WIMPs
the event rate above threshold can be small and have significant
(order of magnitude) uncertainties.  Note also that if an experiment
is only sensitive to high-speed WIMPs then the Earth's orbital speed
must be included, and averaged over, for an accurate calculation of
the event rate~\cite{Morgan:2012sv}.

Consequently Ref.~\cite{Morgan:2012sv} found that the exposure
necessary to accumulate the $10-200$ events required to reject
isotropy depends strongly on the energy threshold and the target
nuclei and WIMP masses. For a spin independent elastic scattering cross-section on the
proton of $\sigma_{\rm p}^{\rm SI} = 10^{-5} \, {\rm pb}$ the
exposure required varies between $10^{2}$ and $>10^{7} \, {\rm kg \,
  day}$. If the minimum speed required to cause a recoil above
threshold, $v_{\rm min}(E_{\rm th})$, is significantly smaller than
the maximum WIMP speed in the lab frame, $v_{\rm max}$, then the
exposure required is fairly modest, $\sim 10-100 \, {\rm kg \, day}$.
This is the case for a low energy threshold and the larger WIMP and
target masses considered.  However as the energy threshold is
increased, and $v_{\rm min}(E_{\rm th})$ approaches $v_{\rm max}$, the
event rate above threshold drops rapidly and the exposure required
increases substantially. The shape of the high-speed tail of the
velocity distribution, and in particular the value of the escape
speed, also have a significant effect on the required
exposure~\cite{Morgan:2012sv}.

\subsection{Experimental prospects}
\label{light-exp}

In this sub-section we consider the experimental possibilities, and
requirements, for a directional low mass WIMP search. The usual
challenges for a Dark Matter experiment probing this regime are
exacerbated for directional searches that, in addition, require
tracking of the very low energy recoils. The most mature current
directional experiments are all based on the low-pressure gas TPC
technology operating at pressures of $\sim 40-100$ Torr, which are optimal
for $m_{\chi} \sim 100 \, {\rm GeV}$ WIMPs. At these pressures, however, recoil
tracks from low mass WIMPs are too short to be resolved, resulting in
little or no sensitivity in the $ m_{\chi} \approx 10 \, {\rm GeV}$ region. 
Nevertheless, with the recent interest in low mass WIMPs, a number of authors~\cite{Martoff,Vahsen}
have studied the question of whether
appropriate optimization of the experiments can lead to directional
sensitivity.

Gas-based TPCs, with their ability to vary the target gas and its
pressure, are ideally suited to address this question. As discussed in
Sec.~\ref{light-intro}, due to the kinematics of elastic scattering
the optimal target nuclei mass is of order the WIMP mass. Fortunately
there exist numerous drift chamber gases, such as ${\rm CO}_2$, ${\rm
  CF}_4$ and ${\rm CS}_2$, with nuclei ideally suited as light WIMP
targets. Additionally, low mass WIMP searches require the directional
energy threshold to be as low as possible due to the steeply falling
recoil energy spectrum. The possibility of gas-based TPCs to operate at arbitrarily low pressures 
could provide the ability to lengthen recoil tracks of a
given energy, thereby lowering the directional energy threshold for a
fixed minimum resolvable track length, $R_{\rm min}$.
%The ability of gas-based TPCs to operate
%at low pressures provides a means of lengthening recoil tracks of a
%given energy, thereby lowering the directional energy threshold for a
%fixed minimum resolvable track length, $R_{\rm min}$. 
The minimum resolvable track length is a 
characteristic of the experimental properties of the detector and can be determined by simulations or direct measurements.

Given $R_{\rm min}$, the optimal pressure can be derived by maximizing the
directional
%, or tracking, 
event rate. The existence of an optimal
pressure is due to two competing effects that conspire to give a maximum
directional event rate. As the pressure is increased both the target
mass and the directional threshold increase. The former increases the
event rate linearly, while the latter decreases it exponentially,
resulting in an optimal pressure where the rate peaks.
 Martoff et al.~\cite{Martoff} used simulations for $R_{\rm min}$ values in the range 
 $0.1-10 \, {\rm  mm}$ to find optimal pressures for Ar, 
 ${\rm CF}_4, {\rm CS}_2$ and Xe for WIMP masses in the range $10$ to $1000 \, {\rm GeV}$. 
Jaegle {\it et al.}~\cite{Vahsen} first ran simulations to determine $R_{\rm min}$ as a function of pressure in ${\rm CF}_4$ and and ${\rm CS}_2$, then followed a similar procedure to find the optimal pressure. 
Ref.~\cite{Martoff} predict a few events per year per cubic meter for  
$m_{\chi} \approx 10 \, {\rm GeV}$ and $\sigma_{\rm p}^{\rm SI} 
\approx 10^{-42} \, {\rm cm}^2$ , while Ref.~\cite{Vahsen} finds a slightly smaller rate. 
  
Phan et al.~\cite{Phan} follow a similar procedure to optimize pressures for light WIMPs, except that their $R_{\rm min}$ is determined from experimental data. The data were taken using a high resolution, high signal-to-noise prototype TPC that measures two dimensional tracks
 in $100$ Torr ${\rm CF}_4$.  With a diffusion of $\sigma \approx 0.4 \, {\rm mm}$, not unreasonable for
 negative-ion TPCs using $\rm{CS}_2$ gas mixtures~\cite{Martoff2}, they measure 2d vector 
 directionality down to track lengths $\sim 0.6 \, {\rm mm }$, which 
 corresponds to a $ 55 \, {\rm keV_r}$ fluorine recoil in $100$ Torr ${\rm CF}_4$. For axial
directionality ({\it i.e.} with no measurement of the sense of the recoil) their minimum resolvable 
track size is $\sim 0.4 \, {\rm mm}$, corresponding  to $\approx 40 \, {\rm keV_r}$. 
Here we use their result of $R_{\rm min} = 0.6 \, {\rm mm}$ for 2d vector directionality to find the optimal pressure using the procedure described above.

\begin{figure}[h]
\begin{center}
\includegraphics[width=8.0cm]{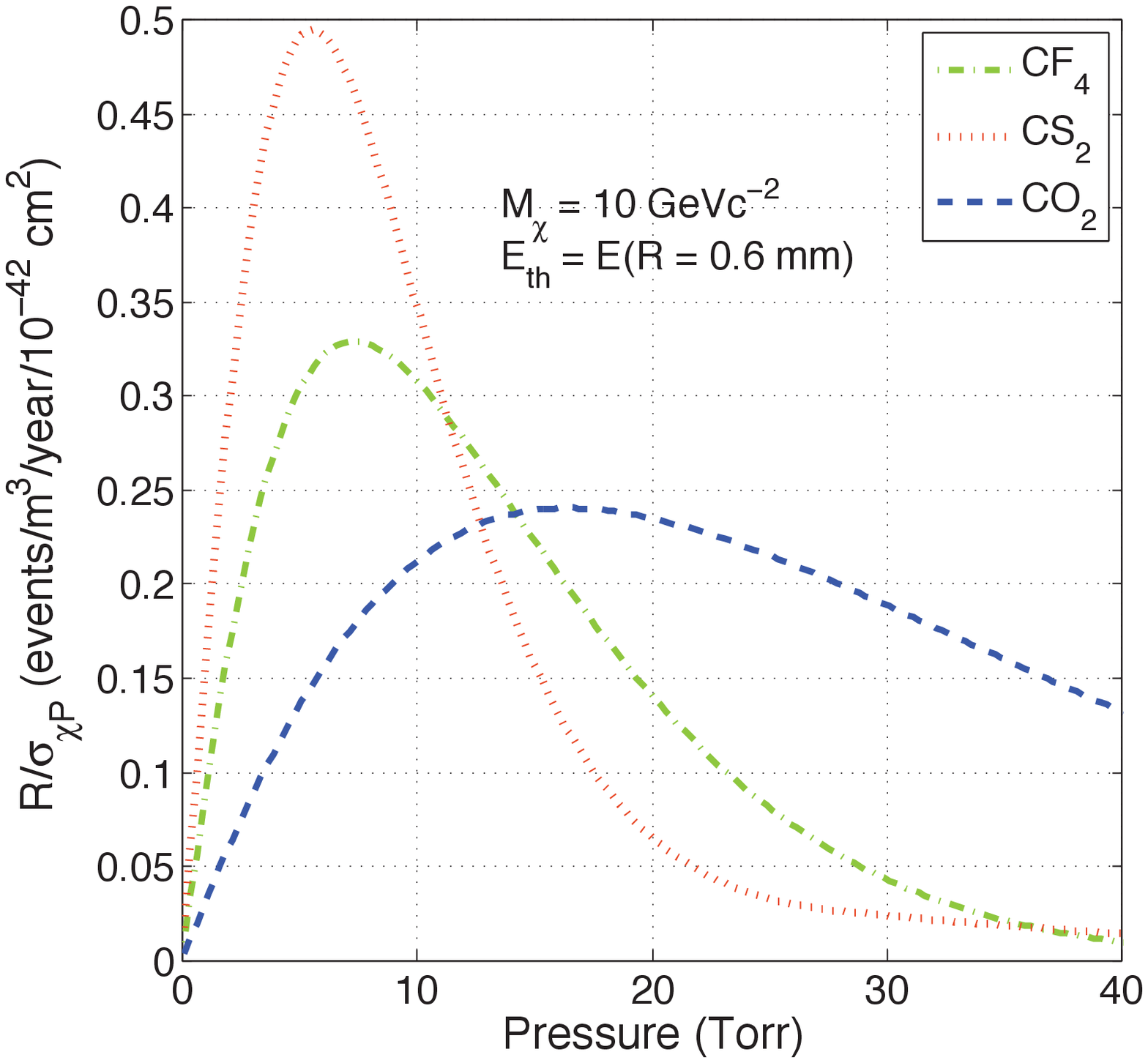}
\includegraphics[width=7.5cm]{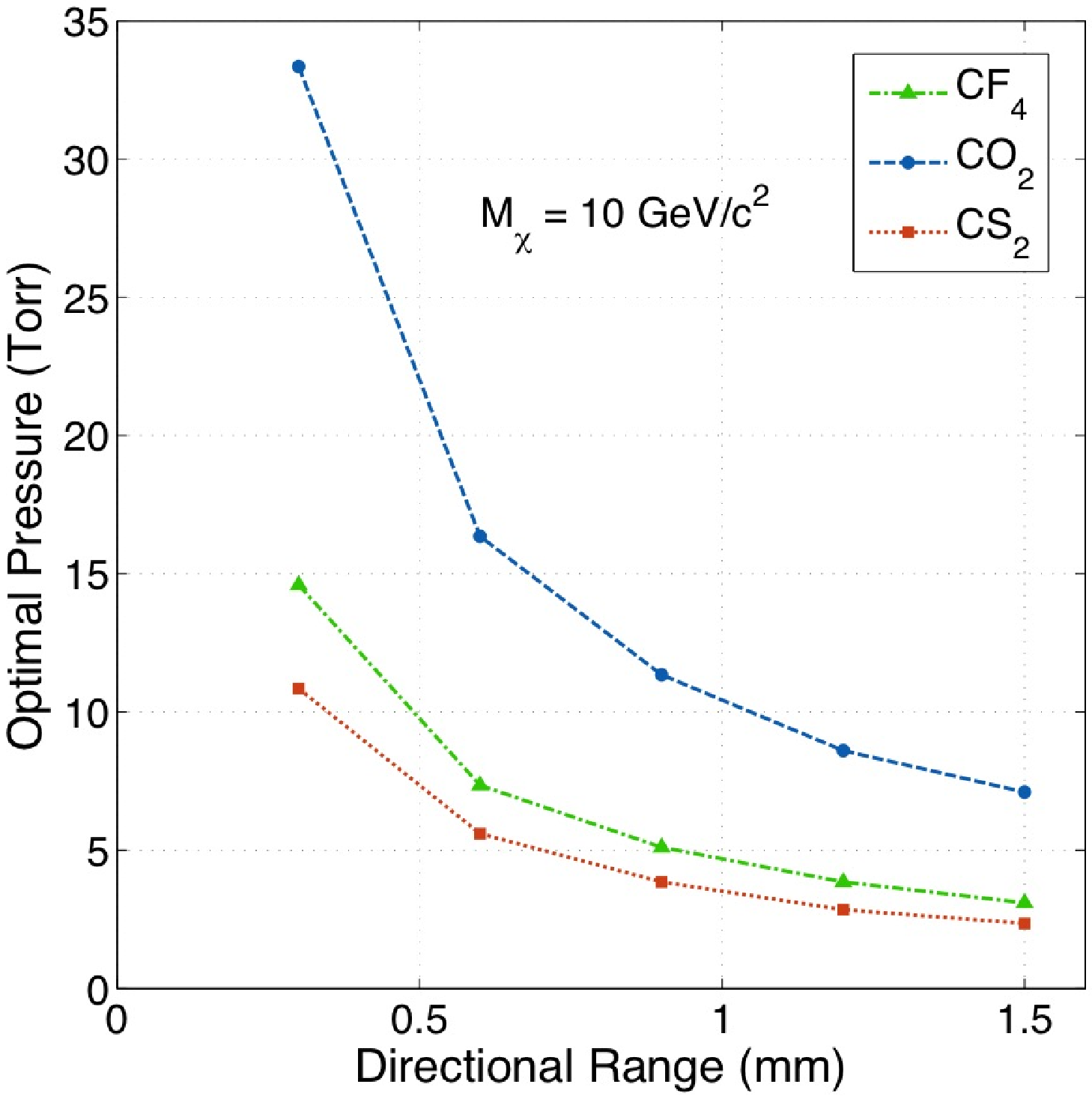}\\
\includegraphics[width=7.5cm]{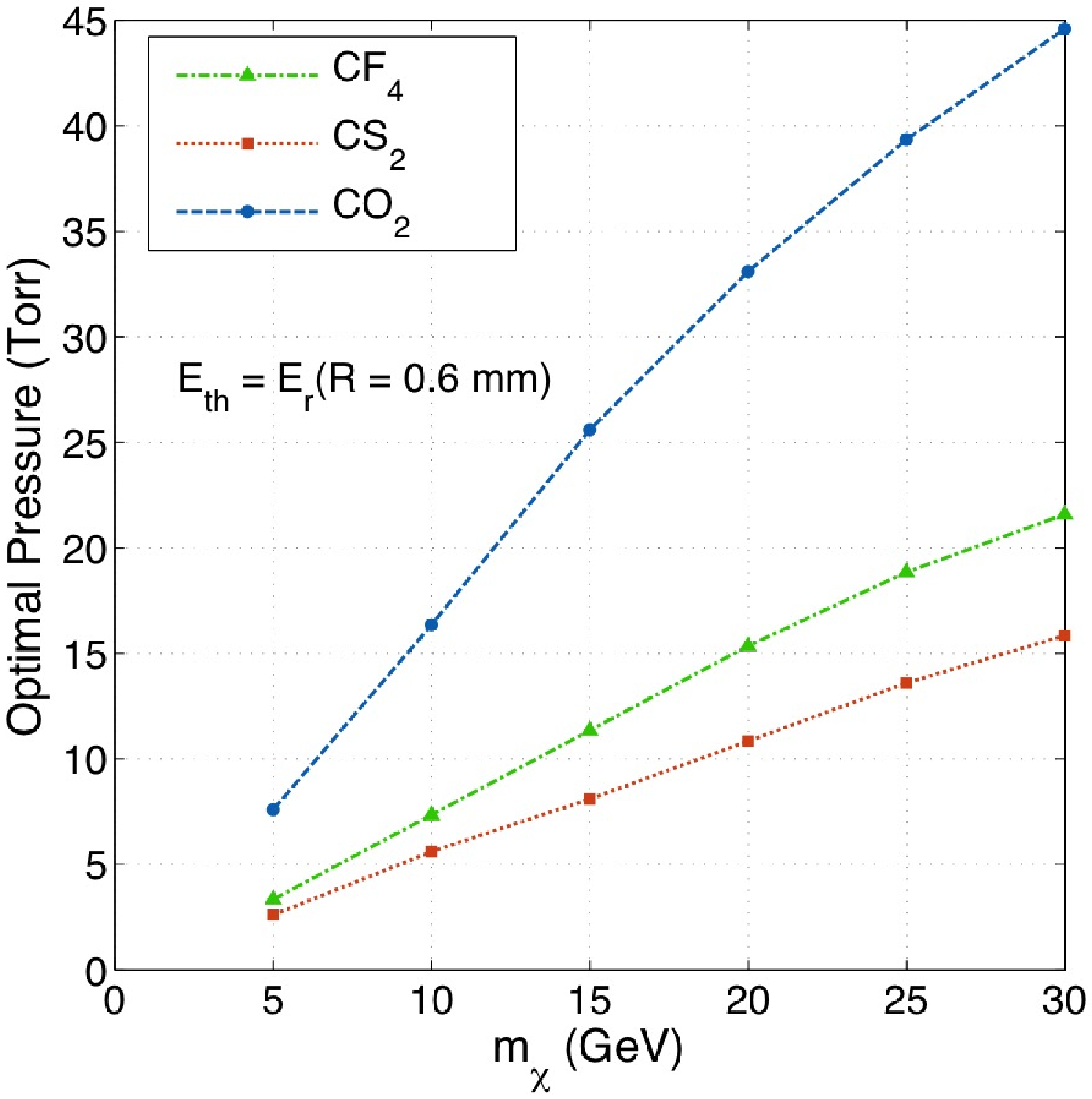}
\includegraphics[width=7.5cm]{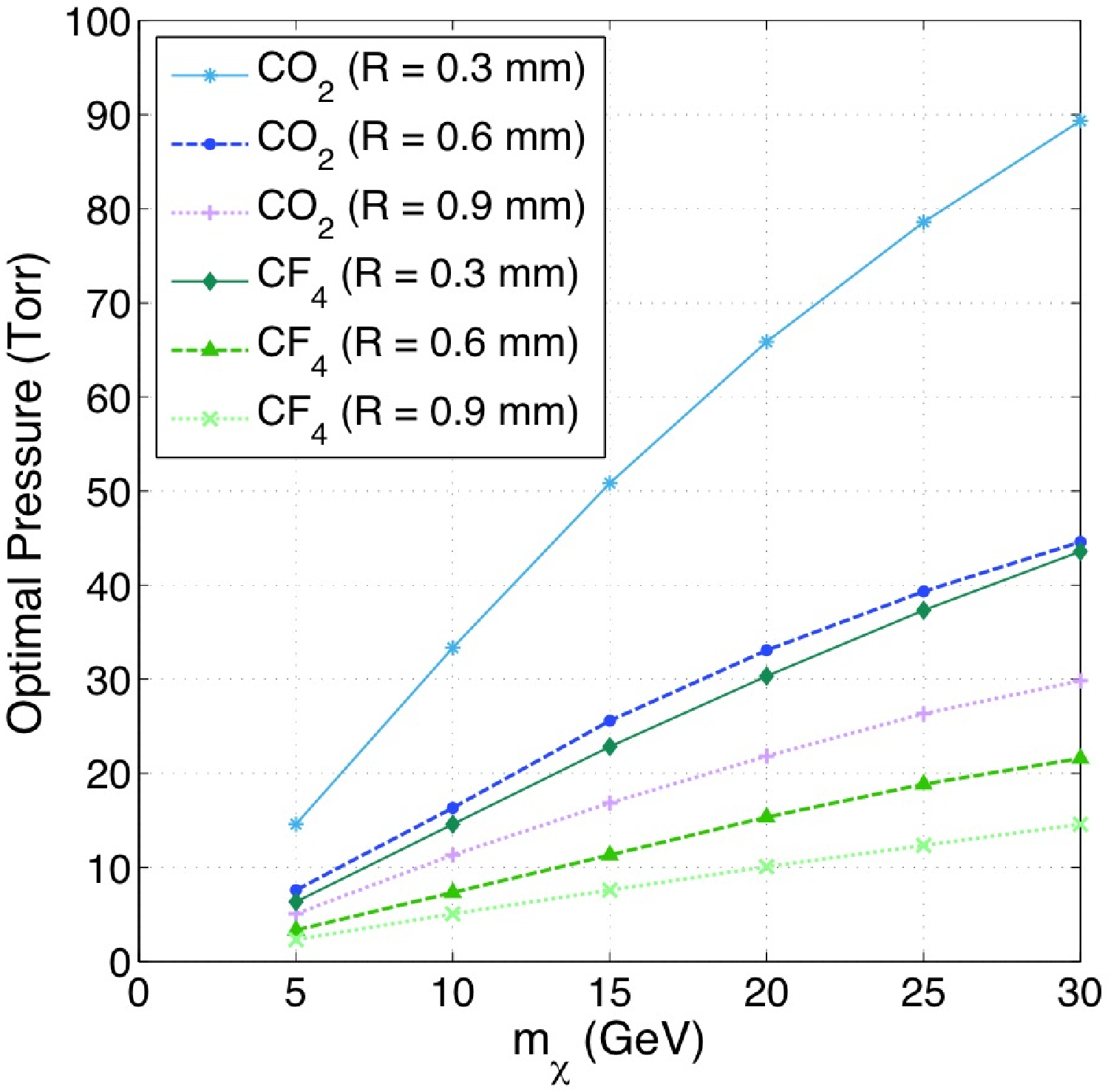}
\caption{Top left panel: The total directional rate, per
  cubic meter, per  year, for  a WIMP with mass $m_{\chi}= 10\,  {\rm
    GeV}$ and spin-independent cross-section $\sigma_{\rm p}=10^{-42} \, {\rm cm}^2$ 
for  CF${}_4$ (green dashed line), CS${}_{2}$ (red dotted) and CO${}_2$
  (blue dotted) as a function of pressure for a minimum track length,
  $R_{\rm min}=0.6 \, {\rm mm}$.
Top right: the optimal pressure as a function of $R_{\rm min}$, for the same gases and WIMP mass. Bottom left: the optimal pressure as a function of WIMP mass, with $R_{\rm min}=0.6 \, {\rm mm}$, and bottom right, the same but with 3 different values of $R_{\rm min}$, as indicated, and 
for CO${}_2$ and CF${}_4$ only.
}
%The final panel shows the total directional rate at the optimal
%pressure for each of the three gases as a function of $R_{\rm min}$.
%In each case the optimal pressure is taken from the previous three plots. 
\label{fig:exp3}
\end{center}
\end{figure}

 The directional rate as a function of pressure is shown in Fig.~\ref{fig:exp3} for three
 gases, ${\rm CO}_2, {\rm CF}_4$ and ${\rm CS}_2$, for a WIMP with $m_{\chi} = 10
\, {\rm  GeV}$ and $\sigma_{\rm p}^{SI}=10^{-42} \, {\rm cm}^2$. These plots show that to detect light WIMPs  a directional detector needs to operate in the $5-20$
 Torr range depending on the choice of light target.  The
 corresponding directional energy threshold is in the range $5-10 \, {\rm keV_r.}$  The resulting
 maximum rates are consistent with those derived in Ref.~\cite{Vahsen} but
 systematically lower than in Ref.~\cite{Martoff}.  
 The top right panel of Fig.~\ref{fig:exp3} show how the optimal pressure depends   on the minimum resolvable track length, for $0.3 \, {\rm mm} < R_{\rm min} < 1.5 \, {\rm mm}$, and WIMP masses in the range $5 - 30 \, {\rm GeV}$.
 These studies show that several cubic meter scale directional experiments, if appropriately
 configured with high resolution readouts, and operating with low
 diffusion, optimized pressures and targets, can probe new parameter space for low mass WIMPs.  
  
 We now describe some of the experimental challenges and steps needed to operate a directional detector at   low pressure 
 and  low energy threshold, as 
 required for a low mass WIMP search. The first set of challenges are related to the
 practical issues of operating gaseous TPCs at the 
 low pressure, $<20$ Torr, see Fig.~\ref{fig:exp3}.  The requirement of high voltages needed
 for maintaining the drift field in the TPC, and for achieving high gas gains in the
 amplification device, is incompatible with stable operation at low pressures.
 Nevertheless, high gas gains in micro-patterned gas
 detectors such as Thick GEMs (THGEMs) have been demonstrated down to $\sim 5$ Torr in
 certain gases~\cite{Weizmann}. Using such devices, Phan et al.~\cite{Phan} have shown
 sufficiently high gas gains with ${\rm CS}_2$ close to $10$ Torr. 
 Stable high voltage operation for sustaining a high drift field over tens of cm remains to be demonstrated.  Another practical challenge at these low pressures is maintaining the low
 diffusion that was critical for Phan et al to achieve their result of $R_{\rm min} \approx
 0.4- 0.6 \, {\rm mm}$.  Whether the thermal diffusion characteristic of negative ion gases, such as ${\rm CS}_2$, extends to low pressures also needs to be tested.
 
The second set of challenges are more fundamental in
 that they relate to the physics that characterizes the track structure of very low energy nuclear recoils. 
 The above extrapolation relies on the assumption
 that a $1 \, {\rm mm}$ long nuclear recoil track at high pressure is essentially
 the same as a lower energy, $1  \, {\rm mm}$ long track at low pressure.  
 There are a number of reasons why this might not be true. One is that, at low energies, there
 is more straggling (fluctuations in energy-loss and range) and larger deviation in the 
 recoil's trajectory from its initial direction due to multiple scattering. 
 In addition, the asymmetry in the ionization along the track, which
 gives rise to the head-tail effect, could be weaker at low energies. Finally, the fraction of
 energy loss into ionization (quenching) is expected to decrease at lower energies, which reinforces the need to achieve high gas gains at low pressures.  Given the steeply falling energy spectrum from low
 mass WIMPs, all of these effects could strongly dilute their directional signal relative to that observed at higher energies in the 100 Torr data from Ref.~\cite{Phan}.  Addressing these issues will require experimental data at low pressures, and will be critical for determining the exposure needed for the directional discovery of low mass WIMPs.
 
The third challenge is related to background discrimination, which also needs to extend down to the $5-10 \, {\rm keV_r}$ range required for directionality. For nuclear recoil energy
 thresholds in this range the detected energy would be in the range $2-4
\, {\rm  keV}_{{\rm ee}}$,  or even lower, since quenching factors are not yet measured at these energies. Thus, experimental 
data are also needed  to investigate discrimination at these low energies, as well as studies of the  background rates in the large detectors required.

\subsection{Conclusions}
\label{light-conc}
In this section we have discussed the theoretical and experimental
challenges for the directional detection of light WIMPs. In principle,
with an ideal detector, only $10-100$ events are required to detect an
anisotropic WIMP signal. However for light WIMPs only high speed WIMPs
in the tail of the speed distribution, can cause nuclear recoils which
are long enough for their direction to be measured. The event rate
above threshold is therefore typically small, and hence the exposure
required to detect a signal can be large. Furthermore there are
significant uncertainties in the shape of the high speed tail of the
WIMP distribution which lead to significant uncertainties in the event
rate above threshold, and consequently the required exposure.

Most directional detectors use gaseous TPCs operating at $\sim 100$ Torr. However at this pressure the track lengths of the low energy nuclear recoils induced by light WIMPs are too short to be resolved.
We have overviewed results from simulations of TPCs at 10 Torr 
and also extrapolations of data from a prototype 2d detector operated with CF${}_4$
at 100 Torr. These studies indicate that cubic meter scale low pressure TPCs could measure the directions of nuclear recoils 
from light WIMPs and probe new regions of WIMP mass--cross-section parameter space.
Finally we discussed the challenges for a directional light WIMP
search in the required low pressure, $10 - 20$ Torr, regime. Namely
the practical challenges in operating a TPC at low pressure, and the
uncertainties in the physics of low energy nuclear and electron
recoils at low pressures and energies.

%
% -------------- probinghalo
%10. Probing halo properties (substructures and constraining halo properties) : F. Mayet and S. Lee  
%\newpage
% Edited by FM 30/11/15 : notation (q-->r)
%   modify "inverse of the direction of Solar motion" to fit convention for skypmaps.
% Edited by FM 27/11/15 : link to def. of lambda (sec. exp)
% Edited by FM 25/11/15 : minor corrections
% Edited by FM 16/11/15 : minor corrections (following comments from AMG and GBG)

% -------------- probinghalo
%10. Probing halo properties (substructures and constraining halo properties) 
%F. Mayet and S. Lee (mayet@lpsc.in2p3.fr,samuelkl@princeton.edu)
%and B. J. Kavanagh (BRADLEY.KAVANAGH@cea.fr)

\section{Probing halo substructure}
\label{sec:probinghalo}

\subsection{Introduction}
The use of directional detection to constrain the astrophysical properties of 
the Dark Matter velocity distribution is an exciting possibility.  Directional detection 
may not only constrain the properties of the halo velocity distribution 
\cite{Billard:2010jh,Lee:2012pf} (see Sec.~\ref{sec:identification}),  
but may also be sensitive to the presence of velocity substructures in the halo.  
As discussed in Sec.~\ref{sec:theo.halo}, simulations suggest that these 
substructures might include Dark Matter tidal streams (spatially localized), debris flows (spatially homogenized), and a 
co-rotating dark disk.  Such components of the local velocity distribution can lead to distinctive 
features in the directional signal \cite{Lee:2012pf,Kuhlen:2012fz,Lisanti:2009vy,Green:2010gw,Billard:2012qu}.  
However, the detectability of these features depends strongly on the unknown properties of the 
substructure---as does the possibility of constraining these properties---and may require a very low energy 
threshold and/or a large exposure.  In the following, we review the sensitivity of directional detection to 
these velocity substructures, focusing on Dark Matter streams in Sec.~\ref{sec:detect-streams} and a dark disk in Sec.~\ref{sec:detect-disk}.

%
%Ciaran

\subsection{Streams}
\label{sec:detect-streams}

As discussed in Sec.~\ref{sec:theo.halo}, tidal streams are a class of substructures caused by the disruption and accretion 
by the Milky Way of stars and Dark Matter from satellite galaxies. The highly localized and anisotropic nature of tidal streams in both position and velocity make them an intriguing prospect for directional detection. A tidal stream such as this in the local Galactic region would manifest itself in the directional signal as a ring feature with increasing angular radius at lower energies (see Sec.~\ref{sec:features}).  Streams produce step-like features to the non-directional energy spectrum~\cite{Freese:2003tt}, as well as non-sinusoidal components to the annual modulation signal~\cite{Savage:2006qr}.

The simplest way to incorporate substructure into a halo model is by combining two parameterized distributions, a background distribution for the smooth halo and an additional component corresponding to the substructure.  When necessary, we shall adopt the Maxwell-Boltzmann velocity distribution for the smooth halo, as deviations from this are not likely to have an appreciable effect on the detection of streams. We further assume that a single tidal stream makes up a fraction $A_S = \rho_\textrm{str}/\rho_0$ of the local Dark Matter density, so that the total velocity distribution of the halo+stream model is given by
\begin{equation} \label{eq:halo_stream_vdf}
	f^\textrm{h+s}_\textrm{g}(\textbf{v}) = (1-A_S)f^\textrm{halo}_\textrm{g}(\textbf{v};\sigma,v_\textrm{esc}) + A_S f^\textrm{str}_\textrm{g}(\textbf{v};\textbf{v}_\textrm{str},\sigma_\textrm{str},v_\textrm{esc}) \, ,
\end{equation}
where,
\begin{eqnarray}  \label{eq:halo_stream_vdf_ind}
  f^\textrm{halo}_\textrm{g}(\textbf{v};\sigma,v_\textrm{esc}) &=& f^{\textrm{MB}}_\textrm{g}(\textbf{v};\sigma,v_\textrm{esc}) \, , \\
  f^\textrm{str}_\textrm{g}(\textbf{v};\textbf{v}_\textrm{str},\sigma_\textrm{str},v_\textrm{esc}) &=& f^{\textrm{MB}}_\textrm{g}(\textbf{v}-\textbf{v}_\textrm{str};\sigma_\textrm{str},v_\textrm{esc}) \,. \quad \quad
\end{eqnarray}
For simplicity, we have taken the velocity distribution of the stream to also be a Maxwell-Boltzmann distribution, but with dispersion $\sigma_\textrm{str}$  and a mean velocity $\textbf{v}_\textrm{str}$ in the Galactic frame.

The full Radon transform for the halo+stream model is therefore
\begin{equation}
	\hat{f}^\textrm{h+s}_\textrm{e}(v_\textrm{min},\hat{\textbf{r}}) = (1-A_S)\hat{f}_\textrm{g}^\textrm{MB}(v_\textrm{min} + \textbf{v}_\textrm{e} \cdot \hat{\textbf{r}},\hat{\textbf{r}}; \sigma, v_\textrm{esc}) 
	+ \, A_S\hat{f}_\textrm{g}^\textrm{MB}(v_\textrm{min} + (\textbf{v}_\textrm{e} - \textbf{v}_\textrm{str}) \cdot \hat{\textbf{r}},\hat{\textbf{r}}; \sigma_\textrm{str}, v_\textrm{esc}) \,,
	  \, 
\end{equation}
where $\hat{f}^\textrm{MB}_\textrm{g}$ is the Radon transform of the Galactic frame Maxwell-Boltzmann distribution. The recoils originating from scattered stream WIMPs are only dependent on the direction of the stream through the angle between the stream velocity and Earth's velocity. Hence, the dependence of the recoil distribution on the stream velocity is completely determined by the stream speed $v_\textrm{str}$ and the angle
\begin{equation}
	\Delta \theta = \cos^{-1}(\hat{\textbf{v}}_\textrm{e} \cdot \hat{\textbf{v}}_\textrm{str})\,.
\end{equation}

Using this halo+stream model to simulate recoil events, we can investigate the ability of directional detectors to characterize tidal streams.  There are two steps in characterizing a stream using directional detection.  The stream must first be detected;
 its parameters, {\it e.g.} density and velocity, can then be measured. Thus, we first consider statistical tests of stream 
detectability (summarizing results from Ref.~\cite{O'Hare:2014oxa}), and 
then discuss parameter estimation of stream properties (summarizing Ref.~\cite{Lee:2012pf}). The model detector used throughout this section is an idealized MIMAC-like detector with a 5 keV threshold. In the case of the parametric test a pixelization of a few degrees was used as a proxy for angular resolution.

\subsubsection{Stream detectability}
We first discuss non-parametric statistical tests of stream detectability.  Using only directional information, these tests can determine whether features inconsistent with a completely isotropic halo can be detected for streams, such as those with velocity distributions 
given by Eqs.~(\ref{eq:halo_stream_vdf}--\ref{eq:halo_stream_vdf_ind}).  Note, however, that they do not require that the form of the stream velocity distribution is known \emph{a priori}.  Such tests have been implemented in previous studies to determine the number of events required to distinguish a WIMP signal from isotropic backgrounds~\cite{Morgan:2004ys}, see Sec.~\ref{sec:discovery}. One advantage of a non-parametric analysis is that it is also not necessary to assume a particular model for the smooth component of the halo.  The results are valid provided the basic hypotheses that define the statistical tests are satisfied.  As such, the results here are valid for any background halo model that is isotropic in the Galactic frame. However, a notable disadvantage of such a non-parametric analysis is that it is less powerful than a parametric analysis (in the sense that, for a given data set, the detection significance will be lower). Hence, we will then consider the use of a profile likelihood ratio test to detect the presence of a stream parametrically.  In this case, the choice of background halo model (and the number of parameters needed to specify it) will affect the shape of the likelihood function; again, however, the effect of deviations from our choice of a simple Maxwell-Boltzmann halo is likely to be small.

\begin{figure}[t]
		  \centering
		  \includegraphics[trim = 0mm 0 0mm 0mm, clip, width=\textwidth]{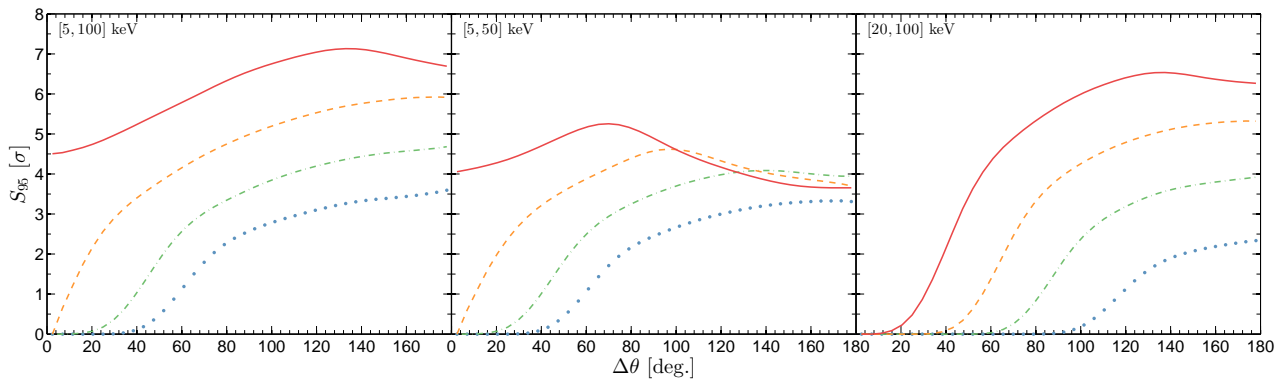}
		  \caption{The significance obtainable by 95\% of idealized experiments, $S_{95}$,  in units of $\sigma$, as a function of the angle $\Delta \theta$ between the Earth and stream velocities. From top to bottom in each panel, the curves correspond to $v_\textrm{str} = 500$ km s$^{-1}$ (red solid), 400 km s$^{-1}$ (orange dashed), 300 km s$^{-1}$ (green dot-dashed), 200 km s$^{-1}$ (blue dotted). The three panels are for energy windows  $[5,100]$ keV, $[5,50]$ keV and $[20,100]$ keV, from left to right.}
		  \label{S95_DeltaTheta}
		\end{figure}

In the following results, shown in Fig.~\ref{S95_DeltaTheta}, the analysis was performed on simulated MIMAC-like experiments with a $^{19}$F target. The recoil sets for each Monte-Carlo experiment were generated assuming an exposure of 10 kg yr and a spin dependent WIMP-proton cross section of $\sigma_p = 10^{-3}$ pb (just below the exclusion limit set by XENON100~\cite{Aprile:2011dd}). The local WIMP density was set to its canonical value 
$0.3 \ {\rm GeV\, cm^{-3}}$, and the WIMP mass was taken to be 50 GeV. For the background halo parameters, the values were set at $v_\textrm{esc} = 533$ km s$^{-1}$~\cite{Piffl:2013mla} and $\sigma = 220/\sqrt{2}$ km s$^{-1}$. The stream velocity dispersion was set to $\sigma_\textrm{str} = 10$ km s$^{-1}$ and its density fraction was set to a slightly large $A_S = 0.1$ (simulations suggest that typical tidal streams are more likely at the 1\% level \cite{Maciejewski:2010gz,Vogelsberger:2008qb}), but the results hold for more realistic stream fractions if correspondingly larger exposure times are assumed. The effects of annual modulation have been ignored in the following results. The experimental backgrounds were generated from an isotropic, energy-flat distribution with a rate of 
$R_\textrm{b} = 10$ kg$^{-1}$ yr$^{-1}$.

%	\subsubsection{Non-parametric tests}

\vspace{0.5cm}

\noindent{\bf Non-parameteric tests}\\	
	The non-parametric tests we consider consist of extracting a statistic that is distributed according to some null distribution in the case that the halo is a smooth and isotropic distribution ($A_S = 0$), and is distributed as an alternative distribution (which can be Monte-Carlo generated using mock data) in the case that the halo possesses an additional stream component ($A_S \neq 0$). A smooth and isotropic halo model will produce a directional signal that has a 
	median direction that corresponds to the Earth's velocity, as well as rotational symmetry around this direction. There are well defined procedures, which are 
	outlined in detail in Ref.~\cite{O'Hare:2014oxa}, that can be followed to test for these two hypotheses. To test the median direction, a $\chi^2$ statistic with a distribution that asymptotes to a $\chi^2_2$ distribution in the null case is extracted. To test the rotational symmetry, we use a modified Kuiper statistic $V^\star$; in the null case, the distribution of this statistic has no analytic form, but can be built by performing the test on any data set with rotational symmetry about the hypothesized direction ({\it e.g.}, isotropic 
	experimental backgrounds). The tests do not perform equally for all stream directions. In particular, the tests return a low significance when the stream is anti-aligned with the Earth velocity, as in this case the null hypotheses of rotational symmetry and inverse-Earth median direction are in fact correct. The number of events from stream WIMPs also plays an important role in determining the significance which can be achieved.  For streams with small $\Delta\theta$, this number is very low, and for low stream speeds or high threshold energies it can even be zero. 

%	\subsubsection{Parametric tests}

\vspace{0.5cm}
\noindent {\bf Parametric tests}\\
		We now consider parametric tests of detectability, assuming that the velocity distribution of the halo+stream is indeed
		given by the parameterized function described by Eqs.~(\ref{eq:halo_stream_vdf}--\ref{eq:halo_stream_vdf_ind}).  The profile likelihood ratio between the null (no stream) and alternative (stream) hypotheses is
		\begin{equation}
			\Lambda = \frac{\mathcal{L}(\hat{\hat{\boldsymbol{\theta}}} , A_S = 0)}{\mathcal{L}(\hat{\boldsymbol{\theta}})} \, ,
		\end{equation}
		where $\hat{\boldsymbol{\theta}}$ are the maximum likelihood estimators for the WIMP and halo parameters in the alternative model, and $\hat{\hat{\boldsymbol{\theta}}}$ are the maximum likelihood estimators evaluated when the stream density fraction is zero, $A_S = 0$. The profile likelihood ratio test statistic is then defined as
		\begin{equation}
			\mathcal{D} = \left\{ \begin{array}{rl}
			-2\ln \Lambda  & \, \, 0\le \hat{A}_S\le1 \,,\\
			0  & \, \, \hat{A}_S<0, \, \, \hat{A}_S>1 \,.
			\end{array} \right. 
		\end{equation}
		The advantage of this test for the detection of streams is that it assumes the null hypothesis is contained within the alternative hypothesis under a certain condition ($A_S = 0$).  Then, according to Wilk's theorem~\cite{Cowan:2010js}, the distribution of the test statistic when the null hypothesis is true becomes a $\chi_1^2$ distribution, so the significance of a particular result $\mathcal{D}_o$ can be written in units of $\sigma$ as $S = \sqrt{\mathcal{D}_o}$. 

		Fig.~\ref{S95_DeltaTheta} shows the detection significance obtainable for $95\%$ of experiments that is inferred with this test, assuming exposure times of 5 kg yr. The test, by virtue of being parametric, performs much better than the non-parametric tests. For example, for a stream with mid-range parameter values (a speed of 300 km s$^{-1}$ and an angle of 90$^\circ$), the tests for the three energy windows detect the stream at 4--5$\sigma$ in 95\% of experiments. For the same exposure, the non-parametric tests only reach a value of $S_{95}$ between 0.1 and 0.2~\cite{O'Hare:2014oxa}. 
		As in Ref.~\cite{Lee:2012pf}, the enhancement in performance is also due to the use of the full energy and direction information, as opposed to the use of only the latter in the non-parametric tests. Furthermore, the parametric tests achieve high significance over a wide range of stream velocities, with the limiting factor being the number of WIMPs coming from the stream. For low values of $\Delta\theta$, where the number of stream WIMPs drops to zero, the significance can be seen to do likewise. There is similarly a dependence on the energy window of the detector, which causes a reduction in the number of stream WIMPs when the stream becomes boosted such that the recoils fall past the maximum of the energy window. This can be seen clearly in the $[5,50]$ keV case. For a stream speed of 500 km s$^{-1}$, the significance begins to decrease for $\Delta\theta > 70^\circ$, and drops by $1.6\sigma$ up to $\Delta \theta = 180^\circ$. However, the significance for faster stream speeds is enhanced over what might be expected simply from the dependence on the number of stream WIMPs. This is due to faster streams becoming more prominent because of the exponential drop off with energy of the event rate for the smooth halo.

To summarize, the results described here demonstrate that there is reasonable prospect for detecting streams in the local Galactic region. However the detection significance achievable is highly dependent on the speed and direction of the stream, with slow streams aligned with the Earth's velocity evading detection in both non-parametric and likelihood analyses. The threshold and maximum energy of the detector also restrict the detection significance as they exclude recoils scattering from the stream WIMPs for very slow and very fast streams.

\subsubsection{Constraining stream parameters}
\label{sec:stream-constrain}

Directional-detection experiments may also be able to reconstruct the parameters of a stream~\cite{Lee:2012pf,O'Hare:2014oxa}.  
Likelihood-based methods~\cite{Lee:2012pf,Billard:2012qu} used to estimate the parameters of the halo velocity distribution 
can be easily extended to models with substructure, if the substructure velocity distributions can also be characterized 
by parameterized functional forms---as we have done for the case of a stream in 
Eqs.~\ref{eq:halo_stream_vdf}--\ref{eq:halo_stream_vdf_ind}.   Here, 5 parameters characterize the stream: the stream fraction $A_S$, its direction in Galactic coordinates ($\ell_S, b_S$), its speed $v_S$, and 
its velocity dispersion $\sigma_S$.  Combining these with parameters that describe the smooth halo, the WIMP mass, 
and the fraction $\lambda$ (defined by Eq.~\ref{eq:def.lambda}), 
Ref.~\cite{Lee:2012pf} considered a halo+stream model with a total of 10 parameters.  In particular, $A_S = 0.1$, $v_S = 510$ km s$^{-1}$, and $\sigma_S = 10$  km s$^{-1}$ were taken for the stream parameters, while typical values were assumed for the smooth halo. 

\begin{figure}[t]
\centering
%$\begin{array}{cc}
\includegraphics[scale=0.6]{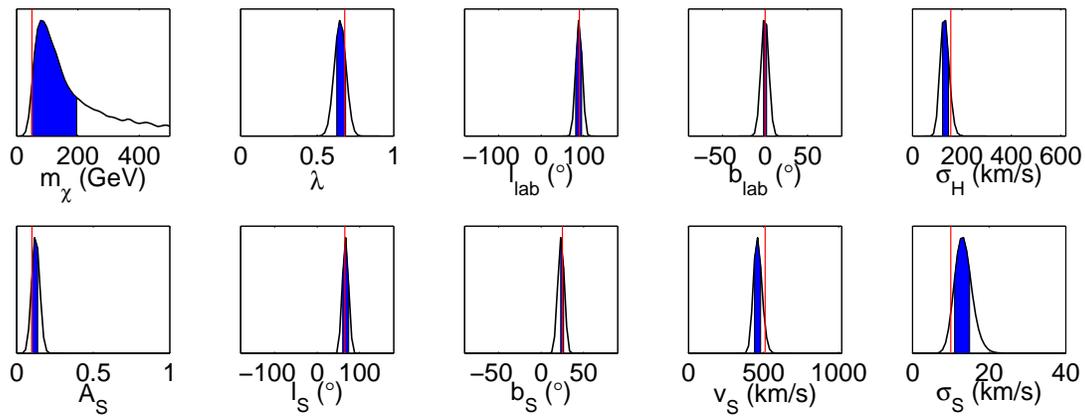}
%\end{array}$
\caption{Posterior probability distributions for the WIMP mass, the signal fraction $\lambda$ (defined by Eq.~\ref{eq:def.lambda}), and the 
SHM+stream model parameters.  Red lines indicate the fiducial parameter values assumed in simulating the data.  
68\% minimum credible intervals are shaded in blue. Fig. from \cite{Lee:2012pf}.}
\label{fig:halo-stream-posteriors}
\end{figure}

Assuming a WIMP mass of $50 \ {\rm GeV}$, a spin dependent WIMP-proton cross section of $\sigma_p = 10^{-3}$ pb, and a 30 kg year exposure for a $CF_4$ detector with a $5$-$\textrm{keV}$ threshold, a corresponding $\sim$650 signal and 
$\sim$300 background events were simulated (giving a signal fraction $\lambda  \approx 68\%$).  Bayesian parameter estimation was performed on this simulated data set, leading to the posterior probability distributions 
presented in Fig.~\ref{fig:halo-stream-posteriors}.  For all parameters, reconstructed values are consistent with their input values.  However, the constraint on the WIMP mass is relatively poor.  Despite this, the stream parameters 
(bottom row of Fig.~\ref{fig:halo-stream-posteriors}) are recovered with fairly good accuracy. This parameter-estimation analysis, along with a similar one carried out 
in Ref.~\cite{O'Hare:2014oxa}, thus demonstrates the possibility of using directional detectors to measure the properties of tidal streams. Finally it is worth mentioning that although this analysis used a 
pixelization of the angular data to provide a proxy for angular resolution, a full treatment of angular smearing would be required to determine how the constraints on stream parameters were affected by a finite angular resolution. However since the angular spread of stream WIMP induced recoils is rather large provided the angular resolution is below a few tens of degrees the stream is likely to be resolved relative to the 
background halo recoils.

\subsection{Dark disk}
\label{sec:detect-disk}

\subsubsection{Constraining dark-disk parameters}
\label{sec:disk-constrain}

The ability of directional-detection experiments to measure the properties of a dark disk was also discussed in Ref.~\cite{Lee:2012pf}.  
As mentioned in Sec.~\ref{sec:theo.halo}, simulations suggest that subhalos merging with the Milky Way may have been disrupted by the baryonic disk, leading to the formation of a co-rotating dark disk.  A parameterized velocity distribution for a halo+disk model can be constructed as in the case of a halo+stream, by simply replacing the stream parameters with the corresponding disk parameters.  Parameter estimation of simulated data sets can then be performed identically.  

\begin{figure}[t]
\centering
\includegraphics[scale=0.6]{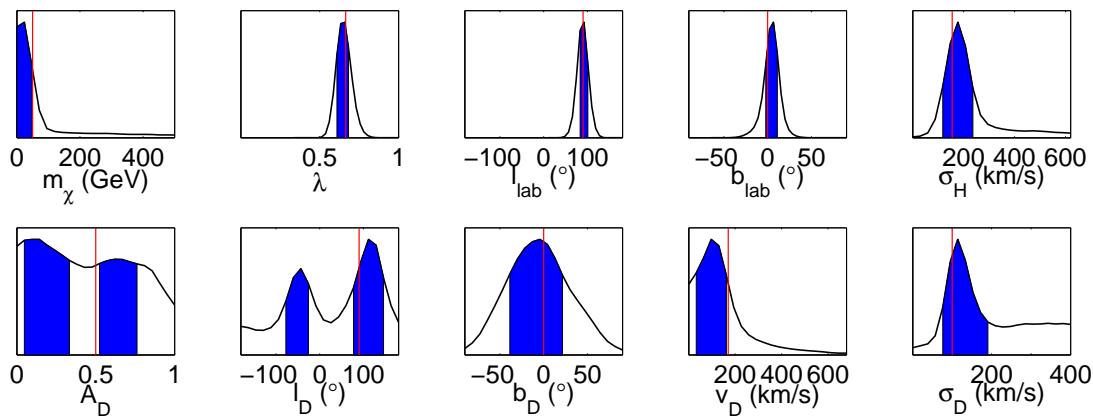}
\caption{Posterior probability distributions for the SHM+disk model parameters, as in 
Fig.~\ref{fig:halo-stream-posteriors}.  Estimation of the disk parameters (bottom row) is limited by the threshold, but significantly improves as the threshold is lowered.
Fig. from Ref. \cite{Lee:2012pf}.}
\label{fig:halo-disk-posteriors}
\end{figure}

Fig.~\ref{fig:halo-disk-posteriors} presents the resulting posterior probability distributions.  Here, both the experimental setup and
values for the WIMP and smooth-halo parameters were assumed to be identical to those in Sec.~\ref{sec:stream-constrain} (although slightly fewer events---a total of $\sim$800---were simulated; see Ref.~\cite{Lee:2012pf} for details).  Disk parameters appropriate for a slowly co-rotating dark disk, $v_D=170$  km s$^{-1}$ (corresponding to a lag speed of $\sim$50 km s$^{-1}$) and $\sigma_D = 100$  km s$^{-1}$, were also chosen.  Furthermore, as in the halo+stream analyses above, a disk fraction slightly larger than those typically found in simulations \cite{Maciejewski:2010gz} of $A_S = 0.5$ was assumed.

It is clear from the poor reconstruction of these disk parameters in Fig.~\ref{fig:halo-disk-posteriors} that measurements of the properties of a dark disk will be challenging.  This is primarily because WIMPs in the dark disk will have low velocities in the detector frame and will induce correspondingly low-energy recoils.  Thus, depending on the dark-disk lag speed, detector energy thresholds may need to be improved before detection of the dark disk is feasible. Note, however, that the parameters of the smooth halo are still constrained consistently with their input values and with a relatively small 
dispersion (Fig.~\ref{fig:halo-disk-posteriors}), even in the presence of a co-rotating dark disk.  The discovery potential of directional-detection experiments is similarly robust to the presence of a dark disk \cite{Billard:2012qu}, which does not significantly affect the experimental reach in the WIMP mass-cross section plane.
\

These studies have demonstrated that constraining simple parameterized halo models is possible even in the presence of substructures. In the most optimistic cases, the parameters of these substructures may also be evaluated accurately.  This is indeed a useful approach for early observations of 
the Dark Matter sky.  However, as directional detectors mature and exposures increase, and if we are ultimately able to observe a large number of WIMP events, other, more sophisticated approaches may be required.  We explore such approaches in the next section.

%
% -------------- decomposition
%11. Decomposing the Velocity Distribution
%F. Mayet and S. Lee (mayet@lpsc.in2p3.fr,samuelkl@princeton.edu)
%and B. J. Kavanagh (BRADLEY.KAVANAGH@cea.fr)
%\newpage
%Edited by FM (30/11/2015) - minor corrections + strong jeans theorem + notations
% Edited by BJK (06/10/2015) - edits to Sec. XII B, including removing Fig. 25
% Edited by BJK (02/10/2015) - edits to intro, response to comments from Ciaran (inc. new fig)
% Edited by FM (17/07/15) - this section is new (material comes from another
%section)

% -------------- decomosition
%11. Decomposing the Velocity Distribution
%F. Mayet and S. Lee (mayet@lpsc.in2p3.fr,samuelkl@princeton.edu)
%and B. J. Kavanagh (bradley.kavanagh@lpthe.jussieu.fr)

\section{Decomposing the velocity distribution}
\label{sec:decomposing}
\subsection{Introduction}

In Sec.~\ref{sec:probinghalo}, we reviewed the sensitivity of directional-detection experiments to halo and substructure velocity distributions that could be described by functional forms with a small number of parameters.  
Three recent studies have shown how to go beyond the assumption of such velocity distributions.  The first expands the velocity distribution in terms of moments of the integrals of motion \cite{Alves:2012ay} (which is possible if certain equilibrium conditions hold). The second and third use Fourier-Bessel expansion \cite{Lee:2014cpa} and 
angular discretization \cite{Kavanagh:2015aqa}, which allow for arbitrary velocity distributions.  These expansions provide a general framework for a detailed characterization of the velocity distribution using directional detection in the post-discovery era, which could be possible once a sufficiently large number of events have been observed.

\subsection{Moment decomposition}
\label{sec:halo.3d}
If the majority of Dark Matter particles in the halo have reached a state of equilibrium, Jeans' Theorem holds to a good
approximation \cite{biney}. It states that  any function of the integrals of motion is a steady-state solution of the collisionless 
Boltzmann
equation. This allows us to infer global properties of the Dark Matter halo 
from an estimation of the integrals of motions in the solar neighborhood. 
Frollowing the strong version of the Jeans' theorem, 
three integrals of motion have to be chosen. The choice made in 
Ref. \cite{Alves:2012ay} is as follows: 
\begin{itemize}
\item The energy $\mathcal{E}$ given by $\mathcal{E}=\frac{v^2-v^2_{\text{esc}}}{2}$.
\item The $z$-component of the angular momentum, $L_z$ that is equal to $r_\oplus v_\phi$ 
in the solar neighborhood, $\phi$ being      the azimuthal angle with respect  to the orbit of the Earth.
\item The magnitude of the angular momentum $L_t$ given by $L_t=r_\oplus v_\theta$ in the 
Solar neighborhood,  $\theta$ being   the angle between the  velocity vectors of the recoil and of the Earth's motion.
\end{itemize}
The assumption is made that   $f(\mathcal{E},L_t,L_z)$ is separable, {\it i.e.} 
$f(\mathcal{E},L_t,L_z) =f_1(\mathcal{E})f_2(L_t)f_3(L_z)$ and the distribution functions of each of the integral 
of motion are given by the following 
special function decompositions   
\begin{eqnarray}
\label{expansion}
f_1(\mathcal{E})&=&\sum_{\ell}c_{P_\ell}\tilde{P}_\ell\left(\frac{\mathcal{E}}{\mathcal{E}_\text{lim}}\right),\nonumber\\
f_2(L_t)&=&\sum_{n}c^t_{F_n}\cos\left(n\pi \frac{L_t}{L_{\text{max}}}\right),\\
f_3(L_z)&=&\sum_{m}c^z_{F_m}\cos\left(m\pi \frac{L_z}{L_{\text{max}}}\right)\nonumber\,,
\end{eqnarray}
where $\tilde{P}_\ell$ are shifted Legendre polynomials, $\mathcal{E}_\text{lim}$ 
is the lowest Dark Matter energy corresponding to the recoil threshold $E_r^\text{th}$ and  
$L_\text{max}\equiv r_\oplus v_\text{esc}$.\\

This parametrization has previously been applied to mock data, simulated for a background-free $\text{CS}_2$ directional detector, 
with recoil energy  between 5 and  15 ${\rm keV}$  and for a ${6 \ \rm GeV}$ WIMP. The double-differential spectrum $\mathrm{d}^2R/\mathrm{d}E_r$d$\Omega_r$, from this simulated data, is then used to reconstruct the decomposition 
coefficients in
order to infer $f_1(\mathcal{E})$, $f_2(L_t)$, and $f_3(L_z)$
and hence $f(\mathcal{E},L_t,L_z)$. 

Ref.~\cite{Alves:2012ay} thus demonstrated that 
a few thousand events are sufficient to allow for  a good measurement of the  Dark Matter 
distribution. The method was applied to a distribution described by the analytic Michie distribution \cite{Michie}, as well as to a distribution extracted from the 
 Via Lactea II N-body simulation \cite{Kuhlen:2008qj}. In the latter case, the input model contains anisotropies and departures from a standard Maxwellian distribution  
 \cite{Lisanti:2010qx}. Within error bars, the distributions of integrals of motion that are evaluated from the directional 
 data are in very good agreement with the input Dark Matter distribution, indicating that this approach can accommodate such deviations from the SHM.
The evaluation of the integrals of motion at a given radius may then be used to infer  the Dark Matter distribution at any radius, 
providing  the distribution is in a steady state of equilibrium. This has been checked on data from the Via Lactea II 
simulation up to larger radii (30 kpc).

%The underlying assumptions of the method have been sucessfully tested. It concerns the fact 
%the energy and angular momentum are integrals of motion and also the separability hypothesis.

In conclusion, directional detection may be used to infer Dark Matter
phase space distribution in the solar neighborhood, through an evaluation of 
the coefficients of its moment decomposition on a model independent basis. As noted in Ref. \cite{Alves:2012ay}, this method 
avoids the use of an analytic Dark Matter model and is hence one step toward a model independent analysis of 
directional data. However, it relies on the assumption of equilibrium and also requires that the WIMP mass is known \textit{a priori}, and can therefore only be applied in the post-discovery era.   

%-----------------------
\subsection{Fourier-Bessel decomposition}
\label{sec:halo.lee}

The moment decomposition requires that the WIMP velocity distribution is a separable function of the integrals of motion.  If 
the velocity distribution cannot be represented as such, it may then be necessary to 
work in a basis that allows for arbitrary velocity distributions. 
This may be the case if the Dark Matter halo contains a significant unvirialized fraction.

One possible basis is given by a generalization of the Fourier-Bessel basis.  Since the Galactic-frame WIMP velocity distribution $g(\mathbf{v})$ is typically truncated at the escape velocity $\vesc$, we consider basis functions that are given by the product of a spherical Bessel function of the first kind $j_l$ and a real spherical harmonic $S_{lm}$, and which vanish for ${v > \vesc}$:
\es{eq:FBfunctiondisc}{
\Psi^n_{lm}(\mathbf{v}) &= 4\pi c_{ln} j_l (u_{ln} v) S_{lm}(\vhat) \theta(\vesc-v)\,.
}
These basis functions are hence labeled by integers $n$, $l$, and $m$.  Here, ${u_{ln} = x_{ln} / \vesc}$, where $x_{ln}$ is the $n$th zero of $j_l$, and ${c^{-1}_{ln} \equiv x_{ln} j_{l+1}(x_{ln})/\sqrt{\pi}}$.  An arbitrary velocity distribution truncated at $\vesc$ can then be expanded as
\es{eq:g-exp}{
g(\mathbf{v}) = \sum_{nlm} \frac{x_{ln}^2 \vesc^{-3}}{(2\pi)^3} \Psi^n_{lm}(\mathbf{v}) g^n_{lm}\,,
}
and is completely characterized by the expansion coefficients $g^n_{lm}$.  Note that the requirement that $g(\mathbf{v})$ is positive definite places a restriction on these coefficients.

Besides the ability to represent arbitrary velocity distributions, another advantage of the Fourier-Bessel basis is that it also admits an analytic expression for the directional recoil 
spectrum ${\rm d}^2R/{\rm d}E_r \dOq$ in terms of the expansion coefficients $g^n_{lm}$.  A detailed derivation of this analytic expression can
be found in Ref.~\cite{Lee:2014cpa}; we simply summarize the result here.

The directional recoil spectrum ${\rm d}^2R/{\rm d}E_r \dOq$ is easily found given the Radon transform $\widehat{f}$ of the lab-frame velocity distribution $f$.  In turn, because the lab-frame distribution $f$ and the Galactic-frame distribution $g$ are simply related by a translation in velocity space, using the translation properties of the Radon transform we can easily find $\widehat{f}$ given $\widehat{g}$.  Furthermore, because of the linearity of the Radon transform, we see that the expansion 
in Eq.~\eqref{eq:g-exp} yields
\es{eq:ghat-exp}{
\widehat{g}(\vq, \uhat) = \sum_{nlm} \frac{u_{ln}^2 \vesc^{-1}}{(2\pi)^3} \widehat{\Psi}^n_{lm}(\vq, \uhat) g^n_{lm}\,.
}
Thus, the net implication is that the directional recoil spectrum ${\rm d}^2R/{\rm d}E_r \dOq$ can be easily computed if the coefficients $g^n_{lm}$ and the Radon transforms $\widehat{\Psi}^n_{lm}$ of the basis functions are known.

The crux of the derivation is then the calculation of $\widehat{\Psi}^n_{lm}$, which can be accomplished by contour integration and the utilization of various mathematical properties of both the Fourier-Bessel basis functions and the Radon transform.  The result is
\es{eq:PsiRT}{
\widehat{\Psi}^n_{lm}(\vq, \uhat) &= 8\pi \vesc^2 i^{-l} c_{ln} x_{ln} j_{l-1}(x_{ln}) K^n_l(\vq/\vesc) S_{lm}(\uhat)\,,
}
where we have defined the integrals
\es{eq:Knl}{
K^n_l(a) &= \int\! dx\, \frac{e^{i a x} j_l(x)}{x^2-x_{ln}^2}\,.
}
This integral can be performed analytically and expressed as a finite sum of polynomial and trigonometric terms; see Ref.~\cite{Lee:2014cpa} for the full result.

The directional recoil spectrum is then given by
\es{eq:key-result}{
\frac{{\rm d}^2R}{{\rm d}E_r \dOq} = \frac{\rho_0 \sigma_\mathrm{N} S(q)}{4\pi^3 \mchi \mu_\mathrm{N}^2 \vesc} \sum_{nlm} i^{-l} c_{ln} x_{ln}^3 j_{l-1}(x_{ln}) K^n_l[(\vq + \mathbf{\vlab}\cdot\qhat)/\vesc] S_{lm}(\qhat) g^n_{lm}\,.
}
Although this result may superficially appear somewhat complicated, it is actually surprisingly straightforward.  Given a velocity distribution characterized by a finite set of expansion coefficients $g^n_{lm}$ (truncated at some maximum $n$ and $l$), we have an analytic result for the directional recoil spectrum as a sum of a finite number of terms composed of elementary functions.  Such a result allows likelihood-based estimation of the expansion coefficients to be performed efficiently, obviating the need for numerical integration to calculate the predicted directional recoil spectrum during computation of the likelihood.

%BJK - 02/10/2015
\subsection{Discretized decomposition}
\label{sec:halo.kavanagh}

As mentioned above, the requirement that the velocity distribution be everywhere positive places a restriction on the possible coefficients  $g_{lm}^n$. This is a consequence of using the spherical harmonics (which can take negative values) as an angular basis, meaning that any method which makes use of a spherical harmonic decomposition will have a similar requirement. Consequently, for a given set of coefficients $g_{lm}^n$, it may be necessary to perform computationally expensive tests to check for unphysical negative values of $f(\mathbf{v})$. Even if such unphysical distributions can be rejected, it is not clear how this problem will affect parameter scans which attempt to reconstruct the coefficients $g_{lm}^n$ and therefore the shape of $f(\mathbf{v})$.

An alternative approach \cite{Kavanagh:2015aqa} is to discretize the velocity distribution in the angular variables, such that $f(\mathbf{v})$ is a function only of $v = |\mathbf{v}|$ within each angular bin. For discretization into $N$ bins, one writes

\es{eq:discretized-f}
{
f(\textbf{v}) = f(v, \cos\theta', \phi') =
\begin{cases}
f^1(v) & \textrm{ for } \theta' \in \left[ 0, \pi/N\right]\,, \\
f^2(v) & \textrm{ for } \theta' \in \left[ \pi/N, 2\pi/N\right]\,, \\
 & \vdots\\
f^k(v) & \textrm{ for } \theta' \in \left[ (k-1)\pi/N, k\pi/N\right]\,, \\
 & \vdots\\
f^N(v) & \textrm{ for } \theta' \in \left[ (N-1)\pi/N, \pi\right]\,, \\
\end{cases}
}
where for simplicity only discretization in the polar angle $\theta'$ is considered. The velocity distribution is now described by $N$ functions of $v$, 
labeled $f^k(v)$. Using this decomposition, $f(\mathbf{v})$ is guaranteed to be everywhere positive (and therefore physical) 
provided that a suitable parametrization of $f^k(v)$ is chosen which is also everywhere positive (for examples, 
see Refs.~\cite{Peter:2011eu,Kavanagh:2012nr, Kavanagh:2013wba}).

Because some directional information is discarded in discretizing the velocity distribution, it is expected that the corresponding Radon transform may not provide a close approximation to the true Radon transform. Instead, the integrated Radon transform (IRT) is considered, obtained by integrating $\hat{f}(v_\mathrm{min}, \hat{\textbf{r}})$ over the same angular bins as above,
\es{eq:IRT}
{
\hat{f}^j(v_\mathrm{min}) = \int_{\phi = 0}^{2\pi} \int_{\cos(j\pi/N)}^{\cos((j-1)\pi/N)} \hat{f}(v_\mathrm{min}, \hat{\textbf{r}})\, \mathrm{d}\cos\theta\mathrm{d}\phi\,.
}
Though using the IRT requires that the data be binned in angle, this should reduce the error induced by using the discretized (rather than full) velocity distribution. In the simplest case, $N=1$, the IRT corresponds to the total non-directional rate and induces no discretization error 
\cite{Gondolo:2002np}. For $N=2$ (capturing a simple forward-backward asymmetry), the IRTs can be written

\begin{align}
\label{eq:N2result}
\hat{f}^1(v_\mathrm{min}) &= 4\pi\int_{v_\mathrm{min}}^{\infty} v \left\{ \pi f^1(v) + \tan^{-1}\left(\sqrt{v^2/v_\mathrm{min}^2-1}\,\right)\left[f^2(v) - f^1(v)\right] \right\} \, \mathrm{d}v \,,\\
\hat{f}^2(v_\mathrm{min}) &= 4\pi\int_{v_\mathrm{min}}^{\infty} v \left\{ \pi f^2(v) + \tan^{-1}\left(\sqrt{v^2/v_\mathrm{min}^2-1}\,\right)\left[f^1(v) - f^2(v)\right] \right\} \, \mathrm{d}v\,.
\end{align}
The angular integrals involved in the Radon transform can be performed analytically, leaving only a single integral over $v$ for each IRT. Furthermore, computation of the IRTs can be extended to arbitrary definitions of the angular bins and to arbitrarily high $N$ (see Appendix B of Ref.~\cite{Kavanagh:2015aqa} for explicit expressions). 

\begin{figure}[t!]
  \centering
  \includegraphics[width=0.49\textwidth]{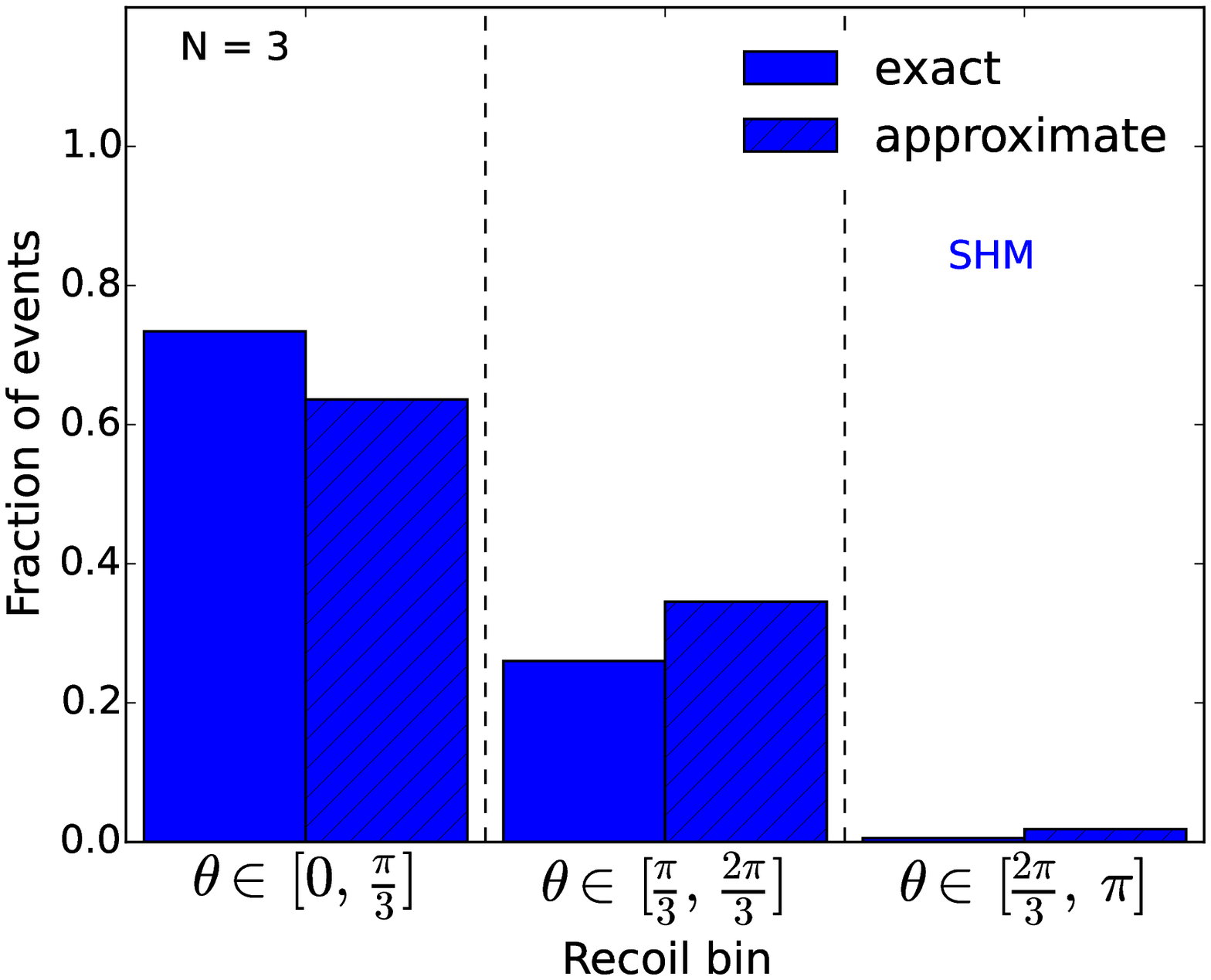}
  \includegraphics[width=0.49\textwidth]{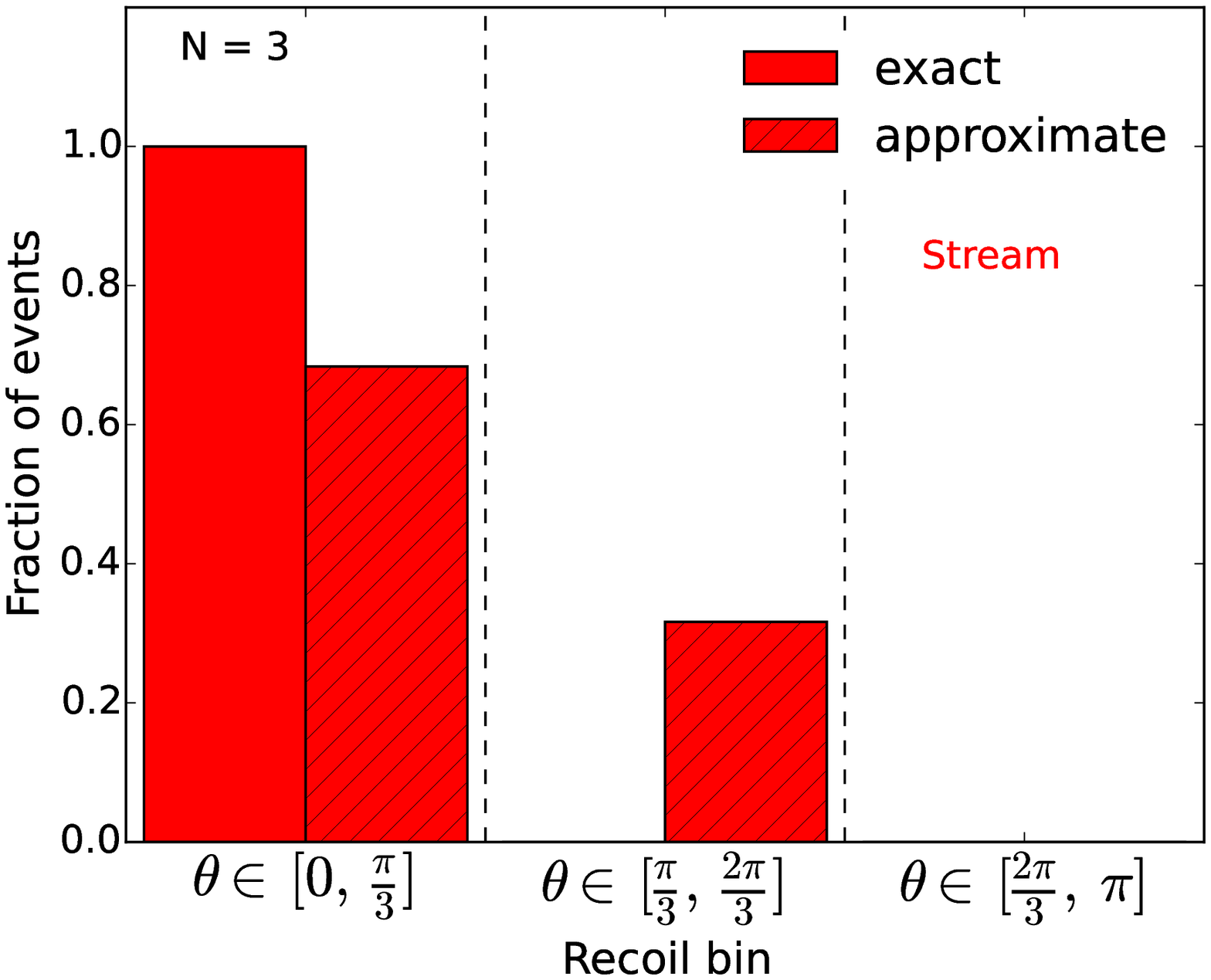}  
\caption{Fraction of events expected in each of $N=3$ angular bins, calculated using the exact velocity distribution (solid bars) and the approximate binned distribution (hatched bars) defined in Eq.~\ref{eq:discretized-f}. We assume a Flourine-based directional detection experiment with a 20 keV threshold and $m_\chi = 50 \,\, \mathrm{GeV}$. Results for the SHM distribution are shown in the left panel and for a stream distribution (with a stream speed of $500 \,\, \mathrm{km} \, \mathrm{s}^{-1}$) in the right panel. The direction of Solar motion is aligned along $\theta = \pi$.}
  \label{fig:CompareN3}
\end{figure}

In order to determine the size of the error induced by the discretization, one can compare the fraction of signal events in each angular bin, calculated using the `approximate' and `exact' IRTs. For a given velocity distribution, the `approximate' IRT is obtained from the discretized velocity distribution as described above, with $f^k(v)$ equal to its angular average over each bin. The `exact' IRT is obtained by integrating the full Radon transform of the underlying distribution over each angular bin. 
Figure~\ref{fig:CompareN3} shows an example of this comparison for the case of $N=3$ bins. For the Standard Halo Model (SHM) distribution, with $v_\mathrm{lab} = 220 \textrm{ km s}^{-1}$ and $\sigma_v = 156 \textrm{ km s}^{-1}$, the discretized $f(\mathbf{v})$ provides a good approximation to the distribution of signal events across the three bins. For example, in the forward-facing bin ($\theta < \pi/3$), the error between the `exact' and `approximate' event numbers is roughly 10\%. For a more extreme distribution, such as a stream ($v_\mathrm{lab} = 500 \textrm{ km s}^{-1}$ and $\sigma_v = 20 \textrm{ km s}^{-1}$), the agreement is poorer. In the forward-facing bin, the discrepancy increases to around 30\%, while in the transverse direction ($\theta \in \left[ \frac{\pi}{3}, \frac{2\pi}{3} \right]$) the discretized approximation significantly over-estimates the scattering rate. 

For smooth distribution functions (such as the SHM), this discrete basis can therefore be used to closely approximate the full velocity distribution using as few as $N=3$ bins, while ensuring that these approximate distributions are everywhere positive and therefore physical. The result is that only 3 free functions of $v$ are required for fitting to data. By choosing some 
suitable parametrization for these 3 free functions, astrophysical uncertainties can be mitigated and information about the shape of $f(\mathbf{v})$ can be extracted from the data. For highly-directional distributions (such as streams), more bins would be required, with the exact number depending on the number of observed events and the angular resolution of the detector.

In practice, it will not be necessary to employ such general formalisms until a large number of directional recoil events has been observed. One might imagine that the use of parameterized functional forms for the velocity distribution and it substructure, as discussed in Sec.~\ref{sec:probinghalo}, will suffice in the early days of the post-discovery era, when fitting a large number of free parameters may be unfeasible. However, after an understanding of the basic features of the velocity distribution is reached, these decomposition methods provide a framework with which a detailed and complete picture of the local Dark Matter sky can be constructed.

%
% -------------- conclusion
%12. Conclusion : all  
%\newpage
%
% -------------- conclusion
%12. Conclusion : all  
\section{Conclusion}
\label{sec:conclusion}

Directional detection is a promising next-generation strategy for the direct detection of Dark Matter. We have reviewed the potential of directional detectors for detecting and characterizing Galactic WIMPs. Thanks to the characteristic features exhibited by the WIMP-induced recoil distribution, mainly the dipole feature but also the ring-like and aberration features, directional detection may be used to genuinely authenticate a Dark Matter signal. Indeed, an unambiguous proof of discovery would be given by a main recoil direction pointing toward the Sun's direction of motion. We have also shown that with a larger exposure, and hence a larger number of WIMP events, directional detection may be used to constrain the properties of both the WIMP particle and the Galactic Dark Matter halo. We have also emphasized the possibility of probing the underlying particle physics model as well as the Dark Matter velocity distribution, including substructures.

The expected reach of directional detection of Dark Matter thus highlights the interest in a worldwide effort toward the construction of large directional detectors.

%
%Acknowledgements
%
%\newpage
\section{Acknowledgements}

J.B.R.B. acknowledges the support of the Alfred P. Sloan Foundation, and the Research Corporation for Science Advancement. 
N.B. acknowledges support from the European Research Council through the ERC starting grant WIMPs Kairos.
G.B.G.  was supported in part by the US Department of Energy under Award Number DE-SC0009937. 
A.M.G. acknowledges funding from STFC and the Leverhulme Trust. 
B.J.K. is supported by the European Research Council (ERC) under the EU Seventh Framework Programme (FP7/2007- 2013)/ERC Starting Grant (agreement n. 278234 -- 'NewDark' project). 
B.J.K. is also supported in part by the John Templeton Foundation Grant 48222.
J.M. acknowledges support from ERC project no. 279980
S.V.  acknowledges support from the U.S. Department of Homeland
Security under Award Number 2011-DN-077-ARI050-03 and the U.S. Department of Energy under Award Numbers DE-SC0007852 and DE-SC0010504.

%\newpage
\appendix
%Edited by FM 30/11/2015 -   notations (q->r)
%Edited by FM 16/11/15 : minor corrections (following comments from AMG and GBG)

\section{Transformation from the detector frame to the Galactic frame}
\label{app-A}

The differential event rate depends on the Radon transform, $\hat{f}$ of the WIMP velocity distribution, which in turn depends on $\hat{\mathbf{r}} \cdot {\bf v}_{\rm lab}$. Hence, in order to compute the differential rate we need to orient the nuclear recoil direction $\hat{\mathbf{r}}$ with respect to ${\bf v}_{\rm lab}$. Here we present complete transformation equations for $\hat{\mathbf{r}}$ and ${\bf v}_{\rm lab}$ to go from the detector frame to the Galactic reference frame. The equations are the same as presented in Appendix A of Ref.~\cite{Bozorgnia:2011vc} except for those describing Earth's revolution.

We define a reference frame fixed to the laboratory and orient its axes so that the $xy$ plane is horizontal, the $x$-axis points North, the $y$-axis points West, and the $z$-axis points to the Zenith. We denote its unit coordinate vectors as $\hat{\cal N}$, $\hat{\cal W}$ and $\hat{\cal Z}$, respectively. The detector is at some orientation in the laboratory. We define the detector frame with $X,Y,Z$ cartesian axes fixed to the detector. The unit coordinate vectors of the detector frame are $\hat{\mathbf{X}}$, $\hat{\mathbf{Y}}$ and $\hat{\mathbf{Z}}$. The transformation between the lab frame and the detector frame is given in Eqs. A1 to A3 of Ref.~\cite{Bozorgnia:2011tk} in terms of ``direction cosines''. $\hat{\mathbf{r}}$ is given in the detector reference frame as $\hat{\mathbf{r}}=r_X ~\hat{\mathbf{X}} +r_Y ~\hat{\mathbf{Y}} +r_Z ~\hat{\mathbf{Z}}$, and we can write it in the lab frame using Eq. A1 of Ref.~\cite{Bozorgnia:2011tk}, $\hat{\mathbf{r}}=r_n ~\hat{\cal N}+r_w ~\hat{\cal W} +r_z ~\hat{\cal Z}$, where
\begin{align}
r_n&=r_X \alpha_X + r_Y \alpha_Y +r_Z \alpha_Z,\nonumber\\
r_w&=r_X \beta_X + r_Y \beta_Y +r_Z \beta_Z,\nonumber\\
r_z&=r_X \gamma_X + r_Y \gamma_Y +r_Z \gamma_Z,
\end{align}
and $\alpha_i$, $\beta_i$ and $\gamma_i$ are the direction cosines between the two sets of cartesian coordinates of the lab and detector frames, for $i=X,Y,Z$. For example $\alpha_Y$ is the cosine of the angle between the $\hat{\cal N}$ and $\hat{\mathbf{Y}}$ directions, and $\beta_Z$ is the cosine of the angle between the $\hat{\cal W}$ and $\hat{\mathbf{Z}}$ directions.

We would like to write $\hat{\mathbf{r}}$ in the Galactic reference frame. We proceed by first writing $\hat{\mathbf{r}}$ in the equatorial frame using Eq. A8 of Ref.~\cite{Bozorgnia:2011tk},
\begin{align}
\hat{\mathbf{r}} &= \left[-r_n \sin(\lambda_{\rm lab})\cos (t^\circ_{\rm lab}) +r_w \sin (t^\circ_{\rm lab}) + r_z \cos(\lambda_{\rm lab})\cos (t^\circ_{\rm lab}) \right] \hat{\bf x}_e \nonumber\\
&+\left[-r_n \sin(\lambda_{\rm lab})\sin (t^\circ_{\rm lab}) -r_w \cos (t^\circ_{\rm lab}) + r_z \cos(\lambda_{\rm lab})\sin (t^\circ_{\rm lab}) \right] \hat{\bf y}_e \nonumber\\
&+\left[r_n \cos(\lambda_{\rm lab}) + r_z \sin(\lambda_{\rm lab}) \right] \hat{\bf z}_e,
\label{q-Equit}
\end{align}
where $\hat{\bf x}_e$,  $\hat{\bf y}_e$, and  $\hat{\bf z}_e$ are the unit coordinate vectors of the geocentric equatorial inertial (GEI) frame: its origin is at the center of the Earth, its $x_e$-axis points in the direction of the vernal equinox, its $y_e$-axis points to the point on the celestial equator with right ascension 90$^\circ$ (so that the cartesian frame is right-handed), and its $z_e$-axis points to the north celestial pole. $t^\circ_{\rm lab}$ is the laboratory Local Apparent Sidereal Time in degrees, and $\lambda_{\rm lab}$ is the latitude of the lab.

The transformation from the Galactic frame to the equatorial frame is given by

\begin{equation}
\left( \begin{array}{c}
\hat{\mathbf{x}}_e \\
\hat{\mathbf{y}}_e \\
\hat{\mathbf{z}}_e \end{array} \right)=
\textbf{A}_\textrm{G}
\left( \begin{array}{c}
\hat{\mathbf{x}}_g \\
\hat{\mathbf{y}}_g \\
\hat{\mathbf{z}}_g \end{array} \right),
\end{equation}
where
\begin{equation}
\textbf{A}_\textrm{G}=
\left( \begin{array}{ccc}
a_x & b_x & c_x \\
a_y & b_y & c_y \\
a_z & b_z & c_z \end{array} \right),
\end{equation}
and $\hat{\mathbf{x}}_g$, $\hat{\mathbf{y}}_g$, and $\hat{\mathbf{z}}_g$ are the unit vectors of the Galactic reference frame. Recall the definition of the Galactic coordinate system: its origin is at the position of the Sun, its $x_g$-axis points toward the Galactic center, its $y_g$-axis points in the direction of the Galactic rotation, and its $z_g$-axis points to the north Galactic pole. Note that these coordinates are related
to Galactic longitude $l$ and latitude $b$ by ($x_g, y_g, z_g$)=($\cos b \cos l, \cos b \sin l, \sin b$). We have $\hat{\mathbf{r}}=r_{xg}~\hat{\bf x}_g + r_{yg}~\hat{\bf y}_g + r_{zg}~\hat{\bf z}_g$, where
\begin{align}
r_{xg} &= r_n \Big(-\big[a_x \cos (t^\circ_{\rm lab}) + a_y \sin (t^\circ_{\rm lab})\big] \sin(\lambda_{\rm lab}) +a_z\cos(\lambda_{\rm lab}) \Big) + r_w \Big(a_x \sin (t^\circ_{\rm lab}) -a_y \cos (t^\circ_{\rm lab}) \Big) \nonumber\\
&+ r_z \Big(\big[a_x \cos (t^\circ_{\rm lab}) +a_y \sin (t^\circ_{\rm lab})\big] \cos(\lambda_{\rm lab}) +a_z\sin(\lambda_{\rm lab}) \Big), \nonumber\\
r_{yg} &=r_n \Big(-\big[b_x \cos (t^\circ_{\rm lab}) + b_y \sin (t^\circ_{\rm lab})\big] \sin(\lambda_{\rm lab}) +b_z\cos(\lambda_{\rm lab}) \Big) + r_w \Big(b_x \sin (t^\circ_{\rm lab}) -b_y \cos (t^\circ_{\rm lab}) \Big) \nonumber\\
&+ r_z \Big(\big[b_x \cos (t^\circ_{\rm lab}) +b_y \sin (t^\circ_{\rm lab})\big] \cos(\lambda_{\rm lab}) +b_z\sin(\lambda_{\rm lab}) \Big),\nonumber\\
r_{zg} &=r_n \Big(-\big[c_x \cos (t^\circ_{\rm lab}) + c_y \sin (t^\circ_{\rm lab})\big] \sin(\lambda_{\rm lab}) +c_z\cos(\lambda_{\rm lab}) \Big) + r_w \Big(c_x \sin (t^\circ_{\rm lab}) -c_y \cos (t^\circ_{\rm lab}) \Big) \nonumber\\
&+ r_z \Big(\big[c_x \cos (t^\circ_{\rm lab}) +c_y \sin (t^\circ_{\rm lab})\big] \cos(\lambda_{\rm lab}) +c_z\sin(\lambda_{\rm lab}) \Big).
\label{qg}
\end{align}

As done in Ref.~\cite{Hipparcos}, the transformation matrix $\textbf{A}_\textrm{G}$ can be found using the definition of the Galactic pole and center in the \textit{International Celestial Reference System} (ICRS). The north Galactic pole can be defined by right ascension $\alpha_G=192^\circ .85948$ and declination $\delta_G=+27^\circ .12825$ in the ICRS. The origin of Galactic longitude is defined by the Galactic longitude of the ascending node of the Galactic plane on the equator of ICRS, which is $l_\Omega=32^\circ .93192$. The angles $\alpha_G$, $\delta_G$ and $l_\Omega$ should be regarded as exact quantities, and they can be used to compute the transformation matrix $\textbf{A}_\textrm{G}$~\cite{Hipparcos},
\begin{equation}
\textbf{A}_\textrm{G}=
\left( \begin{array}{ccc}
-0.0548755604 & +0.4941094279 & -0.8676661490\\
-0.8734370902 & -0.4448296300 & -0.1980763734\\
-0.4838350155 & +0.7469822445 & +0.4559837762 \end{array} \right).
\label{AG-Hipparcos}
\end{equation}

The direction of $\hat{\mathbf{r}}$ in the Galactic rest frame can be specified by the Galactic longitude ($l_r$) and latitude ($b_r$) of $\hat{\mathbf{r}}$,
\begin{equation}
r_{xg}=\cos{b_r}\cos{l_r},~~~~
r_{yg}=\cos{b_r}\sin{l_r},~~~~
r_{zg}=\sin{b_r},
\end{equation}
where $r_{xg}$, $r_{yg}$, and $r_{zg}$ are given in terms of $r_n$, $r_w$, and $r_z$ in Eq.~\ref{qg}.

The velocity of the lab  with respect to the center of the Galaxy can be divided into four components: the Galactic rotation velocity ${\bf v}_c$ at the position of the Sun (or Local Standard of Rest (LSR) velocity), Sun's peculiar velocity ${\bf v}_{s}$ in the LSR, Earth's translational velocity ${\bf v}_{\rm {e, rev}}$ with respect to the Sun, and the  velocity of Earth's rotation around itself ${\bf v}_{\rm {e, rot}}$.
In the following sections, we will write each component of ${\bf v}_{{\rm lab}}$ in the Galactic frame and compute,
\begin{align}
\hat{\mathbf{r}} \cdot{\bf v}_{{\rm lab}}=\hat{\mathbf{r}} \cdot {\bf v}_c+ \hat{\mathbf{r}} \cdot {\bf v}_{s}+ \hat{\mathbf{r}} \cdot {\bf v}_{\rm {e, rev}}+ \hat{\mathbf{r}} \cdot{\bf v}_{\rm {e, rot}}.
\label{qdotVlab}
\end{align}

\subsection{Galactic rotation}

The velocity of the Galactic rotation ${\bf v}_c$ is defined in the Galactic reference frame,
\begin{equation}
{\bf v}_c=v_c \hat{{\bf y}}_g,
\label{GalacticRot}
\end{equation}
where $v_c$ is the Galactic rotation speed (i.e.~the local circular speed), and $\hat{{\bf y}}_g$ is in the direction of the Galactic rotation. We have,
\begin{equation}
\hat{\mathbf{r}} \cdot{\bf v}_c=r_{yg} v_c,
\label{qdotGalacticRot}
\end{equation}
where $r_{yg}$ is given in Eq.~\ref{qg}. The standard value of $v_c$ is 220 km s$^{-1}$ with an order 10\% statistical error~\cite{Bovy:2009dr,Bovy:2012ba,Kerr:1986hz}. Without assuming a flat rotation curve for the MW, there is a larger spread of values for $v_c$~\cite{McMillan:2009yr}.

\subsection{Solar motion}

The Sun's peculiar velocity in the LSR is,
\begin{equation}
{\bf v}_{s}=U \hat{{\bf x}}_g + V \hat{{\bf y}}_g +W \hat{{\bf z}}_g,
\label{Solar}
\end{equation}
and we have,
\begin{equation}
\hat{\bf r} \cdot{\bf v}_{s}=r_{xg}U + r_{yg}V + r_{zg} W.
\label{qdotSolar}
\end{equation}

Ref.~\cite{Schoenrich:2009bx} finds $(U,V,W)_\odot=(11.1^{+0.69}_{-0.75}, 12.24^{+0.47}_{-0.47}, 7.25^{+0.37}_{-0.36})$ km s$^{-1}$, with additional systematic uncertainties $\sim (1,2,0.5)$ km s$^{-1}$. These values are extremely insensitive to the metallicity gradient within the disk. 

\subsection{Earth's revolution}

The velocity of the Earth's revolution around the Sun including first order corrections from the eccentricity of the Earth's orbit and the precession of the equinoxes is given by~\cite{McCabe:2013kea} 
(see also Ref. \cite{Lee:2013xxa} for equivalent results but without including the effect of the precession),
\begin{align}
{\bf v}_{\rm {e, rev}}&=v_{\oplus} \Big( \cos\beta_x \left[ \sin (L-\lambda_x) + e \sin (L + g - \lambda_x) \right] \hat{{\bf x}}_g \nonumber\\
&+ \cos\beta_y \left[ \sin (L-\lambda_y) - e \sin (L+g - \lambda_y) \right] \hat{{\bf y}}_g \nonumber\\
&+ \cos\beta_z \left[ \sin (L-\lambda_z)  - e \sin (L+g - \lambda_z) \right] \hat{{\bf z}}_g\Big),
\label{EarthRev}
\end{align}
where $v_{\oplus}=29.79$ km s$^{-1}$ is the orbital speed of the Earth, $e=0.0167023$ is the ellipticity of the Earth's orbit~\cite{AstroAlmanac}, and $(\beta_i, \lambda_i)$ are the ecliptic latitudes and longitudes of the ($\hat{{\bf x}}_g$,$\hat{{\bf y}}_g$,$\hat{{\bf z}}_g$) axes (given below). The mean longitude of the Sun, $L$, and the mean anomaly of the Earth 
in its orbit around the Sun, $g$ are given by~\cite{AstroAlmanac},
\begin{align}
L &= 279^\circ.344 + 0.9856474 ~ d,\\
g &= 356^\circ.154 + 0.9856003 ~ d.
\end{align}
Here $d$ is the fractional day number from December 31st 2014 at 0 UT (Universal Time in hours). For example, on January 1st 2015 the fractional day number is $d=1$ at 0 UT, and $d=1.5$ at 12:00 UT. For a given calendar date with year $Y$, month $M$, day of month $D$, and time of the day UT (in hours), the fractional day number is given by~\cite{Meeus-2000}
\begin{equation}
d = \lfloor{365.25 ~\tilde{Y}}\rfloor + \lfloor{30.61 ~ (\tilde{M} + 1) }\rfloor + D + \frac{\textrm{UT}}{24} - 736041,
\end{equation}
where the floor function $\lfloor{x}\rfloor$ is the largest integer not greater than $x$. For $M=1$ or 2, $\tilde{Y}=Y-1$ and $\tilde{M}= M+12$. For $M>2$, $\tilde{Y}=Y$ and $\tilde{M}= M$. For example on 7th of March 2015 ($Y=2015$, $M=3$, $D=7$) at 10:00 UT, we have $d=66.42$.

The ecliptic latitudes and longitudes of the ($\hat{{\bf x}}_g$,$\hat{{\bf y}}_g$,$\hat{{\bf z}}_g$) axes are
\begin{align}
(\beta_x, \lambda_x) &= (5^\circ.538, 267^\circ.050) + (0^\circ.013, 1^\circ.397) T ,\\
(\beta_y, \lambda_y) &=  (-59^\circ.574, 347^\circ.546) + (0^\circ.002, 1^\circ.375) T ,\\
(\beta_z, \lambda_z) &=  (29^\circ.811, 180^\circ.234) + (0^\circ.001, 1^\circ.404) T,
\end{align}
where $T=d/36525$ is the epoch of date and accounts for the precession of the equinoxes. Ref.~\cite{McCabe:2013kea} finds that the consequences of ignoring the $T$ dependence of $\beta_i$ for the amplitude and peak time of the annual modulation are very small. The $T$ dependence becomes important over a timescale of decades from the reference time of December 31st 2014 at 0 UT in which $d$ is calculated.

We can compute $\hat{\mathbf{r}} \cdot{\bf v}_{\rm {e, rev}}$ as
\begin{align}
\hat{\mathbf{r}} \cdot{\bf v}_{\rm {e, rev}}&=v_{\oplus} \Big( \cos\beta_x \left[ \sin (L-\lambda_x) + e \sin (L + g - \lambda_x) \right]   r_{xg} \nonumber\\
&+ \cos\beta_y \left[ \sin (L-\lambda_y) - e \sin (L+g - \lambda_y) \right]  r_{yg} \nonumber\\
&+ \cos\beta_z \left[ \sin (L-\lambda_z)  - e \sin (L+g - \lambda_z) \right] r_{zg}\Big).
\label{qdotEarthRev}
\end{align}

\subsection{Earth's rotation}

The velocity of Earth's rotation around its axis is given by
\begin{equation}
{\bf v}_{\rm {e, rot}}=-v_{\rm {e, rot}}^{\rm eq} \cos \lambda_{\rm lab} \hat{\cal W},
\end{equation}
where $v_{\rm {e, rot}}^{\rm eq} $ is the  Earth's rotation speed at the equator, and is defined as $v_{\rm {e, rot}}^{\rm eq} =2 \pi R_{\oplus}/({\rm {1~ sidereal~ day}})$. The Earth's equatorial radius is $R_{\oplus}=6378.137$ km, and one sidereal day is 23.9344696 hr$=86164$ s. Therefore $v_{\rm {e, rot}}^{\rm eq} =0.465102$ km s$^{-1}$.

We use Eq. A8 of Ref.~\cite{Bozorgnia:2011tk} to write $\hat{\cal W}$ in terms of the equatorial frame coordinates,
\begin{equation}
{\bf v}_{\rm {e, rot}}=-v_{\rm {e, rot}}^{\rm eq} \cos \lambda_{\rm lab} \Big(\sin (t^\circ_{\rm lab})\hat{\bf x}_e-\cos (t^\circ_{\rm lab})~\hat{\bf y}_e\Big),
\end{equation}
Using the the Galactic to equatorial transformation, we have
\begin{align}
{\bf v}_{\rm {e, rot}}&=-v_{\rm {e, rot}}^{\rm eq} \cos \lambda_{\rm lab}  \Big\{ \Big(a_x \sin (t^\circ_{\rm lab}) -a_y \cos (t^\circ_{\rm lab})\Big) \hat{{\bf x}}_g\nonumber\\
&+\Big(b_x\sin (t^\circ_{\rm lab}) -b_y \cos (t^\circ_{\rm lab}) \Big) \hat{{\bf y}}_g +\Big(c_x\sin (t^\circ_{\rm lab}) -c_y \cos (t^\circ_{\rm lab})\Big) \hat{{\bf z}}_g \Big\}.
\end{align}
Thus we can compute $\hat{\mathbf{r}} \cdot {\bf v}_{\rm {e, rot}}$ as
\begin{align}
\hat{\mathbf{r}} \cdot{\bf v}_{\rm {e, rot}}&=-v_{\rm {e, rot}}^{\rm eq} \cos \lambda_{\rm lab} \Big\{ \Big(a_x \sin (t^\circ_{\rm lab}) -a_y \cos (t^\circ_{\rm lab})\Big) r_{xg} \nonumber\\
&+\Big(b_x\sin (t^\circ_{\rm lab}) -b_y \cos (t^\circ_{\rm lab}) \Big) r_{yg} +\Big(c_x\sin (t^\circ_{\rm lab}) -c_y \cos (t^\circ_{\rm lab})\Big) r_{zg}\Big\}.
\label{qdotEarthRot}
\end{align}
Inserting $r_{xg}$, $r_{yg}$ and $r_{zg}$ from Eq.~\ref{qg}, we find
\begin{align}
\hat{\mathbf{r}} \cdot{\bf v}_{\rm {e, rot}}&=-v_{\rm {e, rot}}^{\rm eq} \cos \lambda_{\rm lab} \Big\{ \Big([a_y^2 +b_y^2 +c_y^2 -a_x^2 -b_x^2 -c_x^2] \cos (t^\circ_{\rm lab}) \sin (t^\circ_{\rm lab}) \sin \lambda_{\rm lab} \nonumber\\
&+[a_x a_y +b_x b_y +c_x c_y]\left(\cos^2 (t^\circ_{\rm lab}) -\sin^2 (t^\circ_{\rm lab})\right) \sin \lambda_{\rm lab}\nonumber\\
&+ [a_x a_z +b_x b_z + c_x c_z] \sin (t^\circ_{\rm lab}) \cos \lambda_{\rm lab} + [a_y a_z + b_y b_z + c_y c_z] \cos (t^\circ_{\rm lab}) \cos \lambda_{\rm lab} \Big) r_n \nonumber\\
&+\Big([a_x^2+b_x^2+c_x^2]\sin^2 (t^\circ_{\rm lab}) +[a_y^2+b_y^2+c_y^2] \cos^2 (t^\circ_{\rm lab})\nonumber\\
&-2[a_x a_y +b_x b_y +c_x c_y] \cos (t^\circ_{\rm lab}) \sin (t^\circ_{\rm lab}) \Big) r_w \nonumber\\
&+ \Big([a_x^2 +b_x^2 +c_x^2 -a_y^2 -b_y^2 -c_y^2] \cos (t^\circ_{\rm lab}) \sin (t^\circ_{\rm lab}) \cos \lambda_{\rm lab} \nonumber\\
&+[a_x a_y +b_x b_y +c_x c_y]\left(\sin^2 (t^\circ_{\rm lab}) -\cos^2 (t^\circ_{\rm lab})\right) \cos \lambda_{\rm lab}\nonumber\\
&+ [a_x a_z +b_x b_z + c_x c_z] \sin (t^\circ_{\rm lab}) \sin \lambda_{\rm lab} - [a_y a_z + b_y b_z + c_y c_z] \cos (t^\circ_{\rm lab}) \sin \lambda_{\rm lab} \Big) r_z  \Big\}.
\label{qdotEarthRot-expand}
\end{align}
From Eq.~\ref{AG-Hipparcos}, we have $a_i a_j + b_i b_j + c_i c_j=\delta_{ij}$ where $i,j=x,y,z$. Therefore, Eq.~\ref{qdotEarthRot-expand} will be simplified to
\begin{equation}
\hat{\mathbf{r}} \cdot{\bf v}_{\rm {e, rot}}=-r_wv_{\rm {e, rot}}^{\rm eq} \cos \lambda_{\rm lab}.
\label{qdotEarthRot-simplified}
\end{equation}

\subsection{Total Velocity}
Using Eqs.~\ref{qdotGalacticRot}, \ref{qdotSolar}, \ref{qdotEarthRev} and \ref{qdotEarthRot-simplified} we can compute $\hat{\mathbf{r}} \cdot{\bf v}_{{\rm lab}}$ as
\begin{align}
\hat{\mathbf{r}} \cdot{\bf v}_{{\rm lab}}&=r_{xg} \Big(U + v_{\oplus} \cos\beta_x  \left[ \sin (L-\lambda_x) + e \sin (L + g - \lambda_x) \right] \Big)\nonumber\\
&+ r_{yg} \Big(v_c + V + v_{\oplus} \cos\beta_y \left[ \sin (L-\lambda_y) + e \sin (L + g - \lambda_y) \right]  \Big)\nonumber\\
&+ r_{zg} \Big(W + v_{\oplus} \cos\beta_z  \left[ \sin (L-\lambda_z) + e \sin (L + g - \lambda_z) \right]  \Big) -r_wv_{\rm {e, rot}}^{\rm eq} \cos \lambda_{\rm lab},
\label{qdotVlab-2}
\end{align}
where $r_{xg}$, $r_{yg}$, and $r_{zg}$ can be written in terms of $r_n$, $r_w$, and $r_z$ using Eq.~\ref{qg}. We can also write $\hat{\mathbf{r}} \cdot{\bf v}_{{\rm lab}}$ in terms of only  $r_{xg}$, $r_{yg}$, and $r_{zg}$ using Eqs.~\ref{qdotGalacticRot}, \ref{qdotSolar}, \ref{qdotEarthRev} and \ref{qdotEarthRot},
\begin{align}
\hat{\mathbf{r}} \cdot{\bf v}_{{\rm lab}}&=r_{xg} \Big(U + v_{\oplus} \cos\beta_x  \left[ \sin (L-\lambda_x) + e \sin (L + g - \lambda_x) \right]  \nonumber\\
&-v_{\rm {e, rot}}^{\rm eq} \cos \lambda_{\rm lab} \left(a_x \sin (t^\circ_{\rm lab}) -a_y \cos (t^\circ_{\rm lab})\right)\Big) \nonumber\\
&+ r_{yg} \Big(v_c + V + v_{\oplus} \cos\beta_y \left[ \sin (L-\lambda_y) + e \sin (L + g - \lambda_y) \right]  \nonumber\\
&-v_{\rm {e, rot}}^{\rm eq} \cos \lambda_{\rm lab} \left(b_x\sin (t^\circ_{\rm lab}) -b_y \cos (t^\circ_{\rm lab}) \right)\Big)\nonumber\\
&+ r_{zg} \Big(W + v_{\oplus} \cos\beta_z \left[ \sin (L-\lambda_z) + e \sin (L + g - \lambda_z) \right]  \nonumber\\ 
&-v_{\rm {e, rot}}^{\rm eq} \cos \lambda_{\rm lab} \left(c_x\sin (t^\circ_{\rm lab}) -c_y \cos (t^\circ_{\rm lab})\right)\Big).
\label{qdotVlab-galactic}
\end{align}
%

%
%--------------- BIBLIO
% 
%\newpage

\end{document}